\documentclass[11pt,cite,epsfig,psfrag]{article}
\usepackage[utf8]{inputenc}
\usepackage{placeins}
\usepackage{geometry}
 \usepackage{graphicx}
 \usepackage{subfig}
  \usepackage{amsmath} 
  \graphicspath{{images_dyson_achen_massive/}}
  \usepackage[usenames, dvipsnames]{color}
\usepackage{wrapfig}
\usepackage{boxedminipage}
\usepackage{multirow}

\usepackage{fullpage}
\usepackage{amsmath}
\usepackage{amssymb}
\usepackage{graphicx}
\usepackage{mathrsfs}
\usepackage{wrapfig}
\usepackage{boxedminipage}
\usepackage{epsfig}
\usepackage{setspace}
\usepackage{float}
\usepackage{hyperref}
\usepackage{enumerate}
\usepackage{cite}
\usepackage[font=footnotesize,labelfont=rm]{caption}

\usepackage[numbers,sort&compress]{natbib}

\def\be{\begin{equation}}
\def\ee{\end{equation}}
\def\bea{\begin{eqnarray}}
\def\eea{\end{eqnarray}}

\numberwithin{equation}{section}

 \newcommand{\RN}[1]{%
   \textup{\uppercase\expandafter{\romannumeral#1}}%
 }
\begin{document}

\thispagestyle{empty}

\vskip 2cm

\begin{center}
{\Large \bf Static spheres and Aschenbach effect for black holes in massive gravity}
\end{center}

\vskip .2cm

\vskip 1.2cm

\centerline{ \bf
Pavan Kumar Yerra \footnote{pk11@iitbbs.ac.in},
Sudipta Mukherji\footnote{mukherji@iopb.res.in} and
Chandrasekhar Bhamidipati\footnote{chandrasekhar@iitbbs.ac.in}
}

\vskip 7mm

\begin{center}{ $^{1}$ Institute of Fundamental Physics and Quantum Technology, \\ Ningbo University, Ningbo, Zhejiang 315211, China}	
\end{center}

\begin{center}{ $^{1}$ School of Physical Science and Technology, \\ Ningbo University, Ningbo, Zhejiang 315211, China}	
\end{center}

\begin{center}{ $^{1},^{2}$ Institute of Physics, Sachivalaya Marg, Bhubaneswar, Odisha, 751005, India}	
\end{center}

\begin{center}{$^{1},^{2}$ Homi Bhabha National Institute, Training School Complex, \\ Anushakti Nagar, Mumbai,  400085, India}	
\end{center}

\begin{center}{ $^{3}$ Department of Physics, School of Basic Sciences,\\ 
Indian Institute of Technology Bhubaneswar, Bhubaneswar, Odisha, 752050, India}
\end{center}

\vskip 1.0cm
\centerline{\bf Abstract}
\vskip 0.5cm
\noindent
 In this paper, we study the trajectories of massive and massless particles in four dimensional static and spherically symmetric black holes in de Rham-Gabadadze-Tolley (dRGT) massive gravity theory  via phase-plane analysis and point out several novel features. In particular, we show the existence of a static sphere, a finite radial distance outside the black holes in these theories, where a massive particle can be at rest, as seen by an asymptotic zero angular momentum observer.  Topological arguments show that the stable and unstable static spheres, which come in pairs, have opposite charges. In the presence of angular momentum, we first study the behaviour of massless particles and find the presence of stable and unstable photon spheres in both neutral and charged black holes. Subsequently, we study the motion of massive test particles around these black holes, and find one pair of stable and unstable time-like circular orbits (TCOs), such that the stable and unstable TCO's are disconnected in certain regions.
Computing the angular velocity $\Omega_{\text{\tiny CO}}$ of the TCOs, measured by a static observer at rest, shows the unusual nature of its monotonic increase with the radius of TCO, near the location of stable photon sphere. This confirms the existence of Aschenbach effect for spherically symmetric black holes in massive gravity, which was only found to exist in rapidly spinning black holes, with the only other exception being the rare example of gravity coupled to quasi-topological electromagnetism.

\newpage
\setcounter{footnote}{0}
\noindent

\baselineskip 15pt

\section{Introduction}

\noindent
Investigation of orbits of particles is ubiquitous for uncovering novel observable gravitational phenomena and testing various features of compact objects in Einstein gravity as well as modified theories of gravity. Applications range from situations involving strong and weak gravitational effects around black holes, such as ring down in a binary system of black holes\cite{LIGOScientific:2016aoc}, shadows\cite{EventHorizonTelescope:2022wkp,EventHorizonTelescope:2019dse}, and to other interesting phenomena in four and higher dimensions, with potential observational signatures \cite{Grandclement:2014msa,Grould:2017rzz,Teodoro:2020kok,Teodoro:2021ezj,Gibbons:1999uv,Herdeiro:2000ap,Diemer:2013fza,Delgado:2021jxd}. Stable circular orbits of massive particles around black holes can be used to study accretion discs, whereas unstable ones provide information about shadows. Certain physical quantities, such as Energy, angular  momentum and angular velocity are generally examined to extract information about the orbital dynamics~\cite{Stuchlik:1999qk,Claudel:2000yi,Virbhadra:1999nm,Claudel:2000yi,Virbhadra:2002ju,Levin:2008mq,Pugliese:2010ps,Hackmann:2010zz,Villanueva:2013zta,Wei:2017mwc,Chandrasekhar:2018sjg,Lehebel:2022yyz,Bozza:2002zj,Cardoso:2008bp,Hioki:2009na,Collodel:2017end,Ye:2023gmk,Cunha:2020azh,Cunha:2022gde,Wei:2022mzv}. \\

\noindent
Lately, novel methods have been proposed for investigation of orbits of particles around static spherically symmetric, as well as rotating compact objects, accoutred with topological and geometrical techniques\cite{Lehebel:2022yyz,Bozza:2002zj,Cardoso:2008bp,Hioki:2009na,Collodel:2017end,Ye:2023gmk,Cunha:2020azh,Cunha:2022gde,Wei:2022mzv}. In other developments, with a view to unveiling unique signatures which could distinguish rotating objects such as Kerr black holes from other non-Kerr spacetimes (such as boson stars, wormholes etc.,), the phenomena of static spheres has also been studied\cite{Collodel:2017end}. The idea is as follows. In a rotating space-time, particles can have orbits which are non-rotating, co-rotating or counter-rotating, depending on the sign and magnitude of their angular momentum. The later type of orbits are interesting, as they have negative angular momentum and might appear static, to an asymptotically static observer. Of course, the angular momentum of the particle has to be highly fine tuned, in addition to its location being the minimum of the potential. In static and spherically symmetric space-times, obtaining such static spheres is not straightforward, as massive test particles with zero angular momentum generally cannot be at rest around black holes under general conditions. For instance, the angular velocity of the object has to vanish at a finite radius, which does not occur in Einstein gravity coupled to standard matter, even for low black hole spins. However, motivated by recent studies on orbits around various black hole space-times\cite{Lehebel:2022yyz,Bozza:2002zj,Cardoso:2008bp,Hioki:2009na,Collodel:2017end,Ye:2023gmk,Cunha:2020azh,Cunha:2022gde,Wei:2022mzv}, it was shown that such static spheres do exist in a rare example of gravity coupled to quasi-topological electromagnetism \cite{Liu:2019rib,Wei:2023bgp}, which can mimic Dyson sphere \cite{Dyson:1960xib} (originally constructed around stellar structures\cite{Papagiannis,Wright}). Considering that finding such static spheres in spherically symmetric gravitational systems could put to test various modified theories of gravity, it is imperative to look for further examples where such phenomena could occur.\\

\noindent
Another related development which could be worthwhile to explore, especially, for testing modified theories of gravity, concerns an intriguing effect found by Aschenbach\cite{Aschenbach:2004kj}, coming from the study of radial gradient of orbital velocity around rotating black hole space-times. Let us elaborate. Generally, in Newtonian mechanics, or even while studying orbits of massive particles around spherically symmetric and/or rotating black holes, the angular velocity increases monotonically with decreasing radius. This behaviour also continues to hold for small values of rotation $a$ and large orbit radius $r$. What was found in\cite{Aschenbach:2004kj} is that, for rapidly spinning black holes with $a > 0.9953$ (but still below the Thorne limit\cite{Thorne:1974ve} $a=0.998$), the radial gradient of orbital velocity can decrease with decreasing $r$ in a narrow window of orbital radius, before resuming its normal increasing behaviour. This effect is very specific to zero angular momentum observers (ZAMOs) or so called Bardeen observers, and the results appear to be commensurate with the high frequency quasi-periodic oscillations originating in the accretion disk of compact objects, such as from X-ray binaries.  Considering the possibility of experimental observations, the Aschenbach effect has received increased attention with investigations extended to more general Kerr black holes\cite{Stuchlik:2004wk,Mueller:2007tf,Stuchlik:2011sw}, as well as when the particles are charged\cite{Tursunov:2016dss}, and/or rapidly spinning \cite{Khodagholizadeh:2020sex,Vahedi:2021ssf}.\\

\noindent
Now, since the Aschenbach effect occurs for test particles in orbits around rapidly spinning black holes\cite{Aschenbach:2004kj,Stuchlik:2004wk}, as observed by ZAMOs, it was not expected to be present when the black hole is non-rotating (or slowly rotating). At least, this is what can be deduced for spherically symmetric black holes in Einstein gravity. However, recently, in the aforementioned model of Einstein gravity coupled to nonlinear electrodynamics, the Aschenbach effect was found to be present~\cite{Wei:2023bgp}, aided by the occurrence of static spheres in a certain narrow range of charge to mass ratio of the black hole. As emphasised earlier, due to the possibility of potential experimentally observational signatures, the Aschenbach effect should be explored further, in more general settings, which could give rise to clues to the possible nature of the compact object. 
In this paper, we find a novel example where the Aschenbach effect can be present in a theory of massive gravity, which contains static and spherically symmetric black holes, with a rich structure of stable/unstable circular orbits of test particles. Our study is also expected to be helpful while exploring various lower bounds on physical quantities related to orbits of black holes\cite{Hod:2018kql}, particularly, in massive gravity.\\

\noindent
Following are some broad motivations for considering a massive theory of gravity.
Einstein's gravity has had high success with experimental observation of several of its predictions, buoyed further with the most recent evidences coming from the LIGO collaboration~\cite{LIGO2017,deRham2014Review} pertaining to gravitational wave signals. Notwithstanding these triumphs, there continue to be other phenomena of paramount importance, foremost being the accelerated expansion of the universe, and the longstanding cosmological constant problem, among others, necessitating exploration of modified theories of gravity. One of the possibilities explored concerns massive graviton theories, looked into with motivation from hierarchy problems and strong connections with quantum theory of gravity~\cite{MassiveIb,MassiveIc}, which is within the ambit of recent observations~\cite{Abbott}, where novel limits on the graviton mass are obtained. We should mention that, massive theories of gravity have a good history, with the first models being constructed as early as 1939 by Firez and Pauli~\cite{Fierz1939}. These models of course have now undergone considerable modifications, most notably the advent of new massive gravites\cite{BDghost,Newmasssive,dRGTI,dRGTII} which are being investigated actively\cite{NewM1,NewM2,NewM3,NewM4,NewM5,HassanI,HassanII}. Black hole solutions in these theories have interesting thermodynamic phases~\cite{BHMassiveI,BHMassiveII,BHMassiveIII,BHMassiveIV} with wide ranging cosmology/astrophysics applications, which should be helpful in identifying deviations from Einstein's theory~\cite{Katsuragawa,Saridakis,YFCai,Leon,Hinterbichler,Fasiello,Bamba} . There are considerable advantages in constructing massive theories of gravity~\cite{Vegh} (see also the reviews~\cite{Hinterbichler:2011tt, deRham:2014zqa}), which have the necessary ingredients for addressing some of the aforementioned issues as elaborated in~\cite{Gumrukcuoglu,Gratia,Kobayash,DeffayetI,DeffayetII,DvaliI,
DvaliII,Will,Mohseni,GumrukcuogluII,NeutronMass,Ruffini,EslamPanah:2018evk}: to mention one, it may be possible to understand 
the current observations emanating from dark matter
\cite{Schmidt-May2016DarkMatter} and also correlating with the accelerating
expansion of our universe where the requirement of dark energy component could be relaxed
\cite{MassiveCosmology2013,MassiveCosmology2015}. Let us also mention in passing, the continued efforts in embedding massive theories of gravity in the string theory and holographic framework~\cite{MGinString2018,Geng:2020qvw,Geng:2020fxl}. Some recent developments in general theories of gravity are also on the thermodynamic front, where the cosmological constant is taken to a variable, giving rise to the concept of pressure and black hole chemistry program (see\cite{Ahmed:2023snm} and references therein), which has been explored extensively in massive gravity theories~\cite{Cai2015,PVMassV,PVMassIV,Alberte,Zhou,Dehyadegari,Magmass,Dehghani:2019thq,Hendi:2015eca,Akbarieh:2021vhv}. The issue of black hole microstructures in these theories have also been explored from the point of view of thermodynamic geometry with interesting results~\cite{Yerra:2020tzg,Yerra:2020oph,Yerra:2021hnh,Hogas:2021saw,Caravano:2021aum,Chabab:2019mlu,Wu:2020fij}.\\

\noindent
With regard to the massive gravity theories considered in this work, 
we focus on the black hole solutions in the dRGT massive gravity theory in a four dimensional Einstein-Maxwell gravity coupled to nonlinear interaction terms giving the graviton a mass $m_g$~\cite{deRham:2010kj,Ghosh:2015cva,Hendi:2022qgi}. This theory  includes a non dynamical reference metric $f_{\mu\nu}$, which breaks diffeomorphism invariance and may be singular~\cite{Vegh}.  However, the theory has several useful features of stability and absence of ghosts\cite{HZhang}, together with the presence of black holes\cite{PVMassI,PVMassII,PVMassIII,PVMassIV,PVMassV}. In particular, this set up is well suited for holographic discussions of momentum dissipation, as the framework in~\cite{Vegh} describes a class of strongly coupled quantum field theories with broken translational symmetry, due to the Lorentz breaking graviton mass term.
The later aspect is helpful, as due to this, the mass of the graviton can play a role similar to the one played by lattice in the holographic models of conductors.  The conductivity in general possesses a Drude peak which in the massless gravity limit tends to a delta function. Thus, non-linear massive gravity theories require an auxiliary reference metric 
(for giving mass to gravitons) to yield a Boulware-Deser ghost-free theory~\cite{deRham2014Review,Hinterbichler:2011tt,HassanII}. In principle, a special theory of massive gravity for each choice of reference metric can be constructed, which is non-degenerate~\cite{HassanII}. In the context of holographic duality, massive gravity theories in AdS space with a singular (degenerate) reference metric are helpful to model 
a class of strongly interacting quantum field theories having broken translational symmetry (that is, momentum dissipation). For instance, one can build a holographic model for normal conductors (with momentum dissipation) and finite DC conductivity~\cite{Vegh,Davison:2013jba}, in contrast with massless gravity theories,  such as,  Einstein and with higher derivatives, with infinite DC conductivity ~\cite{Hartnoll:2008vx,Hartnoll:2008kx,Gregory:2009fj,Barclay:2010up}. Furthermore,  massive gravity theories with the feature of a singular reference metric can effectively capture
different phases of known condensed matter systems having broken translational symmetry, such as solids, liquids and (perfect)
fluids~\cite{Alberte,Alberte:2015isw,Alberte:2017oqx}, with interesting applications to cosmological~\cite{Zhang:2019oes} and other holographic situations~\cite{Vegh,Davison:2013jba,Blake:2013bqa,Cao:2015cza,Baggioli:2014roa}.  Of course, for a more viable description of massive gravity, the reference metric needs to be invertible and dynamical, which is the program pursued in bimetric gravity theories~\cite{Schmidt-May2016DarkMatter,MassiveCosmology2013,MassiveCosmology2015,Hogas:2021saw,Caravano:2021aum} (see also~\cite{Gialamas:2023aim,Gialamas:2023lxj,Gialamas:2023fly}). We should mention that, there are alternatives to using a reference metric $f_{\mu\nu}$, namely, the St\"{u}ckelberg fields~\cite{dRGTII}, where the holographic conductivity can be studied by regarding the fields as dynamical\cite{Alberte}. In the St\"{u}ckelberg formalism of massive gravity, the diffeomorphism invariance seems to be restored. The methods and techniques developed in this work though are very general, and can be carried over to study these interesting massive gravity theories.\\

\noindent
Rest of the paper is organised as follows. In section-(\ref{2}), we present a brief overview of the static and spherically symmetric black holes in massive gravity in four dimensions. These black holes, often endowed with multiple horizons, have rich geometric structures.  Consequently, motions of massive and massless particles around these black holes possess diverse trajectories, including that of the static spheres. To explore the nature of these  trajectories, in section-(\ref{3}), we develop a phase-plane analysis of the system. We find the fixed points and complete  phase portraits, dictated by those fixed points. Owing to the fact that the black holes in massive gravity have several free parameters, our study in this section is not exhaustive. We fix some of these parameters and vary the rest in a way that captures qualitatively different geodesics. Subsequent section is devoted to examining one class of these geodesics associated with the massive particles, namely the  static spheres. 
In section-(\ref{4.1}), we present the conditions for the existence of circular orbits for massive test particles when the orbital angular momentum vanishes, and then proceed to count the number of such static spheres, based on topological arguments. Exploiting the presence of static spheres and stable photon spheres in the massive gravity, in section-(\ref{4.3}), we study the Aschenbach effect, where the angular velocity of a time-like circular orbit is found to be increasing with its radius coordinate under certain conditions.  Finally, we conclude in section-(\ref{5}).

\section{Static \& spherically symmetric black holes in massive gravity} \label{2}
We consider the dRGT massive gravity theory, where the action for a four dimensional Einstein-Maxwell gravity coupled to nonlinear interaction terms giving the graviton a mass $m_g$, can be written as~\footnote{We set the Newtons gravitational constant $G$ and the speed of light $c$ to be $G=c=1$.}~\cite{deRham:2010kj,Ghosh:2015cva,Hendi:2022qgi}:
	\begin{equation}\label{Act:dRGT}
	I=~\frac{1}{16\pi}\int d^{4}x \sqrt{-g}\Big(\mathcal{R} - \mathcal{F} +m_{g}^{2}\, \mathcal{U}(g,\Phi^{a})\Big)\,.
	\end{equation}
Here, $\mathcal{R}$ is the  Ricci scalar, and the Maxwell invariant $\mathcal{F} = F_{\mu\nu}F^{\mu\nu}$  with $F_{\mu\nu}= \partial_\mu A_\nu - \partial_\nu A_\mu$ is given in terms of the gauge potential $A_\mu$.  $\mathcal{U}$ is the effective potential for graviton, and it has the following
expression in four dimensional spacetime: 
\begin{equation*}
	\mathcal{U}(g,\Phi^{a})=~\mathcal{U}_{2}+\alpha_{3}\mathcal{U}_{3}+\alpha_{4}\mathcal{U}_{4},
\end{equation*}	
where, $ \alpha_{3}$ and $\alpha_{4} $ are dimensionless parameters which will be replaced by two new parameters, $ \alpha$ and $\beta $, such that $ \alpha_{3}=\frac{\alpha-1}{3} $ and $ \alpha_{4}=\frac{\beta}{4}+\frac{1-\alpha}{12} $. 
The $\mathcal{U}_i$'s can be expressed as
\begin{eqnarray}
\mathcal{U}_{2} &\equiv &[\mathcal{K}]^{2}-[\mathcal{K}^{2}],  \notag \\
\mathcal{U}_{3} &\equiv &[\mathcal{K}]^{3}-3[\mathcal{K}][\mathcal{K}^{2}]+2[%
\mathcal{K}^{3}],  \notag \\
\mathcal{U}_{4} &\equiv &[\mathcal{K}]^{4}-6[\mathcal{K}]^{2}[\mathcal{K}%
^{2}]+8[\mathcal{K}][\mathcal{K}^{3}]+3[\mathcal{K}^{2}]^{2}-6[\mathcal{K}%
^{4}],
\end{eqnarray}%
in which
\begin{equation*}
\mathcal{K}_{\,\,\,\nu }^{\mu }=\delta _{\nu }^{\mu }-\sqrt{g^{\mu \sigma
	}f_{ab}\partial _{\sigma }\Phi ^{a}\partial _{\nu }\Phi ^{b}},
\end{equation*}%
where $f_{ab}$ is a reference metric, while,  $[\mathcal{K}]=\mathcal{K}%
_{\,\,\,\mu }^{\mu }$ and $[\mathcal{K}^{n}]=(\mathcal{K}^{n})_{\,\,\,\mu
}^{\mu }$ denote the traces.  Here $ \Phi^{a} $ are the St\"{u}ckelberg scalars introduced to preserve the general covariance of the theory.
\vskip 0.2cm
\noindent The above action admits static and spherically symmetric black hole solutions with the line element and reference metric, given respectively as:
\begin{eqnarray}\label{eq:bh_metric}
ds^2 &=& -f(r)dt^2 +\frac{dr^2}{f(r)} +r^2(d\theta^2 + \sin^2\theta d\varphi^2)\, , \\
f_{\mu \nu } &=& diag\left( 0,0,h^{2},h^{2}\sin ^{2}\theta \right) ,
\end{eqnarray}
where $h$ is a positive constant. Using an ansatz for the gauge
potential   $A_\mu = \big(A(r),0,0,0 \big)$, the lapse function $f(r)$ can be obtained to be
	\begin{equation}\label{eq:f_gen}
	f(r)=~1-\frac{2M}{r}+\frac{Q^{2}}{r^{2}}+\frac{\Lambda}{3}r^{2}+\gamma r +\zeta \, ,
	\end{equation}
	where $ M $ is the mass and $Q$ is the charge of the black hole, and the other parameters are defined as
		\begin{eqnarray}
	\Lambda &= & 3m_{g}^{2}(1+\alpha+\beta)\, , \nonumber	\\
	\gamma &=& -h m_{g}^{2}(1+2\alpha+3\beta)\, , \nonumber	\\
		\zeta &=& h^{2}m_{g}^{2}(\alpha+3\beta) \,.
	\end{eqnarray}
	It was noted in~\cite{Ghosh:2015cva} that, in certain parameter space of $(\alpha,\beta)$, the charged (neutral) black hole can have at most  four (three) event horizons. The asymptotic structure of the solution is determined by the signs of $\Lambda$ and $\gamma$. The parameter $\Lambda$ plays the role of cosmological constant. In the absence of graviton mass ($m_g=0$), one can recover the  standard Reissner-Nordstrom black hole solution. It should be mentioned that for a suitable choice of parameters $(\alpha, \beta)$, and a reference metric, one can have de-Sitter/anti-de Sitter solutions, together with global monopole solutions. For further details regarding the possible space of solutions, one can refer~\cite{Ghosh:2015cva}.  
\section{Phase portraits and circular orbits of massive and massless particles} \label{3}
In this section, we aim to classify the trajectories of  massless as well as 
 massive test particles around  neutral and charged  black holes of massive gravity and, subsequently, study related  phase portraits. Owing to the fact that
the geometry has too many free parameters, namely, $M, Q, m_g, h, \alpha, \beta$, a comprehensive
study of phase portraits will be beyond the scope of this section. In what follows, we 
fix $M, Q, m_g$ and $h$ and  vary
 $\alpha, \beta$. The ranges are  chosen appropriately to capture 
 distinct trajectories of the test particles. \\
 
\noindent
We start with the geodesic equation for the line element~\eqref{eq:bh_metric}, that is
\begin{equation}
\epsilon =-g_{\mu\nu}\dot{x}^\mu \dot{x}^\nu,
\end{equation}
where  the dot  denotes a derivative with respect to an affine parameter which we will call $\tau$, and  $\epsilon$ takes values $1(0)$, for massive (massless) test particles.
Considering the equatorial geodesics with $\theta = \pi/2$, and employing the conserved quantities  associated with the symmetries of the spacetime (i.e., the energy $ E = - g_{tt}\dot{t}$, and the orbital angular momentum $L = g_{\varphi \varphi} \dot{\varphi}$ of the test particle~\footnote{For massive test particle, $E$ and $L$ denote the conserved quantities per unit mass of the test particle.}), the above geodesic equation reduces to
\begin{equation}\label{eq:geo_gen}
\dot{r}^2 + V_{\text{eff}} = 0,
\end{equation}
where the effective potential $V_{\text{eff}}$ is,
\begin{equation}\label{eq:veff_gen}
 V_{\text{eff}} = f(r)\bigg(\frac{L^2}{r^2} + \epsilon\bigg) - E^2.
\end{equation}
\noindent
In the next two subsections, we examine (\ref{eq:geo_gen}), first for the photon around neutral and charged black hole, and then repeat the same exercise for a massive particle. 

\subsection{Fixed points and phase portraits for massless particle}\label{3.1}

To study photon trajectories around a neutral black hole, we start with explicit form of (\ref{eq:geo_gen}),
\begin{equation}
\dot r^2 = E^2 -\Big(1 - \frac{2M}{r} + \frac{\Lambda}{3} r^2 + \gamma r + \zeta \Big)\frac{L^2}{r^2}.
\end{equation}
Since
\begin{equation}
\frac{d\phi}{d\tau} = \frac{L}{g_{\phi\phi}} = \frac{L}{r^2},
\end{equation}
we write
\begin{eqnarray}
\Big(\frac{dr}{d\phi}\Big)^2 &=&\Big(\frac{dr}{d\tau}\Big)^2 \Big(\frac{d\tau}{d\phi}\Big)^2\nonumber\\
&=& \frac{r}{L^2} \Big( E^2 r^3 - (r - 2 M + \frac{\Lambda}{3} r^3 + \gamma r^2 + \zeta r)L^2\Big).
\end{eqnarray}
It is useful to introduce a new variable $x = 1/r$. The governing equation now takes a simple form
\begin{equation}
\Big(\frac{dx}{d\phi}\Big)^2 = \frac{E^2}{L^2} - (x^2 - 2 M x^3 +\frac{ \Lambda}{3} + \gamma x + \zeta x^2).
\label{eq:dxdp}
\end{equation}
Differentiating once more, we get
\begin{equation}
\frac{d^2x}{d\phi^2} = - (1 + \zeta) x + 3 M x^2 - \frac{\gamma}{2}.
\end{equation}
It is convenient to rewrite the above equation as a pair of first order differential equations
\begin{eqnarray}\label{deq}
&&\frac{dx}{d\phi} = z,\nonumber\\
&& \frac{dz}{d\phi} =  - (1 + \zeta) x + 3 M x^2 - \frac{\gamma}{2}.
\label{eq:int}
\end{eqnarray}
The appearance of the phase portrait is controlled by the fixed points. 
These points are obtained by setting the right hand sides to zero. We get, therefore,
\begin{eqnarray}\label{fpeqone}
&&A: ( z,  x) = (0, \frac{1}{6M}(1 + \zeta - \sqrt{1 + 6 M\gamma + 2 \zeta + \zeta^2})),~~{\rm and} \nonumber\\
&&B: ( z, x) = (0, \frac{1}{6M}(1 + \zeta + \sqrt{1 + 6 M \gamma + 2 \zeta + \zeta^2})),
\label{eq:fp2}
\end{eqnarray}
as the two fixed points. Note that the reality condition implies
\begin{equation}\label{real_cond}
1 + 6 M \gamma + 2 \zeta + \zeta^2 \ge 0.
\end{equation}
In terms of parameters $(\alpha, \beta)$ it takes the form
\begin{equation}\label{realcondb}
1 + 2 h^2 m_g^2 (\alpha + 3 \beta) + h^4 m_g^4 (\alpha+3 \beta)^2 - 6 h M m_g^2(1 + 2 \alpha + 3 \beta)\ge 0.
\end{equation}
Solving (\ref{eq:fp2}) and (\ref{eq:dxdp}) simultaneously, we can easily find the energies associated with
the fixed points in terms of $M, L, \gamma$ and $\zeta$.\\

\begin{figure}[h!]
\centering
\includegraphics[width=5in]{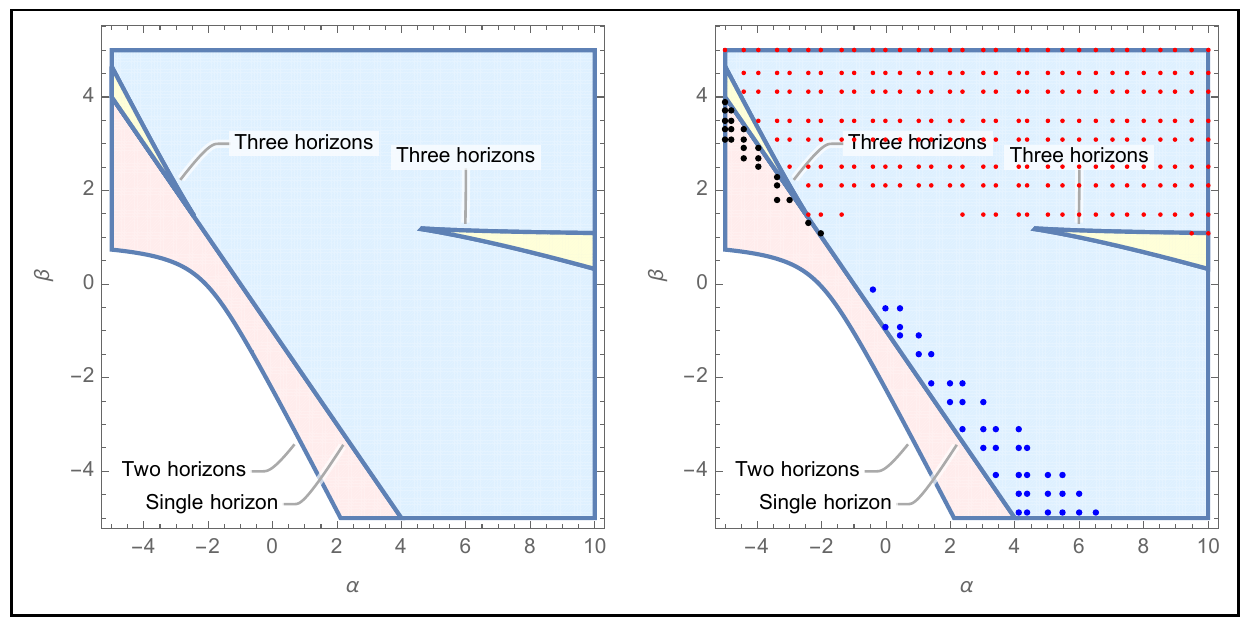}
\caption{
Left:  Number of horizons as we vary
$\alpha$ and $\beta$  while keeping $M = m_g= h = 1, Q=0$ .  There could be at most  three horizons.
Right: Locations and the nature of the fixed points are shown. Region marked with blue dots represents
parameter values for which the fixed point $B$ is outside the horizon. Similarly, for the one marked by the
black dots, fixed point $A$ is outside the outer horizon. Finally both $A,B$ are both outside in the region marked by the
red dots. For other regions with no dots, either fixed points  are not real or lie behind the 
outer horizon. These plots are consistent with the constraint given in (\ref{realcondb}).}
\label{fig:horizon}
\end{figure}

\begin{figure}[h!]
\centering
\includegraphics[width=4in]{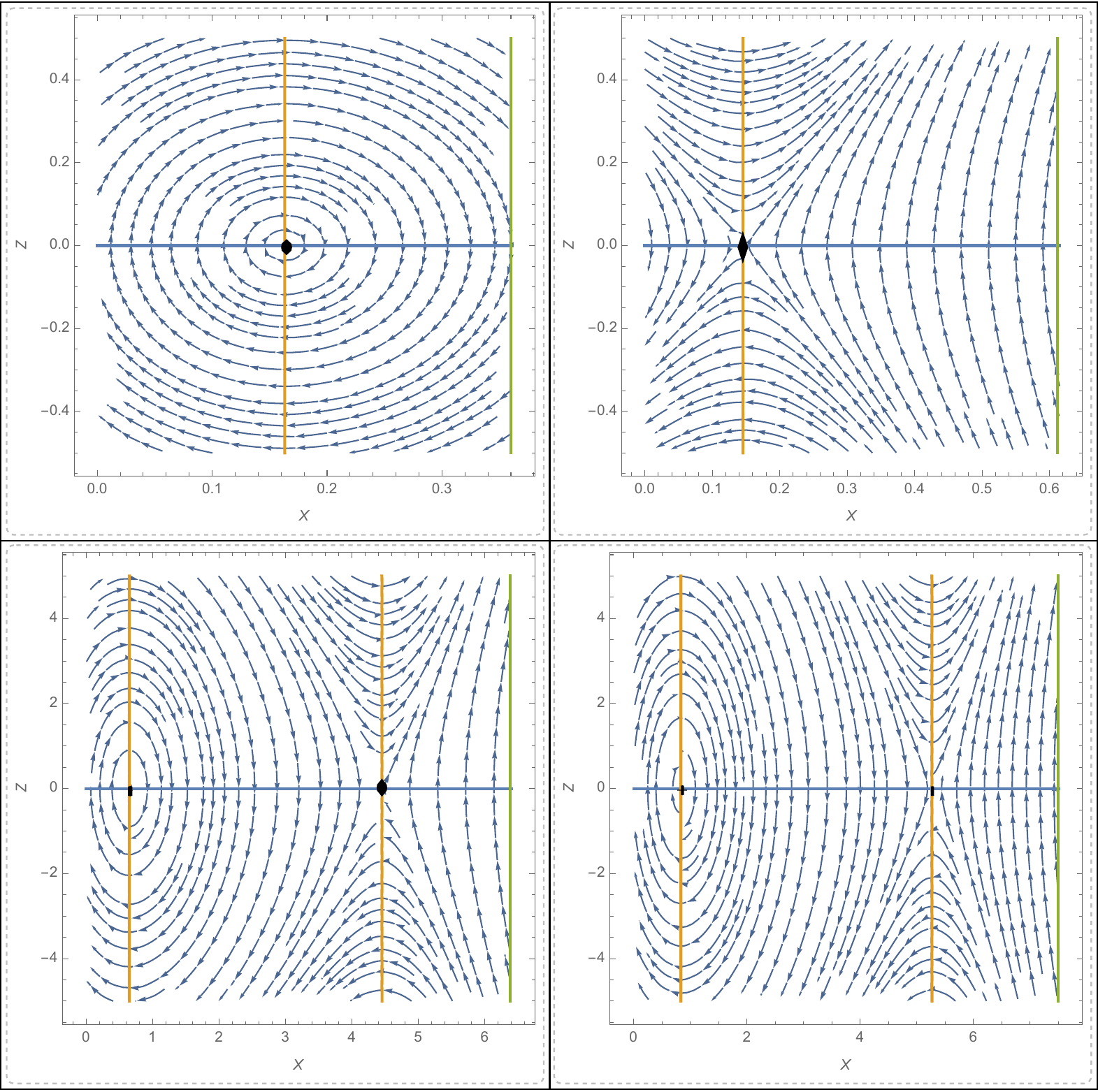}
\caption{The figures show the phase plane portraits of massless particle outside an  uncharged black hole. We vary 
$(\alpha, \beta)$ and keep $M = h = m_g =1, Q=0$. Top left: the fixed point outside the horizon is a center producing
closed orbits. We have chosen $(\alpha, \beta) = (-4, 2.9)$. This point lies in the black-dotted region in Fig.
(\ref{fig:horizon}). 
Top right: the fixed point now is a saddle which lies in the blue-dotted region of Fig.(\ref{fig:horizon}). We have taken
$(\alpha, \beta) = (2,-2)$. Bottom row: the plots are for $(\alpha, \beta) = (2, 4.1)$ (left) and $(8, 3.1)$ (right), chosen from the red-dotted region of Fig. (\ref{fig:horizon}). In this region,  both the fixed points are outside with the saddle 
being nearer to the horizon. In all these figures, the vertical lines drawn at extreme right represent the 
location of the outer horizon.} 
\label{fig:trajectories}
\end{figure}

\noindent
Since we also  require one or both the fixed points to lie outside the black hole horizon(s), additional constraints on the parameters appear. This is what we do next. Keeping $M, m_g, h$ fixed and varying $\alpha$,  $\beta$, we look for the number of horizons as well as the  fixed points outside the black hole. Our finding is summarised in Fig. (\ref{fig:horizon}).
\vskip 0.2cm
\noindent Natures of the fixed points are encoded in the  Jacobian matrix following from (\ref{deq}). This is given by
\begin{equation}
J = \begin{bmatrix} 
\frac{\partial x'}{\partial x}&\frac{\partial x\prime}{\partial z}\\
\frac{\partial z'}{\partial x}&\frac{\partial z'}{\partial z}\\ 
\end{bmatrix} = \begin{bmatrix} 
0&1\\
-(1+\zeta) + 6 M  x &0\\ 
\end{bmatrix} .
\end{equation}
Evaluating at the fixed points $A$ or $B$ gives
\begin{equation}
J  = \begin{bmatrix} 
0&1\\
\mp \sqrt{1 + 6 M \gamma + 2 \zeta + \zeta^2}&0\\ 
\end{bmatrix} \, ,
\end{equation}
respectively. Eigen values of this matrix determine the nature of the orbits. These are given by
\begin{eqnarray}
&&{\rm at} ~A:~ \pm i (1 + 6 M \gamma + 2 \zeta + \zeta^2)^{\frac{1}{4}}\nonumber\\
&&{\rm at} ~B:~\pm (1 + 6 M \gamma + 2 \zeta + \zeta^2)^{\frac{1}{4}}.
\end{eqnarray}
While purely imaginary values (as the quantity under fourth  root is
positive (\ref{real_cond})) indicate that $A$ is a center, $B$ produces a saddle point. The complete phase portraits are obtained then by integrating (\ref{eq:int}) and shown in Fig. (\ref{fig:trajectories}). 

\begin{figure}[h!]
\centering
\includegraphics[width=3in]{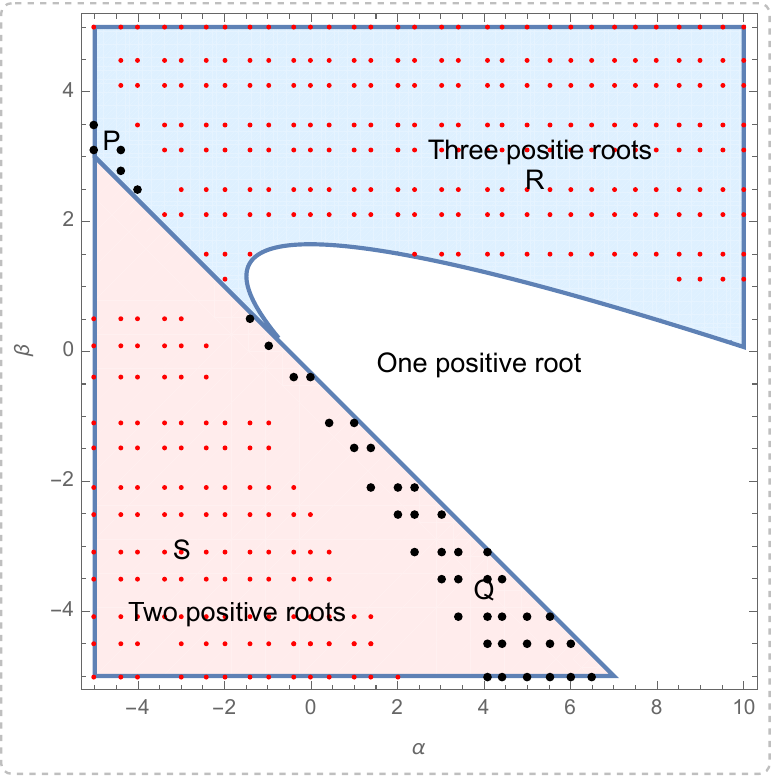}
\caption{Number of positive
roots of the right equation of (\ref{eq:root_Q})  are shown in the  
$(\alpha, \beta)$ plane using three distinct colours.  Further, the regions with black dots $(P ~{\rm and}~Q)$  hold only one  of the fixed points outside 
the outer horizon. While in region $P$, the node is a center, a saddle appears at $Q$. Two fixed points are located outside the outer horizon in the other two  dotted regions $(R ~{\rm and}~S)$. In $R$, the saddle node is closer to the horizon and in $S$, the center.  In the rest of the parameter space, either no fixed point arises outside the outer horizon or they are complex. We have set $M = h = m_g = 1$, and $Q = 0.2$.
}
\label{fig:fp_Q}
\end{figure}

\noindent
Now we move to the charged black hole. Except that the charge parameter makes the computation a bit messy, the analysis is same as before. We would therefore keep our discussion brief. The phase-plane equations  now become 
\begin{eqnarray}
&&\frac{dx}{d\phi} = z,\nonumber\\
&& \frac{dz}{d\phi} = -(1+\zeta) x + 3 M x^2 - 2 Q^2 x^3 - \frac{\gamma}{2}.
\label{fpeq}
\end{eqnarray}
The fixed points are represented by the solutions of 
\begin{equation}\label{eq:root_Q}
z =0,~{\rm and}~ -(1+\zeta) x + 3 M x^2 - 2 Q^2 x^3 - \frac{\gamma}{2} = 0.
\end{equation}
The second equation could have at most three positive roots, providing us at most three fixed points. 
We also require to find the fixed points that lie outside the outer most horizon. Both are shown in Fig. (\ref{fig:fp_Q}). 
\\

\begin{figure}[h!]
\centering
\includegraphics[width=4in]{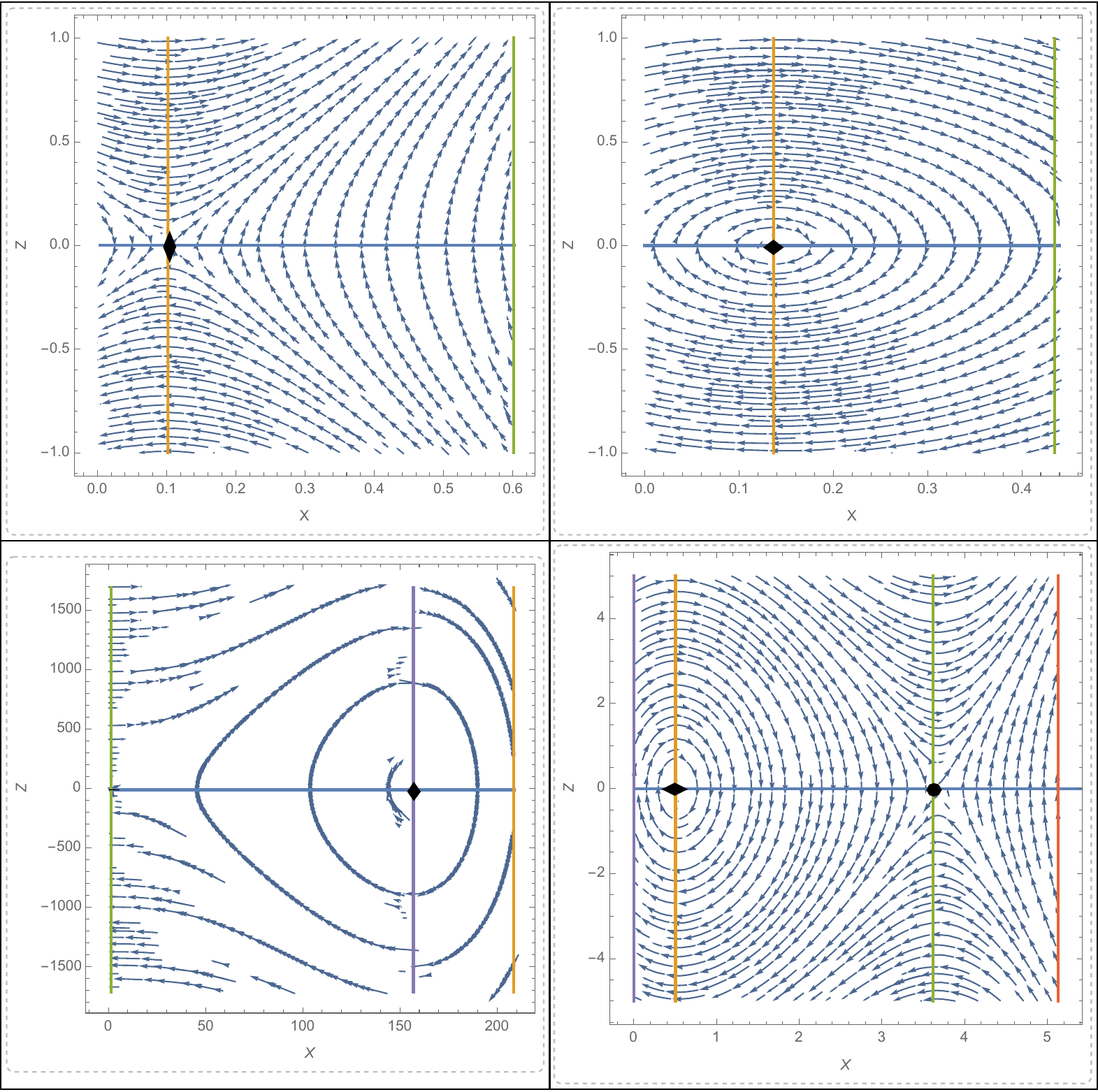}
\caption{These figures show the phase plane portraits of massless particle outside a charged black hole as we vary $(\alpha, \beta)$ keeping $M = h = m_g =1$ and $Q = 0.2$. Top left: The fixed point outside 
the horizon is a saddle with $\lambda_\pm = \pm 2.625$ for $(\alpha, \beta) = (5,-4.1)$. This is chosen from the region $Q$ of figure (\ref{fig:fp_Q}). Top right: the node is now a center with $\lambda_\pm = \pm 2.254 i$. 
The values of $(\alpha, \beta) = (-4.4, 3.1)$ lie in $P$ of figure (\ref{fig:fp_Q}) . 
Bottom left: it  is for 
$(\alpha, \beta) = (-4, -4.1)$.  
While the node closer to the horizon is a center with $\lambda_\pm = \pm 22.56 i$, the  other one is a saddle with $\lambda_\pm = \pm 4.32$. This illustrates the region $R$ of figure (\ref{fig:fp_Q}).
Bottom right: Here  $(\alpha, \beta) = (-1, 4.1)$. Nearer one to the horizon
is now a saddle point with $\lambda_\pm = \pm 1.63$ , and the one further away
is a center, with $\lambda_\pm = \pm 3.05 i$ . This is typical of region $S$ of  figure (\ref{fig:fp_Q}). As we enter region $S$ from $R$, we see the locations of the center and the saddle are swapped.}
\label{fig:phase_por}
\end{figure}

\noindent
We now proceed to examine the nature of these fixed points.
It is best done by linearising the system of equations around a  fixed point $(z_0=0,x_0)$,  where $x = x_0$ is a solution of the
second equation of (\ref{eq:root_Q});
to leading order, we get from (\ref{fpeq}),
\begin{eqnarray}
&&\frac{d\delta x}{d\phi} = \delta z,\nonumber\\
&& \frac{d\delta z}{d\phi} = -(1 - 6 M x_0 + 6 Q^2 x_0^2 + \zeta)\delta x.
\end{eqnarray}
Here $\delta z$ and $\delta x$ represent small disturbances around the fixed point.
The Jacobian matrix is then computed as 
\begin{equation}
J  = \begin{bmatrix} 
0&1\\
-(1 - 6 M x_0 + 6 Q^2 x_0^2 + \zeta)&0\\ 
\end{bmatrix} 
\end{equation}
The eigenvalues of $J$ are
\begin{equation}
\lambda_\pm = \pm \sqrt{-1+ 6 M  x_0 - 6 Q^2 x_0^2 - \zeta}.
\end{equation}
The eigen values depend on $\alpha, \beta$ (through $x_0$ and $\zeta$ ) and also on $M, Q$. Clearly, there are
two possibilities for $\lambda_\pm$. Either both are real, differing only by the sign in front, or, 
both are purely imaginary, equal in magnitude and opposite in sign. First fixed point represents a saddle and the other is that of a center. Having gotten the possible nature of the fixed points from the local analysis, we numerically find the complete phase portrait. 
As we vary $\alpha, \beta$ keeping rest of the parameters fixed,  we see either a single (saddle or center)
or both (a saddle and a center) may appear outside the outer horizon. In fact, two fixed points, that is the center and the saddle could swap their positions outside the horizon as we vary the parameters. The details of the portraits are discussion in Fig. (\ref{fig:phase_por}). 

\subsection{ Fixed points and phase portraits for massive particle }\label{3.2}

While working with massive particles, two changes occur in the geodesic equation (\ref{eq:veff_gen}).
First is that $\epsilon$ now takes the value one and the dependence of $L$, the angular momentum per
unit mass, gets retained. In what follows, we will fix a representative value of $L$ and obtain the 
phase portraits.\\

\begin{figure}[h!]
\centering
\includegraphics[width=5in]{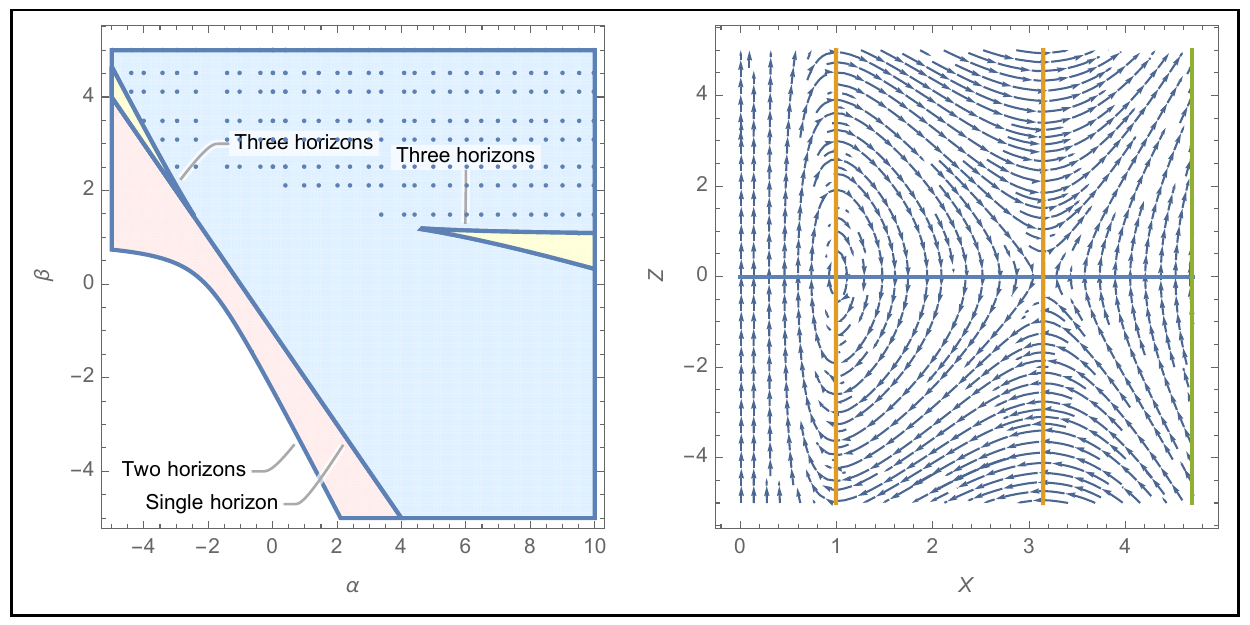}
\caption{These figures show the nodes and corresponding phase portrait for a massive particle around an uncharged black hole for $M =1, L = .5, m_g = h = 1, Q=0$. Left: Dotted region represents the values of $\alpha, \beta$ for which nodes (one center and a saddle) appear outside the horizon. We also note that for these values of the parameters, the black hole possesses only a single horizon. Right:  phase portrait, representing the center and the saddle outside the horizon (shown by the extreme right vertical line). This is drawn for  $\alpha = 4.1, \beta = 2.5$.}
\label{fig:massiveq0}
\end{figure}

\noindent
For  a neutral black hole, massive particle follows
\begin{equation}
\dot r^2 + \Big(1 - \frac{2 M}{r}  + \frac{\Lambda}{3} r^2 + \gamma r + \zeta\Big) \Big(\frac{L^2}{r^2}+1\Big) - E^2 = 0.
\end{equation}
This can be brought to the form
\begin{eqnarray}
&& \frac{d x}{d\phi} = z, \nonumber\\
&& \frac{dz}{d\phi} = -(1 + \zeta) x + 3 M x^2 - \frac{\gamma}{2} + \frac{1}{L^2}\Big(M + \frac{\Lambda}{3 x^3} + \frac{\gamma}{2  x^2}\Big).
\label{eq:diffeq1}
\end{eqnarray}
We take $(z=0, x = x_0)$
as a fixed point solution. To find the nature of this node, we linearise the above equations to get
\begin{eqnarray}
&& \frac{d \delta x}{d\phi} = \delta z, \nonumber\\
&& \frac{d\delta z}{d\phi} = \Big(-1 - \alpha -3\beta +6 M x_0+ \frac{x_0(1 + 2 \alpha + 3 \beta)
-3 (1 + \alpha + \beta)}{L^2 x^4}\Big)\delta x \, . \nonumber
\end{eqnarray}
 The eigenvalues of the Jacobian matrix are then
\begin{equation}
\lambda_{\pm} = \pm \frac {1}{L x_0^2} \sqrt{-3(1 +  \alpha  +\beta) + x_0(1+ 2 \alpha + 3\beta)
- L^2 x_0^4(1+ \alpha + 3\beta) + 6 L^2 M x_0^5}.
\end{equation}
Since under the square root  all the quantities are real, $\lambda_\pm$ could be either real or
purely imaginary numbers. Therefore, the nodes are either saddle point or center. We also require the
nodes to be outside the outer horizon of the black hole. In the following, we set fixed values of 
$M, L$ and study (\ref{eq:diffeq}) numerically to arrive at the phases portraits.
This is shown in figure (\ref{fig:massiveq0}).

\begin{figure}[h!]
\centering
\includegraphics[width=5in]{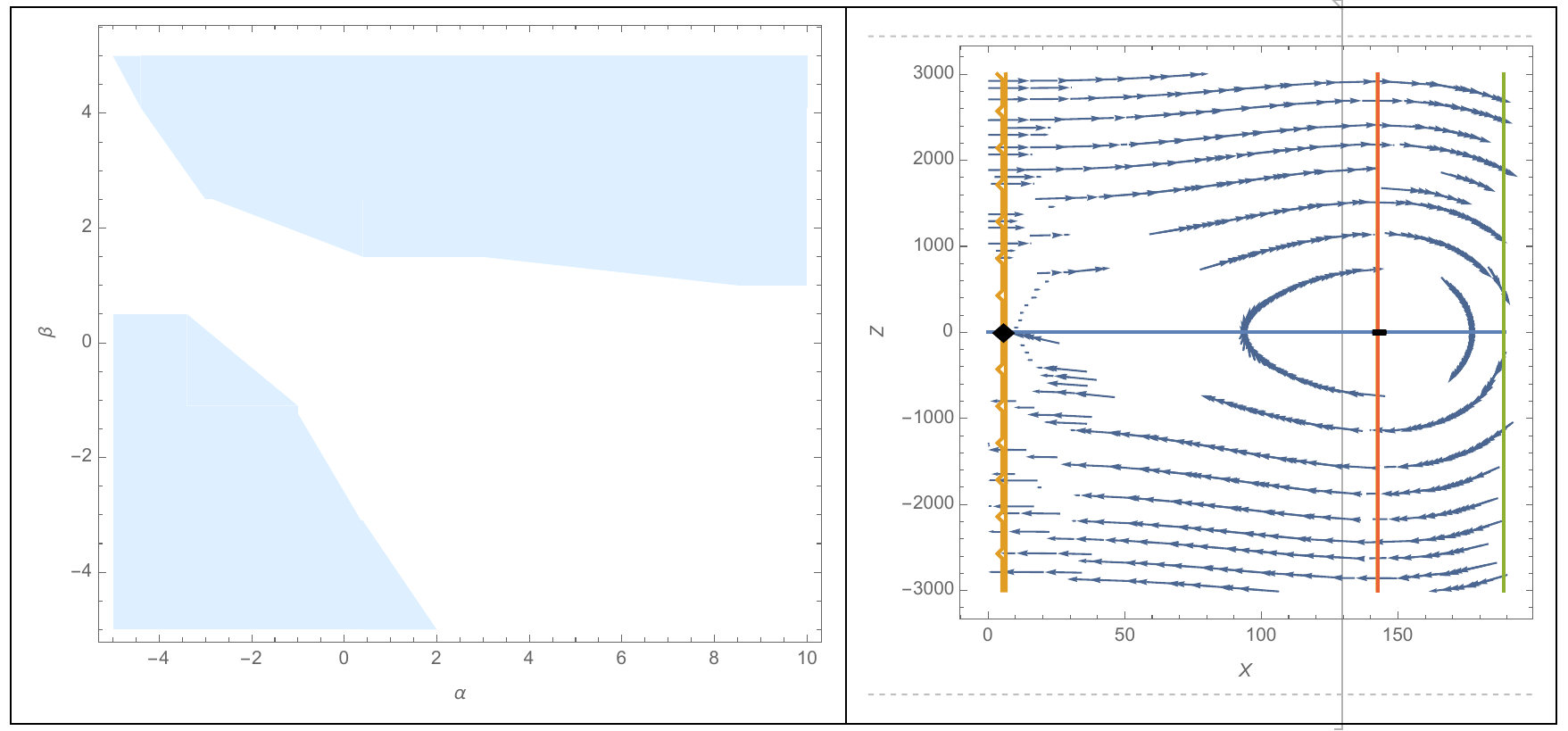}
\caption{These figures show representative phase plane portrait  of  massive particle outside a
charged black hole for $M=1, Q = 0.2, L = .5, m_g = h =1$. Left: 
A  pair of nodes appear outside the outer horizon in the blue region -- one is a saddle and
the other being a center. Right: Typical portrait for $\alpha = 8, \beta = 4.1$. Nodes are marked with black dots. The vertical line on the right is the location of the outer horizon.}
\label{fig:massiveq1}
\end{figure}

\noindent
Finally we examine the allowed  trajectories of a massive particle around a charged black hole. The analogue of (\ref{eq:diffeq1}) are
\begin{eqnarray}\label{eq:diffeq}
&&\frac{dx}{d\phi} = z,\nonumber\\
&&\frac{dz}{d\phi} =  -(1 + \zeta)x + 3 M x^2 - 2 Q^2 x^3 - \frac{\gamma}{2}
+ \frac{1}{L^2} \Big(M + \frac{\Lambda}{3 x^3} - Q^2 x + \frac{\gamma}{2}\Big).\nonumber\\
\end{eqnarray}
Expanding around a fixed point $(z =0, x =x_0)$, where $x_0$ is a solution of the equation
${dz}/{d\phi} =0$, we find the eigen values of the Jacobian matrix as 
\begin{eqnarray}
\lambda_\pm = &&\pm \frac{1}{Lx_0^2}\Big[-3(1 +  \alpha  +\beta) + x_0(1+ 2 \alpha + 3\beta)
- L^2 x_0^4(1+ \alpha + 3\beta) \nonumber\\
&&~~~~+ 6 L^2 M x_0^5 - Q^2 x_0^4- 6 L^2 Q^2 x_0^6\Big]^{\frac{1}{2}}.
\end{eqnarray}
As earlier, these eigen values could either 
be real or purely imaginary. Hence the nodes could be either centers or saddle points. With these inputs coming from the local analysis, we find the phase portraits numerically. 
Figure (\ref{fig:massiveq1}) provides the trajectories emanating from (\ref{eq:diffeq}).


\section{Static spheres} \label{4}
Having obtained the phase portraits of massive as well as massless particles around the neutral and the charged black holes in massive gravity,  in this section, we specialise on  the circular orbits of a massive particle. In particular, we focus on  the static spheres -- these are the circular orbits  with vanishing orbital angular momentum \cite{Wei:2023bgp}. The associated angular velocity  of a particle executing such a trajectory,  measured by a distant static observer, vanishes since
\begin{equation}\label{eq:omega_gen}
\Omega = \frac{\dot{\varphi}}{\dot{t}} =\frac{f(r)L}{r^2 E}.
\end{equation}
\noindent
 
 \subsection{Existence of static spheres}\label{4.1}
 
\begin{figure}[h!]
	
	{\centering
		
		\subfloat[]{\includegraphics[width=2.7in]{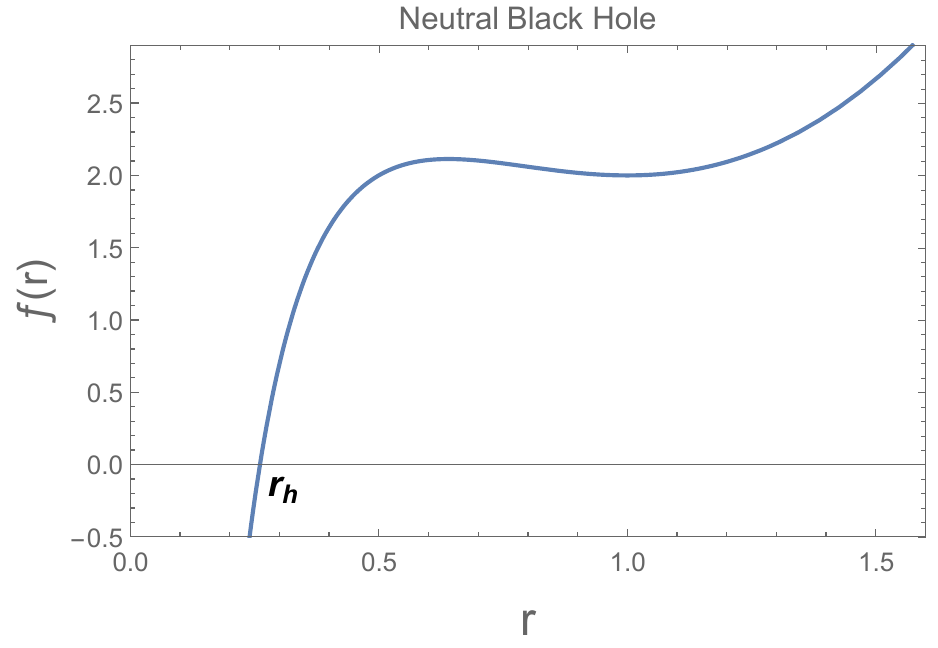}}\hspace{1.4cm}	
		\subfloat[]{\includegraphics[width=2.7in]{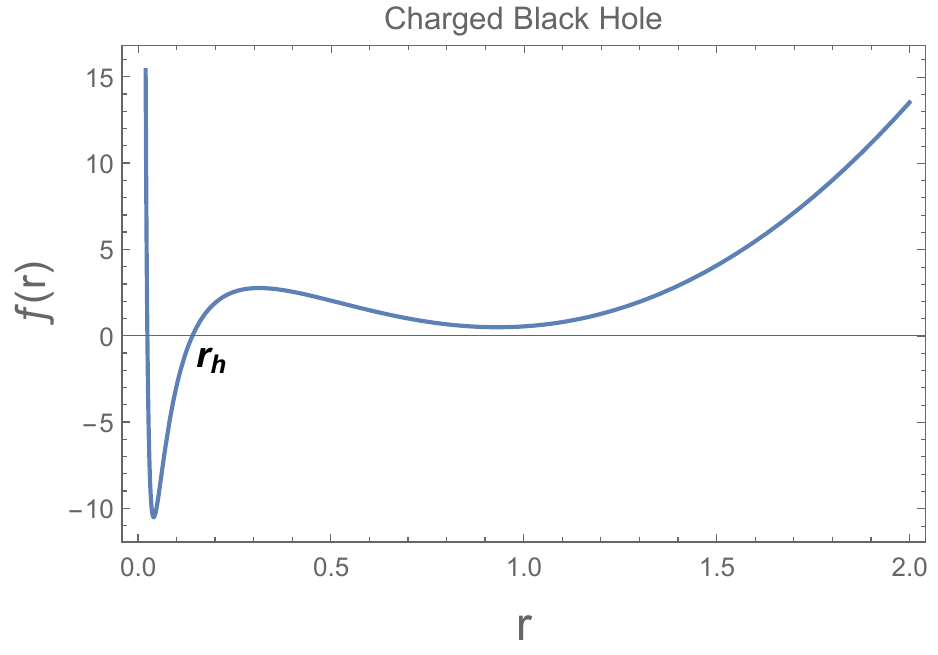}}	
		
		\caption{\footnotesize  Behavior of the lapse function $f(r)$. (a) for a neutral black hole with  single horizon at $r_h = 0.2610$ (here, we set $M=m_g=h=1, Q=0, \alpha =0, \beta =3$). (b) for a charged black hole with two horizons (inner and outer) where the outer horizon is located at $r_h = 0.1414$  (here, we set $M=m_g=h=1, Q=0.2, \alpha =10, \beta =1.5$).}
\label{fig:f_both}	}
\end{figure}

\begin{figure}[h!]
	
	{\centering
		
		\subfloat[]{\includegraphics[width=2.7in]{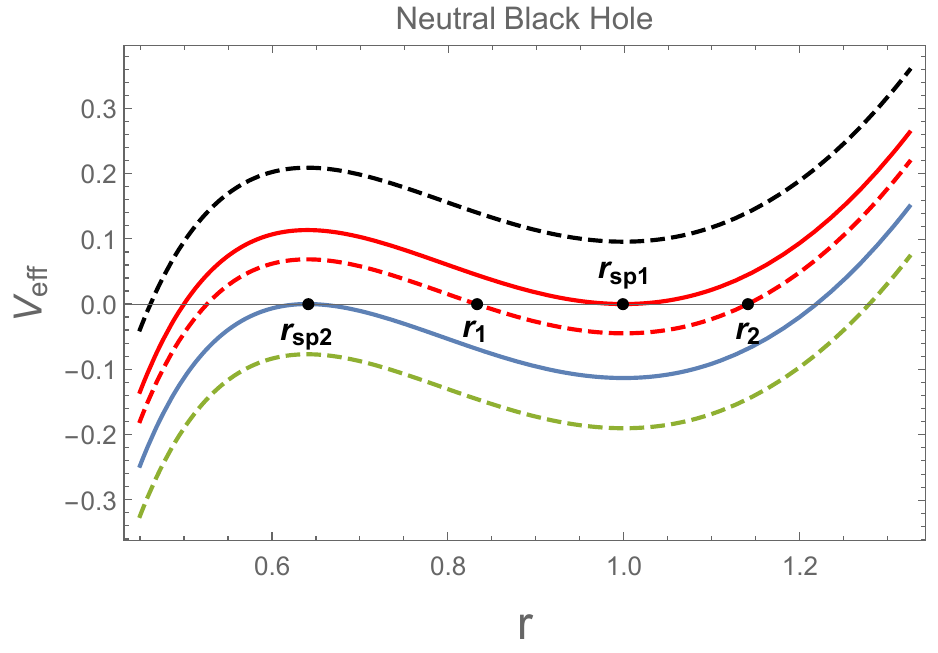}}\hspace{1.4cm}	
		\subfloat[]{\includegraphics[width=2.7in]{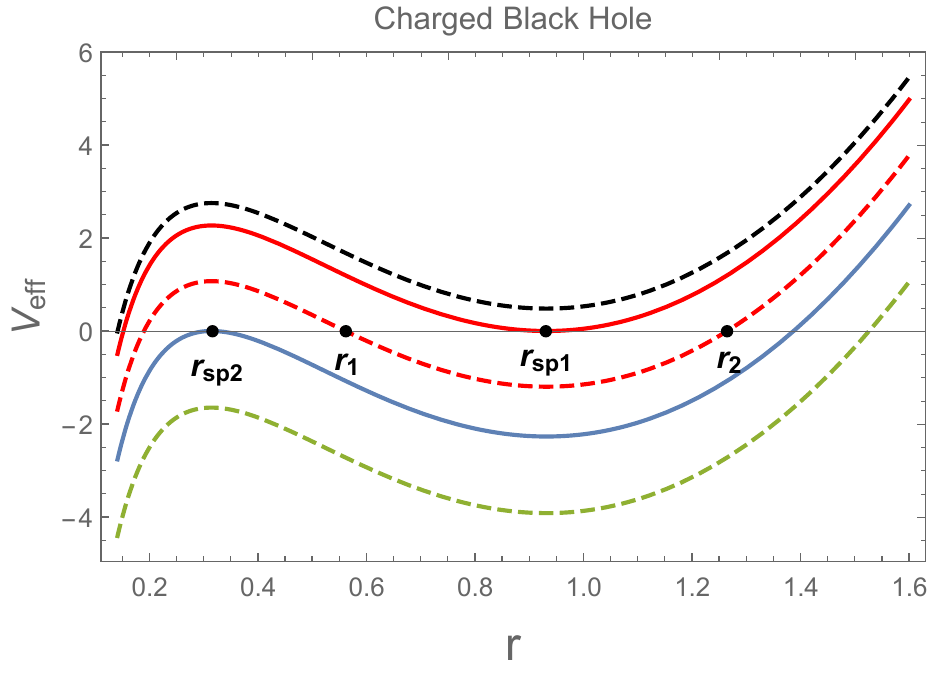}}	
		
		\caption{\footnotesize  Behavior of the effective potential~\eqref{eq:veff_static} for various energies $E$, of the massive test particle.  (a) for neutral black hole background (here, we set $M=m_g=h=1, Q=0, \alpha =0, \beta =3$).  $E = 1.38,1.414 \, (E_{\text{sp1}}), 1.43, 1.454 \, (E_{\text{sp2}}), 1.48$ from top to bottom. The stable static sphere is located at $r_{\text{sp1}} = 1$, while the unstable static sphere is  located at  $r_{\text{sp2}} = 0.6404$. (b) for charged black hole background (here, we set $M=m_g=h=1, Q=0.2, \alpha =10, \beta =1.5$).  $E = 0.1, 0.7014 \, (E_{\text{sp1}}), 1.3, 1.66185 \, (E_{\text{sp2}}), 2.1$ from top to bottom. The stable static sphere is located at $r_{\text{sp1}} = 0.93182$, while the unstable static sphere is  located at  $r_{\text{sp2}} = 0.3146$.}
		\label{fig:veff_static_both}	}	
\end{figure}

\begin{figure}[h!tbp]
		{\centering
		
		\subfloat[]{\includegraphics[width=2.62in]{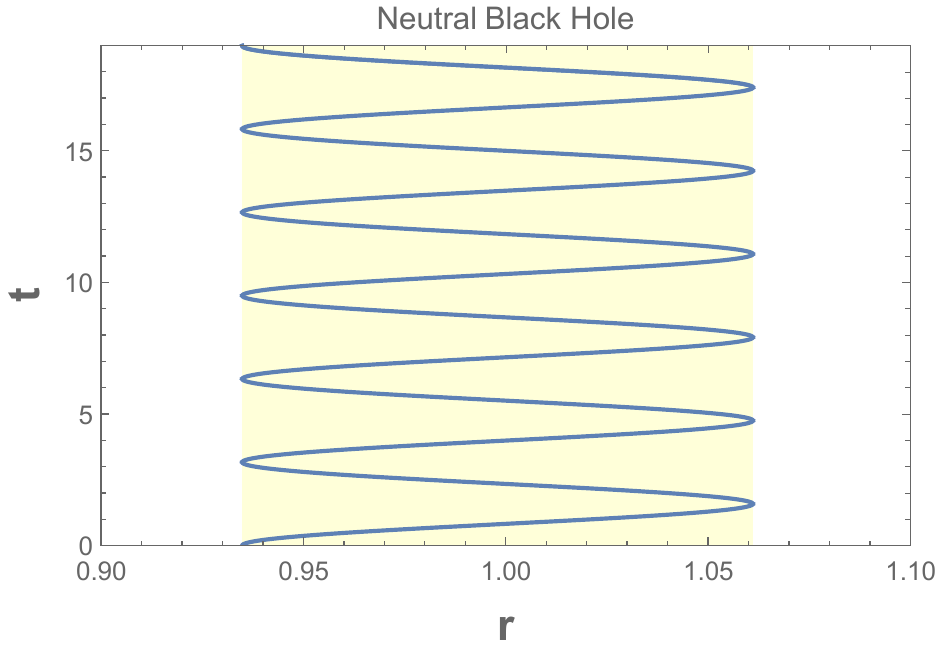}}\hspace{0.75cm}	
		\subfloat[]{\includegraphics[width=2.62in]{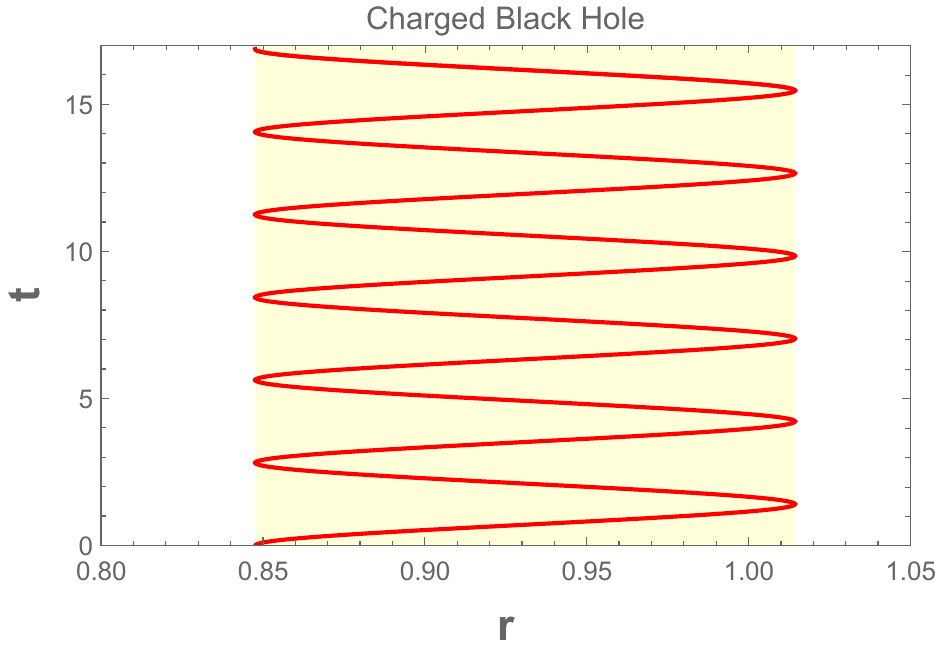}}\hspace{1.5cm}
		\subfloat[]{\includegraphics[width=2.62in]{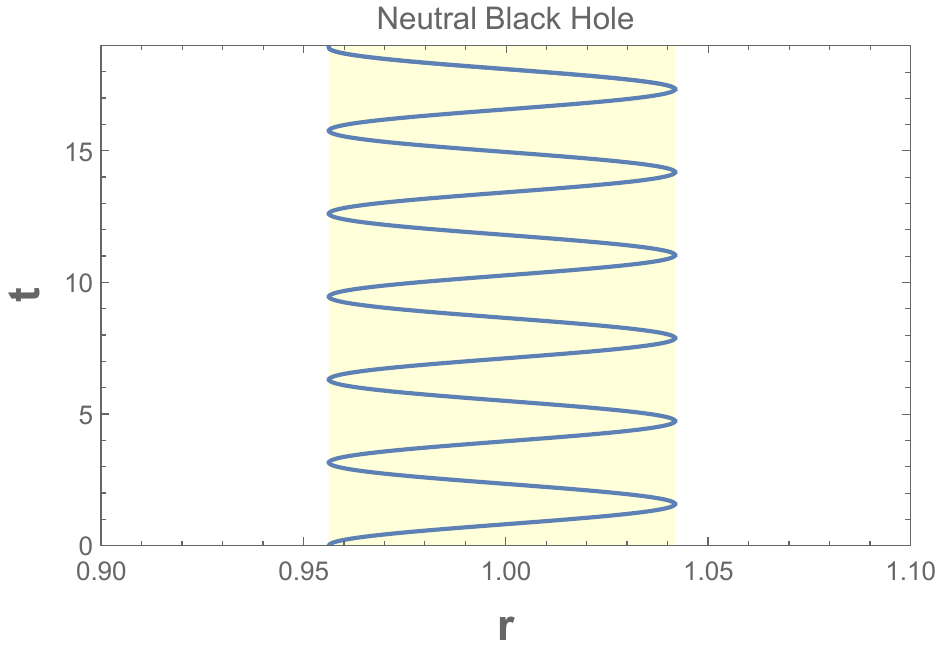}}\hspace{0.75cm}
		\subfloat[]{\includegraphics[width=2.62in]{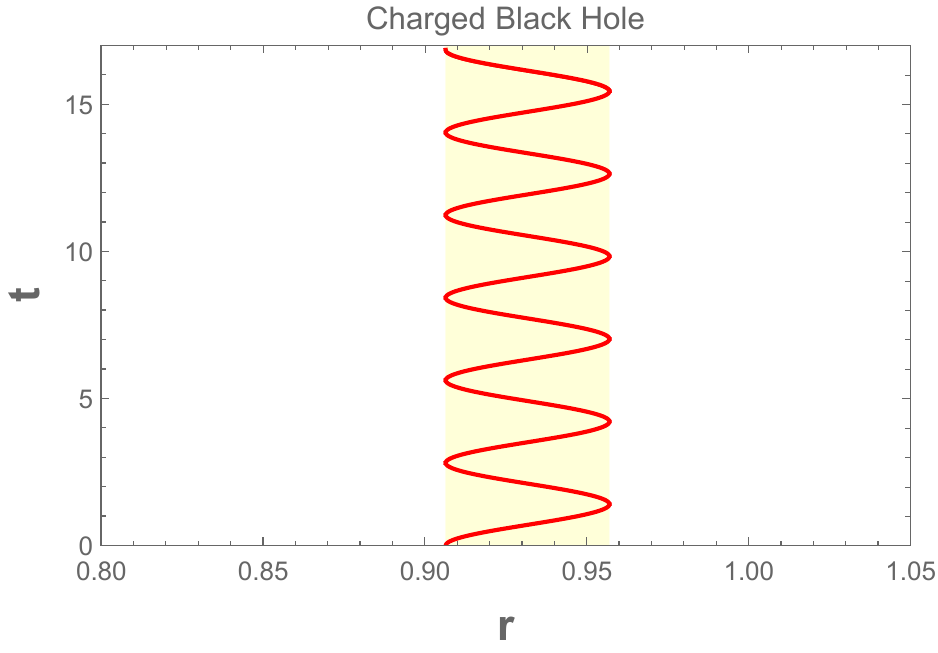}}\hspace{1.5cm}
		\subfloat[]{\includegraphics[width=2.62in]{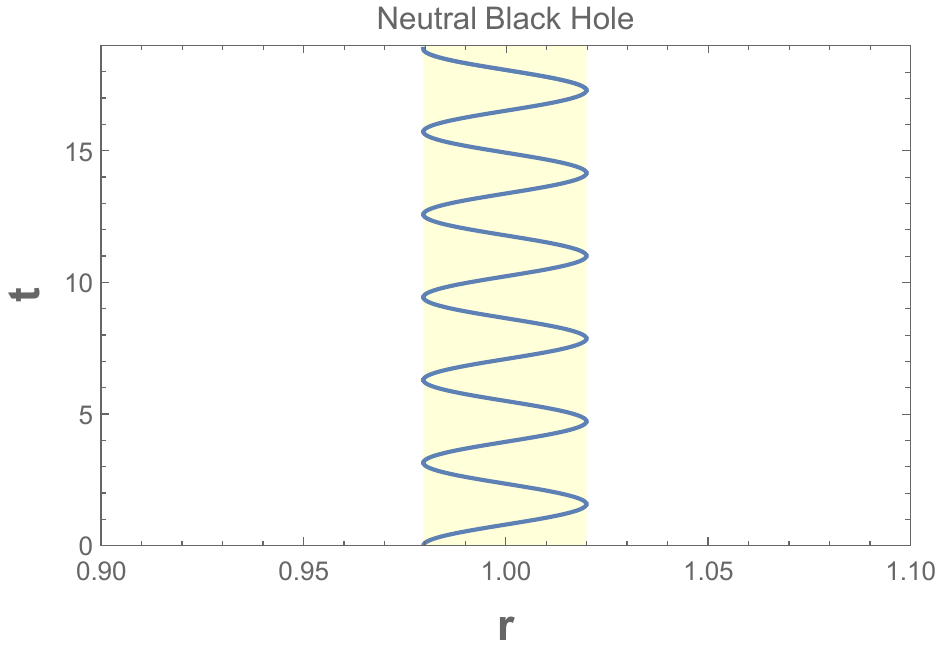}}\hspace{0.75cm}
		\subfloat[]{\includegraphics[width=2.62in]{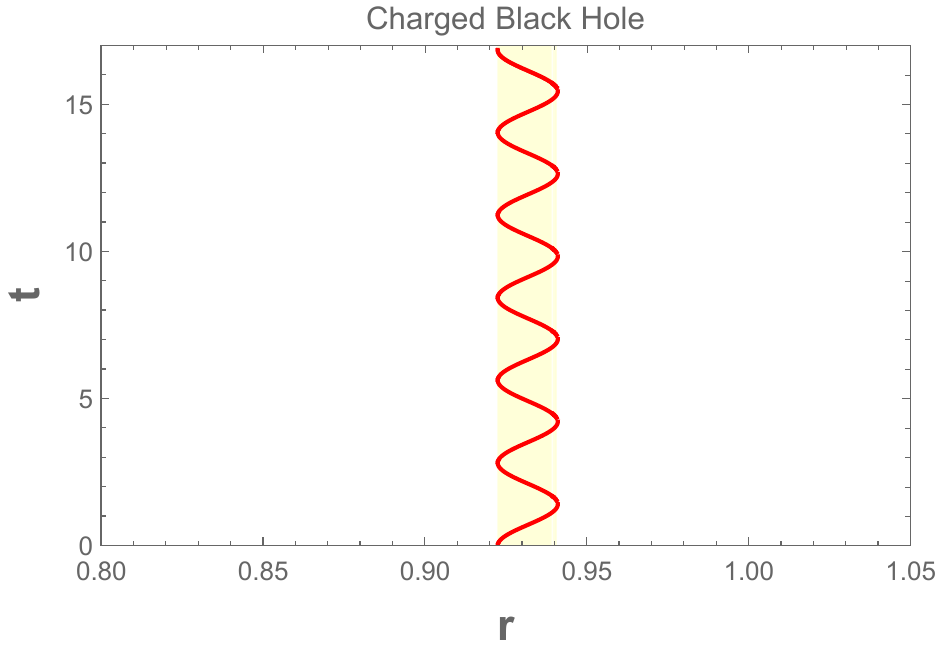}}\hspace{1.5cm}
		\subfloat[]{\includegraphics[width=2.62in]{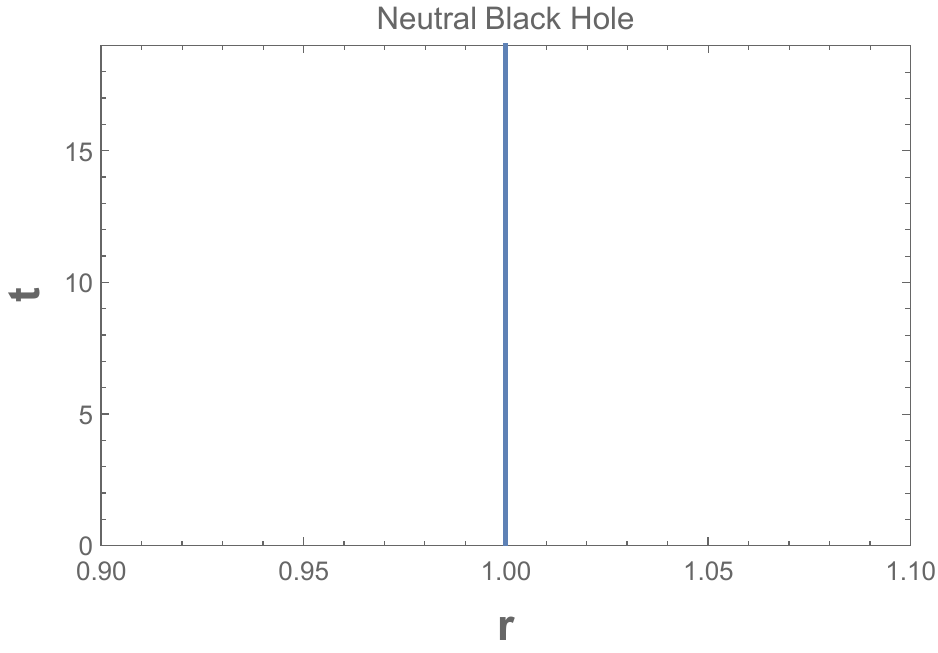}}\hspace{0.75cm}
		\subfloat[]{\includegraphics[width=2.62in]{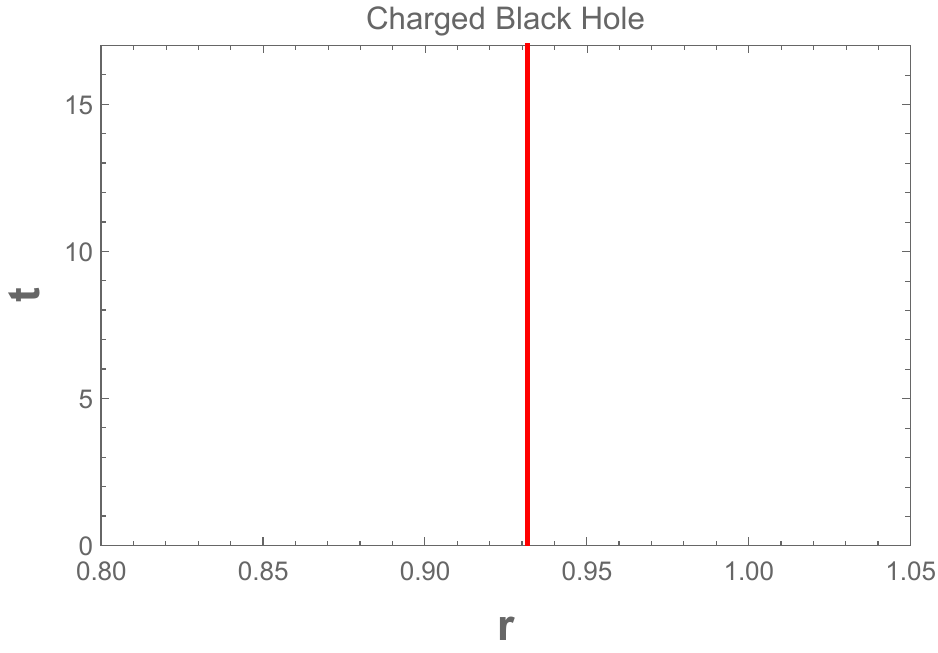}}	
		
		\caption{\footnotesize  Radial motion of the massive test particle. Left panel for neutral black hole background (here, we set $M=m_g=h=1, Q=0, \alpha =0, \beta =3$).  (a) $E = 1.4170$, (c) $E =1.4155$, (e) $E =1.4145$, and (g) $E = E_{\text{sp1}}=1.414$. Right panel for charged black hole background (here, we set $M=m_g=h=1, Q=0.2, \alpha =10, \beta =1.5$). (b) $E = 0.75$, (d) $E =0.706$, (f) $E =0.702$, and (h) $E = E_{\text{sp1}}=0.7014$. When $E=E_{\text{sp1}}$, the radial motion is just a horizontal line denoting a static orbit with vanishing angular momentum.}
		\label{fig:radial_both}	}	
\end{figure}

It is instructive to first have a look at the effective potential (\ref{eq:veff_gen}) as much can be understood from it. It reads, after setting $L =0$,
\begin{equation}\label{eq:veff_static}
V_{\text{eff}} = f(r) - E^2.
\end{equation}
Further, the conditions for the circular orbits  to occur are   $V_{\text{eff}} = V^{'}_{\text{eff}}=0 $, where prime denotes derivative with respect to $r$. These two equations  together give the energy and the radius 
$(r_{\text{sp}})$ of the static sphere as
\begin{eqnarray}
&&E = \sqrt{f(r_{\text{sp}})}\, , \\
&&f^{'}(r_{\text{sp}}) =  0. 
\end{eqnarray}
For the above conditions to make sense, the lapse function $f(r)$ needs to be positive, and should necessarily have extremal points. For the present case, there exist a wide range of parameter values, for which the lapse function $f(r)$,  given in (\ref{eq:f_gen}), fulfils all the requirements. For brevity, we directly present the behaviour of $f(r)$ in Fig. (\ref{fig:f_both}) for some choices of parameters.  
\vskip 0.2cm
\noindent Further, the effective potential (\ref{eq:veff_static}) plotted in Fig. (\ref{fig:veff_static_both}) for both neutral and charged black holes, shows the existence of one stable and one unstable static sphere at $r_{\text{sp1}}$ and $r_{\text{sp2}}$, for fixed energy $E_{\text{sp1}}$ and $E_{\text{sp2}}$, respectively.
\vskip 0.2cm 
\noindent The motion of the massive test particle, possessing the energy $ E \in (E_{\text{sp1}}, E_{\text{sp2}})$, will be bounded between the two turning points $r_1$ and $r_2$  of the potential, see the Fig. (\ref{fig:veff_static_both}). This is a straight back-and-forth motion between $r_1$ and $r_2$, as there is no angular motion, on account of vanishing angular velocity. 
From the radial motion, plotted in $r-t$ plane in  Fig. (\ref{fig:radial_both}), one can see that the radial distance of the motion  between $r_1$ and $r_2$ narrows as the energy $E$ decreases. In fact, when $E = E_{\text{sp1}}$, we see that $r_1 =r_2 =r_{\text{sp1}}$. Thus, the radial distance remains unchanged with the coordinate time, indicating that there is a static sphere at $r=r_{\text{sp1}}$.\\

\begin{figure}[h!]
\centering
\includegraphics[width=5in]{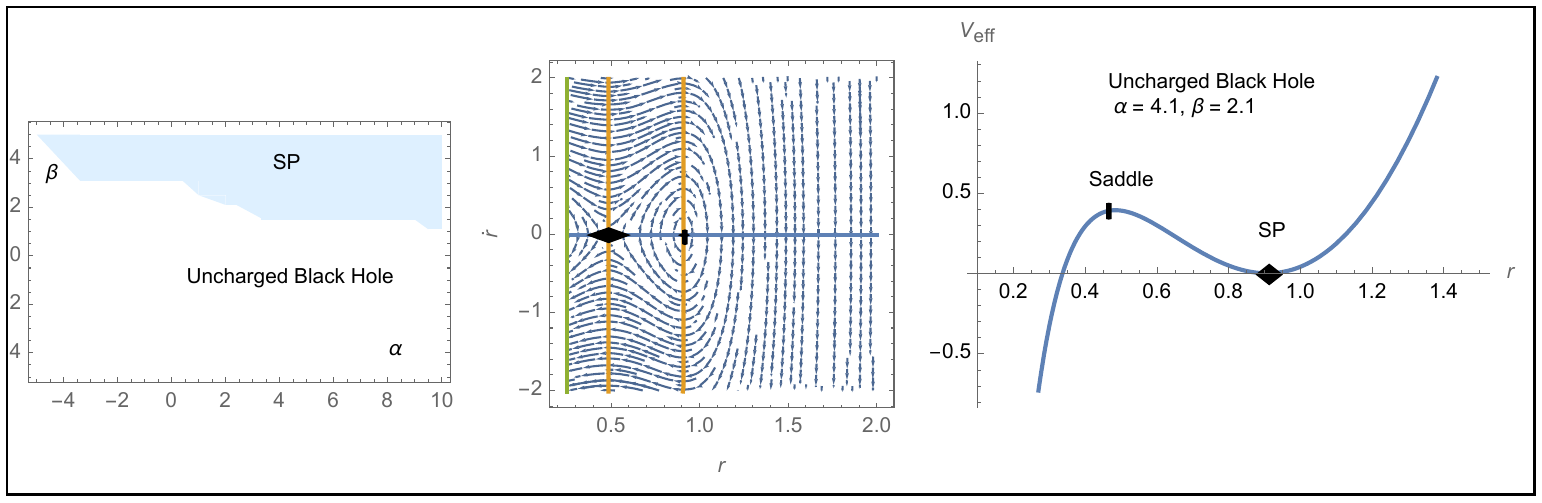}
\caption{Static spheres around an uncharged black hole. We have set $M = m_g = h =1, Q=0$.
Left: Region in blue shows the values of $\alpha, \beta$ for which static spheres appear. Ranges of $\alpha, \beta$ are chosen only for the representative purpose. Middle: Fixed points, denoted in black, and the vector fields around those are shown. The chosen values of $(\alpha, \beta)$ are $(4.1,2.1)$. The vertical line on the extreme left is the location of the outer horizon. Right: Shape of the effective potential following from (\ref{eq:veff_static}) for $(\alpha, \beta) = (4.1,2.1)$. The energy associated with the particle is $E = 1.03$.}
\label{fig:threetogetherq0}
\end{figure}

\begin{figure}[h!]
\centering
\includegraphics[width=5in]{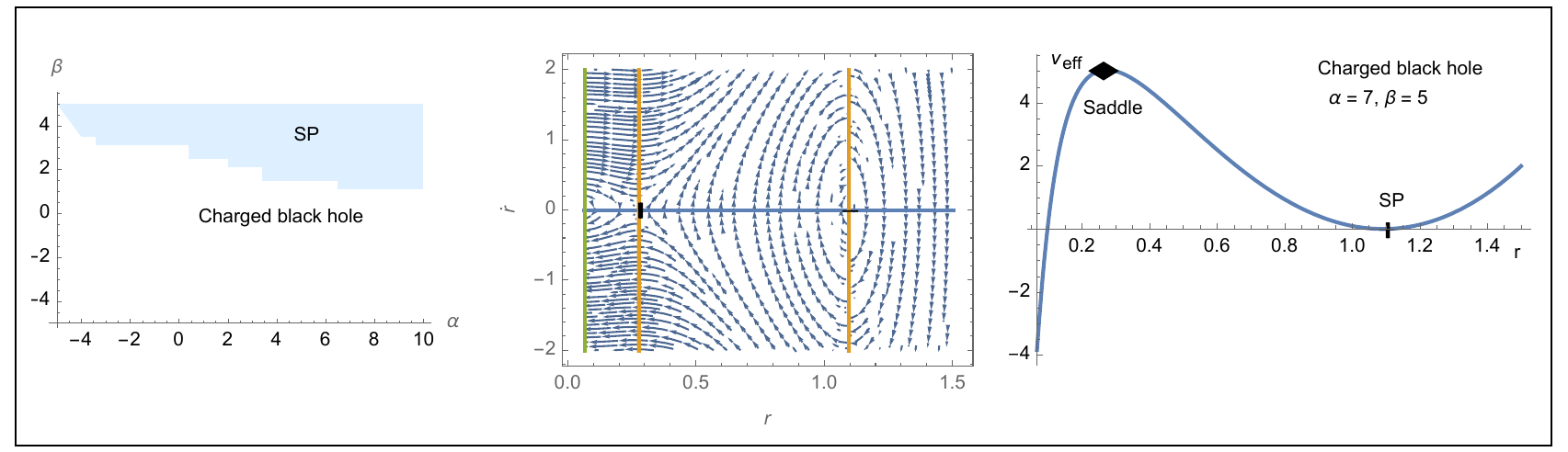}
\caption{Static spheres around a charged black hole. We have set $M = m_g = h =1$ and $Q=0.2$.
Left: Regions in blue show the values of $\alpha, \beta$ for which static spheres occur. Ranges of $\alpha, \beta$ are chosen only for the representative purpose. Middle: Fixed points, denoted in black, and the vector fields around those are shown. The chosen values of $(\alpha, \beta)$ are $(7,5)$. Shape of the effective potential following from (\ref{eq:veff_static}) for $\alpha = 7$ and $\beta = 5$. The energy associated with the particle is $E = 1.98$.}
\label{fig:staticcharged}
\end{figure}

\noindent
Further insights into the nature of the static spheres can be gained from a direct study of the dynamical equation 
that follows from (\ref{eq:geo_gen}). To this end, we differentiate this equation to get
\begin{equation}
\ddot r = \frac{1}{6 r^3} \Big(-6M r + 6 Q^2 - 2 \Lambda r^4 - 3 \gamma r^3\Big).
\end{equation}
It is convenient to convert  this to a set of two first order differential equations as
\begin{eqnarray}\label{eq:static}
&&\dot r = z,\nonumber\\
&&\dot z = \frac{1}{6 r^3} \Big(-6M r + 6 Q^2 - 2 \Lambda r^4 - 3 \gamma r^3\Big).
\end{eqnarray}
Fixed points are obtained by setting $\dot r = \dot z = 0$. These conditions are equivalent to
equating $V_{\rm eff} = V_{\rm eff}' = 0$.
Figures (\ref{fig:threetogetherq0}) and (\ref{fig:staticcharged}) describes our results that we obtained numerically analysing equations (\ref{eq:static}) around neutral and charged black holes, respectively.

\subsection{Number of static spheres} \label{4.2}
We have seen in the previous subsection that, for both neutral and charged black holes, there exists one pair of static spheres (stable and unstable), where we handpicked certain values of massive gravity parameters and used them in the lapse function $f(r)$, in eqn.~\eqref{eq:f_gen}. 
However, for arbitrary values of massive gravity parameters, it is important to know, under what conditions in the region outside the black hole (neutral or charged), do the static spheres exist.
This question can be answered by studying the topological properties of the static spheres, which can exist for the lapse function in eqn.~\eqref{eq:f_gen}, following~\cite{Wei:2023bgp}.    
\begin{figure}[H]	
	{\centering

		\subfloat[]{\includegraphics[width=1.7in]{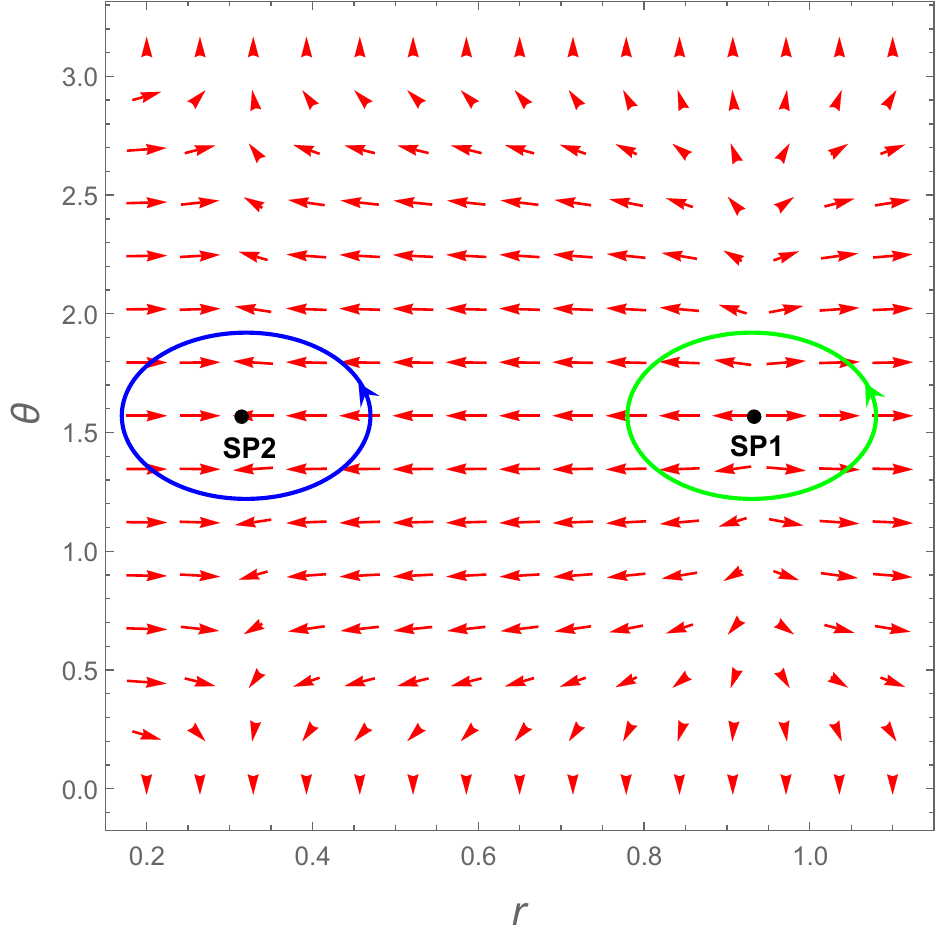}\label{fig:vecplot}}\hspace{0.2cm}
		\subfloat[]{\includegraphics[width=1.8in]{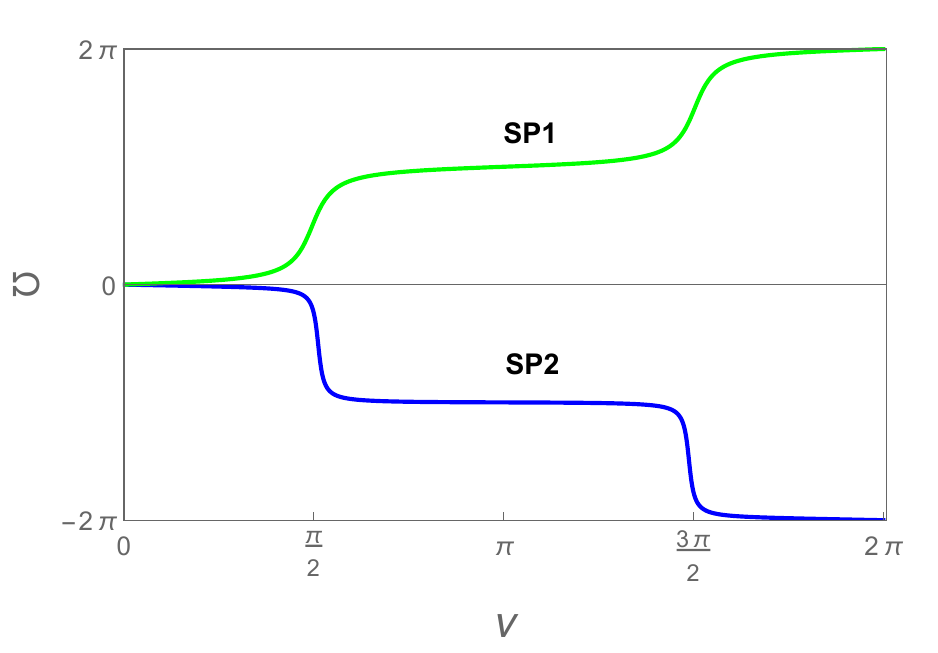}\label{fig:omegaplot}}	
		\subfloat[]{\includegraphics[width=2.6in]{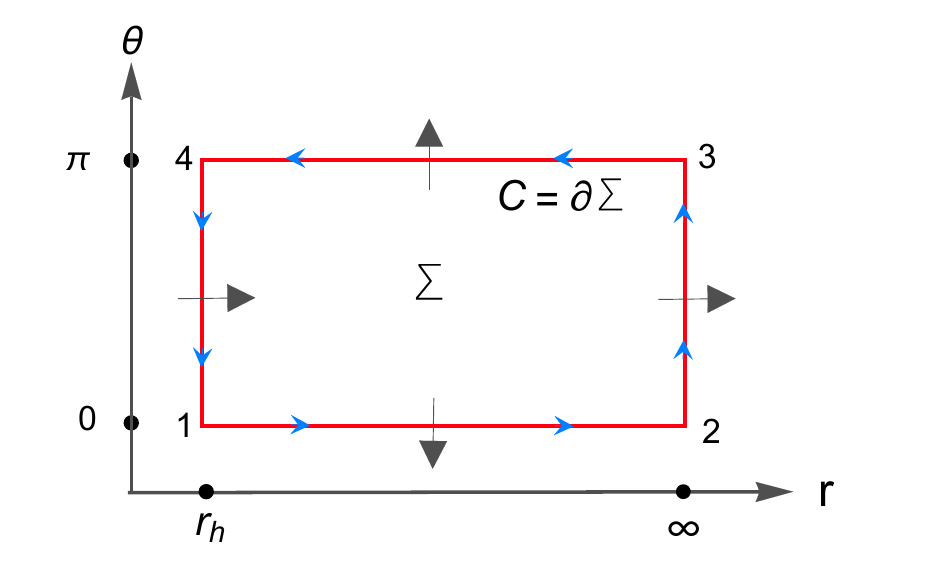}\label{fig:Gtopo}}
		\caption{\footnotesize (a)  The normalised vector field $n$, in $\theta -r$ plane, vanishing at the stable (SP1) and unstable (SP2) static spheres. (b)  The deflection angle $\mho (\text{v})$ of the vector field $\phi$, giving the winding number +1 for stable (SP1) static sphere, and -1 for unstable (SP2) static sphere. Here, plots (a) and (b) are shown for charged black hole case (similar results can be deduced for neutral black hole case, which we do not show). Here, we set $M=m_g=h=1, Q=0.2, \alpha =10, \beta =1.5$. (c) The  rectangular contour $C$ denotes the entire boundary of the
			parameter space $\Sigma$. The black arrows show the direction of the vector field $\phi$ along the boundary.}
		}	
\end{figure}
\vskip 0.2cm \noindent
Using the lapse function $f(r)$, we define a vector field $\phi (\phi^r,\phi^\theta)$ as~\cite{Wei:2023bgp}:
\begin{equation}
\phi^r = \frac{\partial f(r)}{\partial r}\, , \quad \phi^\theta = -\frac{\text{cos}\theta}{\text{sin}^2\theta}\,.
\end{equation}
This vector field $\phi$ vanishes exactly at the location of static spheres. One can assign a  topological charge (i.e., winding number $w$) for each static sphere by defining a topological current $j^\mu$, which is identically conserved, i.e., $\partial_{\mu}j^{\mu}=0$, given by~\cite{Wei:2023bgp}:
\begin{eqnarray}
j^{\mu}=\frac{1}{2\pi}\epsilon^{\mu\nu\rho}\epsilon_{ab}\partial_{\nu}n^{a}\partial_{\rho}n^{b},
\quad \mu,\nu,\rho=0,1,2, \, \, \text{and} \, \, a, b = 1,2,
\end{eqnarray}
where  $n=(\frac{\phi^r}{||\phi||}, \frac{\phi^\theta}{||\phi||})$ is the normalized vector field.
A stable static sphere  possesses the winding number +1, whereas for an unstable static sphere it is -1 (see Fig.~\ref{fig:vecplot} and Fig.~\ref{fig:omegaplot}). 
\vskip 0.2cm
\noindent
We, now compute the total winding number $W$, which is the sum of  the winding numbers of all the static spheres existing in the entire parameter space $\Sigma$ outside the horizon.
For this, consider a closed rectangular contour $C$, which is piece-wise smooth and positive
oriented,   that encloses the entire parameter space $\Sigma$, as shown in Fig.~\ref{fig:Gtopo}. Then, the computation of the deflection $\mho$ of the vector field $\phi$ along the contour $C$, gives the total winding number within $C$ as:
\begin{equation}\label{eq:W_int}
W = \sum_i w_i = \frac{1}{2\pi} \oint_C d\mho,
\end{equation}
where we define,
\begin{eqnarray}
\phi^r &=& ||\phi|| \, \text{cos}\mho, \\
\phi^\theta &=& ||\phi|| \, \text{sin}\mho. \label{eq:def_sin}
\end{eqnarray}
We can write the integration in eqn.~\eqref{eq:W_int} as
\begin{equation}
 \frac{1}{2\pi} \oint_C d\mho = \frac{1}{2\pi} \bigg(\int\limits_{r_{\rm h}}^{\infty} d\mho \Big\vert_{ \theta =0} + \int\limits_{0}^{\pi} d\mho \Big\vert_{ r =\infty} +\int\limits_{\infty}^{r_{\rm h}} d\mho \Big\vert_{ \theta =\pi} + \int\limits_{\pi}^{0} d\mho \Big\vert_{r =r_{\rm h}}\bigg),
\end{equation}
which can be computed as follows~\cite{Wei:2023bgp,Cunha:2020azh}:
\vskip 0.3cm 
\noindent \underbar{\textbf{At $\theta = 0, \pi.$}} 
\vskip 0.2cm 
\noindent
 We have, from eqn.~\eqref{eq:def_sin}, $\mho = \text{arcsin}\big(\frac{\phi^\theta}{||\phi||} \big)$, which becomes
\begin{equation}
 \mho\vert_{\theta =0, \pi} = \text{arcsin}\big(\frac{\phi^\theta}{||\phi^\theta||} \big) = \left\{
 \begin{aligned}
   \text{arcsin}(-1) & = -\frac{\pi}{2}, \quad \text{for} \quad \theta =0. \\
  \text{arcsin}(+1) & =+ \frac{\pi}{2}, \quad \text{for} \quad \theta =\pi. 
 \end{aligned}
 \right.
\end{equation}
This shows that the direction of the vector field $\phi$ is downwards at $\theta = 0$, and upwards at $\theta = \pi$. Further, we get $\int\limits_{r_{\rm h}}^{\infty} d\mho \Big\vert_{ \theta =0} = \int\limits_{\infty}^{r_{\rm h}} d\mho \Big\vert_{ \theta =\pi}= 0 $ (as $d\mho \big\vert_{\theta =0, \pi} = 0$).
\vskip 0.3cm 
\noindent
\noindent \underbar{\textbf{At $r = r_{\rm h}$.}}
\vskip 0.2cm 
\noindent
For the lapse function eqn.~\eqref{eq:f_gen}, we have $f(r_{\rm h}) = 0$ at horizon $r_{\rm h}$, and in order for the static spheres to exist, one should have $f(r>r_{\rm h}) > 0 $. Near the horizon, this implies $ \frac{\partial f(r)}{\partial r} = \phi^r  > 0$, indicating that the direction of the vector field $\phi$ at $r=r_{\rm h}$, is rightward.
\vskip 0.2cm 
\noindent
Since, the vector field $\phi$ is changing the direction from upwards (i.e., $\mho = +\frac{\pi}{2}$) at $\theta = \pi$, to downwards  (i.e., $\mho = -\frac{\pi}{2}$) at $\theta = 0$,
along the line segment $4 \rightarrow 1$ of the contour $C$ (in clock wise direction), we get
\begin{equation}
 \int\limits_{\pi}^{0} d\mho \Big\vert_{ r =r_{\rm h}} = \mho \Big\vert_{\theta =0}-\mho \Big\vert_{\theta =\pi} = -\frac{\pi}{2}-\frac{\pi}{2}=-\pi. 
\end{equation}
\vskip 0.3cm 
\noindent
\noindent \underbar{\textbf{At $r =\infty$.}}
\vskip 0.2cm 
\noindent
 For the lapse function in eqn.~\eqref{eq:f_gen}, one can check that $f(r)$ at $r\rightarrow \infty$, obeying $f(r > r_{\rm h}) > 0$, ensures  that  $\frac{\partial f(r)}{\partial r} = \phi^r$ is positive. Thus,  the direction of the vector field $\phi$ at $r \rightarrow \infty$ is rightward.
\vskip 0.2cm 
\noindent
Since, the vector field $\phi$ is changing direction from downwards (i.e., $\mho = -\frac{\pi}{2}$) at $\theta = 0$, to upwards  (i.e., $\mho = +\frac{\pi}{2}$) at $\theta = \pi$,
along the line segment $2 \rightarrow 3$ of the contour $C$ (in anti-clock wise direction), we get
\begin{equation}
\int\limits_{0}^{\pi} d\mho \Big\vert_{ r =\infty} = \mho \Big\vert_{\theta =\pi}-\mho \Big\vert_{\theta =0} = \frac{\pi}{2}+\frac{\pi}{2}= +\pi. 
\end{equation}
Therefore, we get the total winding number $W = \frac{1}{2\pi} (0+\pi+0-\pi)=0$. 
This shows that if static spheres exist, the stable and unstable static spheres always come in pairs. 

\subsection{Aschenbach effect} \label{4.3}

\begin{figure}[H]	
	{\centering

		\subfloat[]{\includegraphics[width=2.7in]{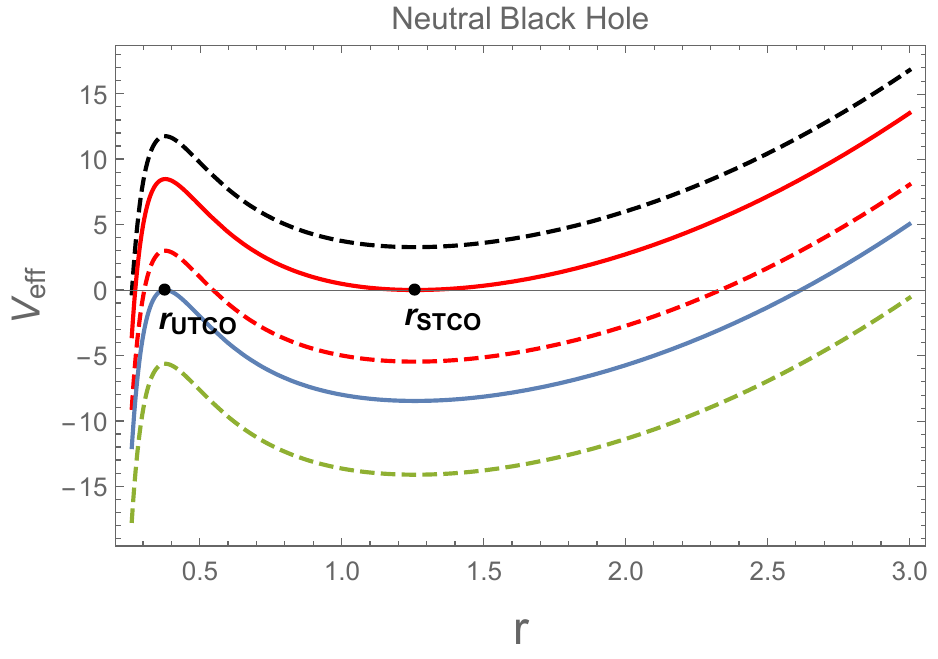}}\hspace{1cm}
		\subfloat[]{\includegraphics[width=2.7in]{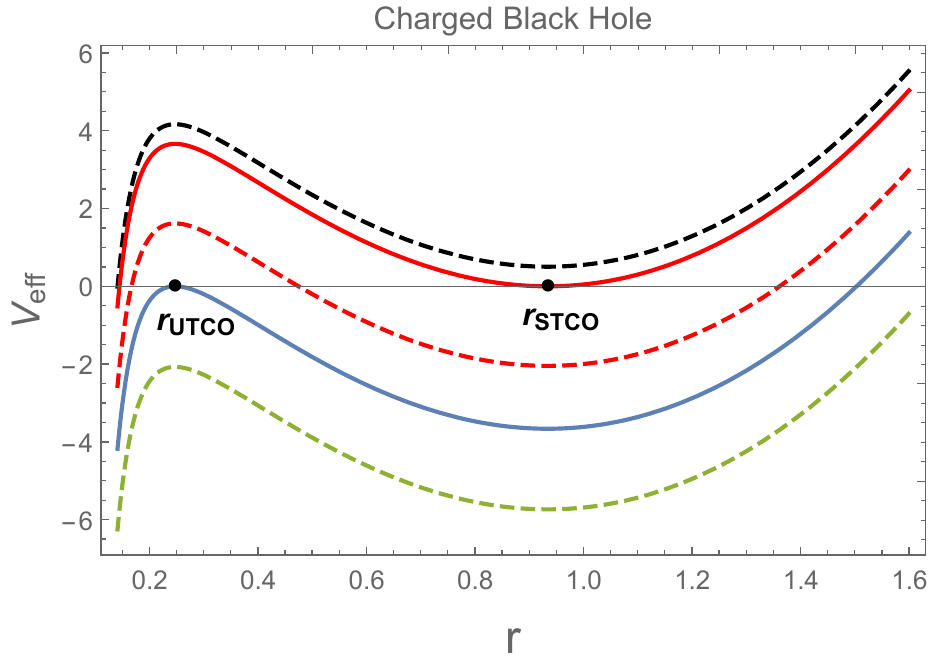}}	
		
		\caption{\footnotesize Plot of effective potential of a massive test particle having 
 fixed angular momentum $L$ and for various energies $E$, showing the existence of  stable timelike circular orbit ({\scriptsize STCO}) and unstable timelike circular orbit ({\scriptsize UTCO}).  (a) for neutral black hole: where, $M=m_g=h=1, Q=0, \alpha =0, \beta =3$,
$ L=1$, and  $ E=0.5, 1.87 \, (E_{\rm {\tiny STCO}}), 3, 3.464 \, (E_{\rm {\tiny UTCO}})$, and $4.2$ from top to bottom.  $(r_{\rm {\tiny STCO}}, r_{\rm {\tiny UTCO}}) = (1.258, 0.379)$.  (b) for charged black hole: where, $M=m_g=h=1, Q=0.2, \alpha =10, \beta =1.5$,
 $ L=0.2$, and  $E=0.1, 0.7173 \, (E_{\rm {\tiny STCO}}), 1.6, 2.0437 \, (E_{\rm {\tiny UTCO}})$, and $2.5$ from top to bottom.  $(r_{\rm {\tiny STCO}}, r_{\rm {\tiny UTCO}}) = (0.9341, 0.2475)$.}
		\label{fig:veff_tco_both}	}	
\end{figure}

\begin{figure}[H]	
	{\centering

		\subfloat[]{\includegraphics[width=2.7in]{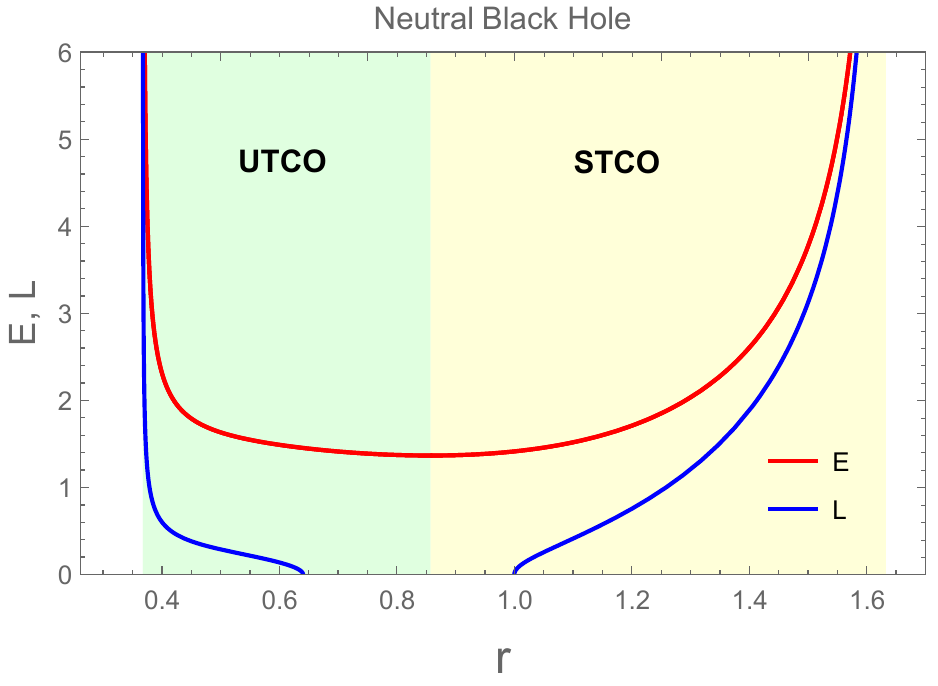}}\hspace{1cm}
		\subfloat[]{\includegraphics[width=2.7in]{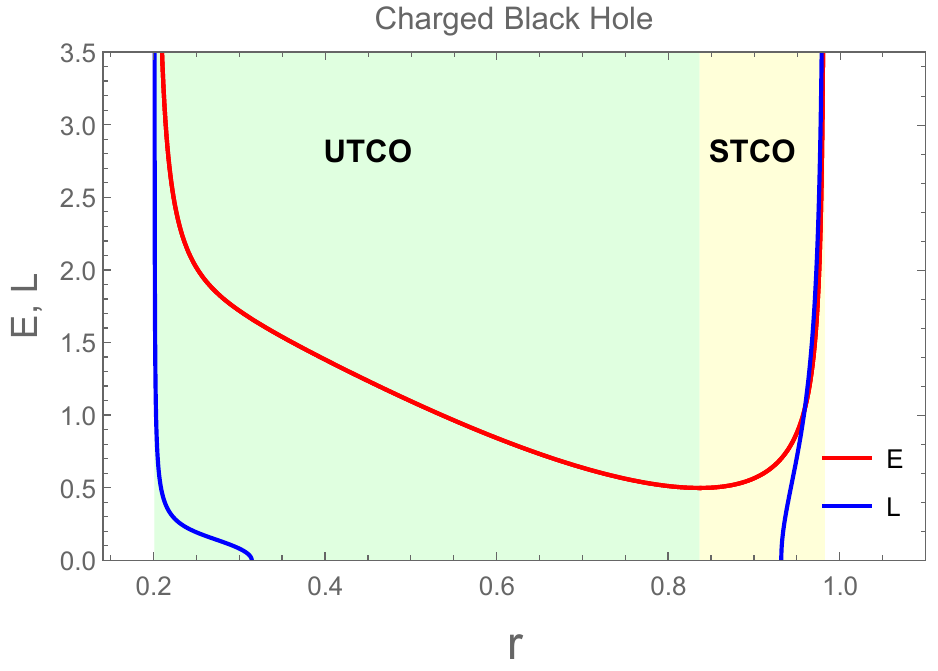}}	
		
		\caption{\footnotesize The energy $E$ and the angular momentum $L$ of the TCOs.  The
			regions in light green and yellow colors are for the unstable TCO (UTCO) and stable TCO
			(STCO) regions, respectively. (a) for neutral black hole background (here, we set $M=m_g=h=1, Q=0, \alpha =0, \beta =3$). (b) for charged black hole background (here, we set $M=m_g=h=1, Q=0.2, \alpha =10, \beta =1.5$). The angular momentum curve is disconnected in the region where it takes imaginary values. $E$ and $L$ diverge at the location of photon spheres i.e., at  $ (r_{\rm sps}, r_{\rm ups}) = (1.632, 0.3675) $ for neutral black hole and at $ (r_{\rm sps}, r_{\rm ups}) = (0.982761, 0.201191) $ for charged black hole.}
		\label{fig:EL_both}	}	
\end{figure}

\begin{figure}[H]	
	{\centering

		\subfloat[]{\includegraphics[width=2.7in]{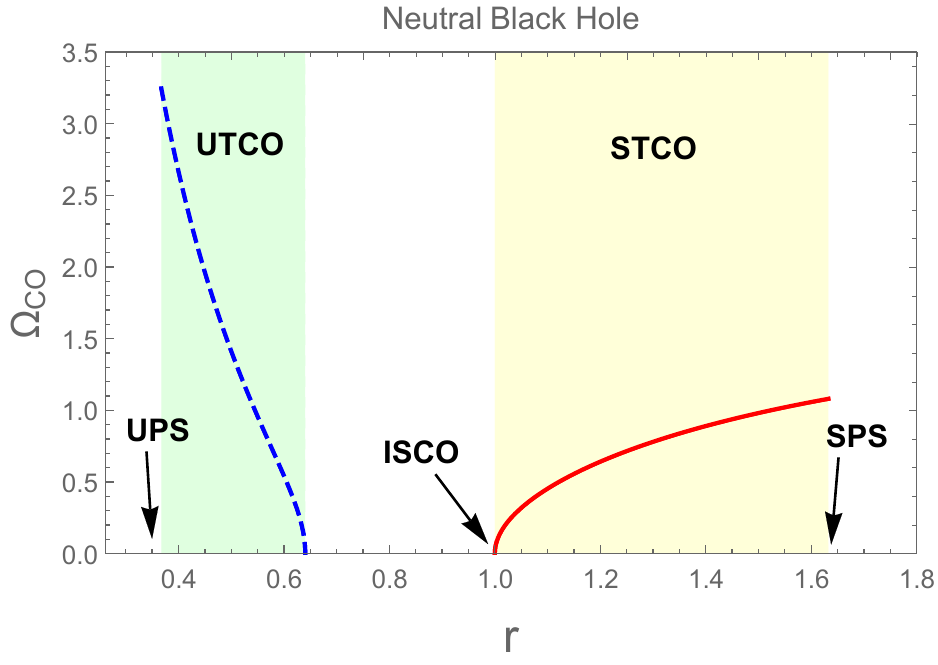}}\hspace{1cm}
		\subfloat[]{\includegraphics[width=2.7in]{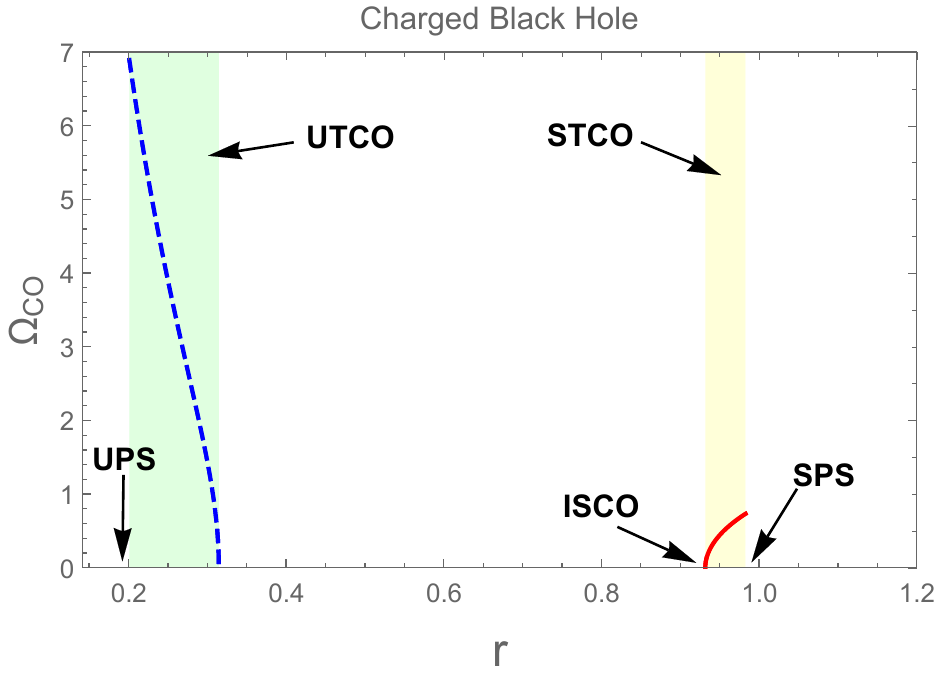}}	
		
		\caption{\footnotesize The orbital angular velocity profile $\Omega_{\text{\tiny CO}}$ of the TCOs.  The
			regions in light green and yellow colors are for the unstable TCO (UTCO) and stable TCO
			(STCO) regions, respectively.   ‘‘ISCO’’ denotes the innermost stable TCO. ‘‘UPS’’ and ‘‘SPS’’ denote the unstable and stable photon spheres. (a) for neutral black hole background, where we set $M=m_g=h=1, Q=0, \alpha =0, \beta =3$, and take $r_{\rm h} = 0.2610,  r_{\rm ups}=0.3675, r_{\text{\tiny ISCO}}=1, r_{\rm sps}=1.632$.  (b) for charged black hole background, where we set $M=m_g=h=1, Q=0.2, \alpha =10, \beta =1.5$, and take $r_{\rm h} = 0.1414,  r_{\rm ups}=0.201191, r_{\text{\tiny ISCO}}=0.93182, r_{\rm sps}=0.982761$.}
		\label{fig:ach_both}	}	
\end{figure}

As noted earlier~\cite{Wei:2023fkn}, the presence of  static spheres and  stable photon spheres around a static and spherically symmetric black hole may indicate the occurrence of a peculiar phenomenon called \textit{Aschenbach effect}, where the angular velocity of a time-like circular orbit is found to be increasing with its radius coordinate. 
\begin{figure}[H]	
	{\centering

		\subfloat[]{\includegraphics[width=2.7in]{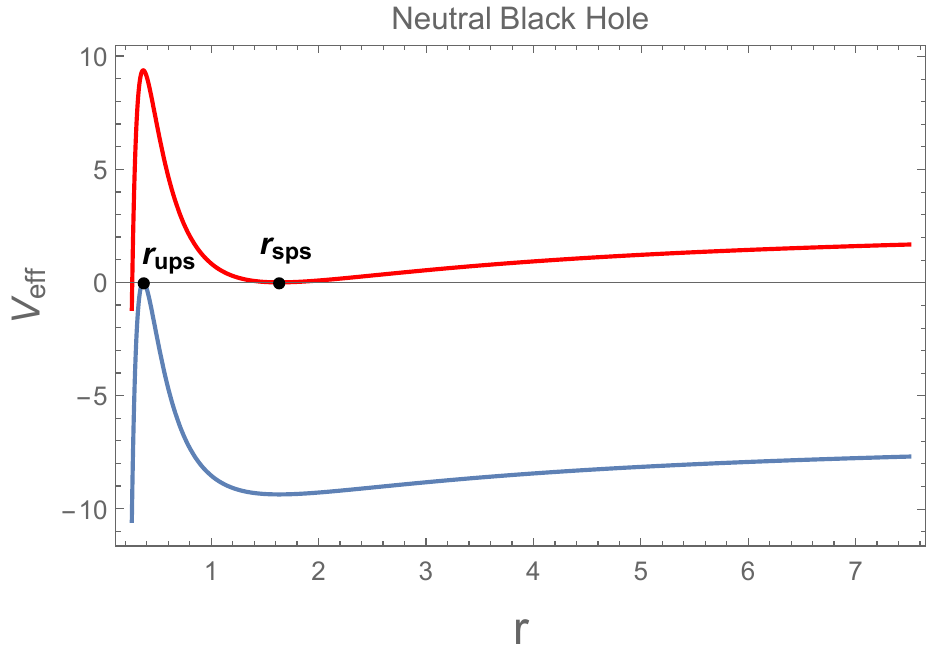}}\hspace{1cm}
		\subfloat[]{\includegraphics[width=2.7in]{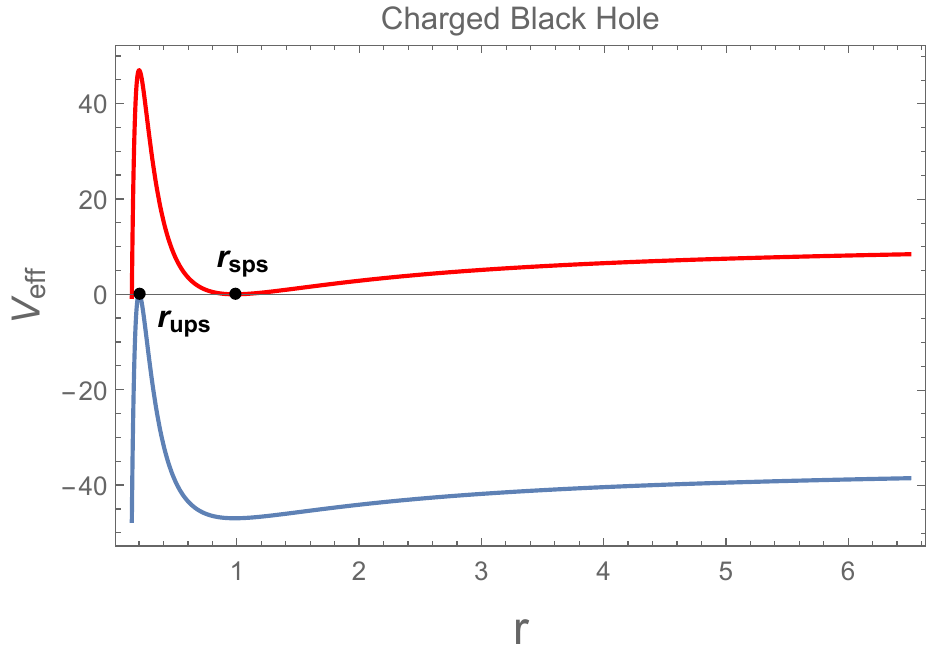}}	
	
		\caption{\footnotesize Plot of effective potential of a photon with fixed angular momentum $L$ and for various energies $E$, showing the existence of  stable photon sphere (sps) and unstable photon sphere (ups), outside the horizon.  (a) for neutral black hole (where we set $M=m_g=h=1, Q=0, \alpha =0, \beta =3$). Red curve for $ E=E_{\rm sps}=1.08028L$, and blue curve for $ E=E_{\rm ups} = 3.24603L$.  $(r_{\rm sps}, r_{\rm ups}) = (1.632,0.3675)$.  (b) for charged black hole (where we set $M=m_g=h=1, Q=0.2, \alpha =10, \beta =1.5$). Red curve for $ E=E_{\rm sps}=0.73281L$, and blue curve for $ E=E_{\rm ups} =  6.8925L$. $ (r_{\rm sps}, r_{\rm ups}) = (0.982761, 0.201191) $. All plots are shown for $L =1$. }
\label{fig:veff_photon_both}	}	
\end{figure}
\vskip 0.2cm
\noindent
 Before looking for \textit{Aschenbach effect}, the presence of static spheres around the black holes in massive gravity noted in the last subsection, motivates us to search also for the presence of stable photon spheres, which might be precursors for such an effect.
 For this, we write the  effective potential~\eqref{eq:veff_gen} for photons (i.e., massless particles) as,
 \begin{equation}
 V_{\text{eff}} = f(r)\Big(\frac{L^2}{r^2}\Big)-E^2,
 \end{equation}
Its behaviour, for the case of fixed angular momentum $L\neq 0$ and for various energies $E$, can be seen in Fig.~\ref{fig:veff_photon_both}, for both  neutral and charged black holes.
We find that, for a given angular momentum $L$, there exist a pair of stable and unstable photon spheres for both neutral and charged black holes. These occur at $(r_{\rm sps},r_{\rm ups}) = (1.632, 0.3675)$  for neutral black hole, and $ (r_{\rm sps}, r_{\rm ups}) = (0.982761, 0.201191) $ for charged black hole.
\vskip 0.2cm
\noindent 
 Thus, due to the existence of one stable photon sphere outside the horizon of neutral and charged black holes, in what follows, we focus on the circular geodesics of a massive test particle~\footnote{See~\cite{Panpanich:2019mll,Hendi:2022qgi} for some possible null and timelike trajectories for our system.} (i.e., timelike circular orbits (TCOs)) in order to examine the occurrence of  \textit{Aschenbach effect}.    
\vskip 0.2cm
\noindent 
The plot of effective potential~\eqref{eq:veff_gen} for a massive test particle is given in Fig.~\ref{fig:veff_tco_both}. We see that for a given angular momentum $L$, there exist a pair of stable and unstable TCOs, such that the stable one is the outer TCO and the unstable one is the inner TCO, which exist in both neutral and charged black hole backgrounds. 
\vskip 0.2cm
\noindent
 One can obtain the expressions for the energy $E$ and angular momentum $L$ of a TCO, using $V_\text{eff} =0$, and $V^{'}_\text{eff} = 0$, given by~\cite{Wei:2023fkn}:
\begin{eqnarray}
E &=& f \, \sqrt{\frac{2}{2f-rf^{'}}}\, , \label{eq:E of TCO}\\
L &=& r^{3/2} \, \sqrt{\frac{f^{'}}{2f-rf^{'}}}\,. \label{eq:L of TCO}
\end{eqnarray}
 The behaviour of $E$ and $L$ of time-like circular orbits, for both neutral and charged black holes in massive gravity theory  is shown in Fig.~\ref{fig:EL_both}.  One can see that  the angular momentum  curve is disconnected in a certain region where it becomes imaginary, implying that the corresponding TCOs should be excluded.   
Therefore, the stable and unstable TCO regions are disconnected, where the inner most stable TCO (ISCO) is located at $r_{\text{\tiny ISCO}} = 1$ for neutral black hole, and  $r_{\text{\tiny ISCO}} = 0.93182$ for charged black hole. We also note that, in both neutral and charged black hole backgrounds,  the marginally stable TCO lies in the region corresponding to imaginary $L$, and thus it does not coincide with the ISCO.    Further, we observe from Fig.~\ref{fig:EL_both},  that both the energy and angular momentum of the TCO diverge at the location of photon spheres.
\noindent 
Now, using the eqns.~\eqref{eq:omega_gen},~\eqref{eq:E of TCO} and~\eqref{eq:L of TCO}, we compute the angular velocity $\Omega_{\text{\tiny CO}}$ of a TCO, measured by a static observer at rest, given by~\cite{Wei:2023fkn}:
\begin{equation}
\Omega_{\text{\tiny CO}} = \sqrt{\frac{f^{'}}{2r}},
\end{equation} 
 where, its behaviour can be seen in Fig.~\ref{fig:ach_both}. One immediately notes that, only in the stable TCO region, the angular velocity $\Omega_{\text{\tiny CO}}$ increases with the radius of the TCO. Moreover, this increasing behaviour of $\Omega_{\text{\tiny CO}}$ occurs near the stable photon sphere (SPS), and also $\Omega_{\text{\tiny CO}}$ maintains finite values at the location of photon spheres, as noted in~\cite{Wei:2023fkn}. 
 Therefore, we confirm that the \textit{Aschenbach effect} can be observed in both neutral and charged black hole backgrounds in massive gravity.
\section{Conclusions} \label{5}
In this paper, we considered the 4-dimensional static and spherically symmetric black holes in dRGT massive gravity theory, where the charged (neutral) black holes can admit at most four (three) event horizons in certain parameter space $(\alpha, \beta)$ of the theory.
In these static and spherically symmetric black hole backgrounds, we studied  geodesics of massive and massless test particles with particular attention to the circular geodesics.\\

\noindent
Owing to the structure of the black holes in the massive gravity, the particle trajectories around these black holes could be diverse in nature. Therefore, via a phase-plane analysis, we first examined the fixed points and the phase portraits of these
particles. We found that the fixed points, that lie outside the outer horizon, are either centers or  saddle-nodes or  both. Consequently, the trajectories offer closed orbits, including the circular one. Subsequently, we focussed on the possible existence of circular orbits for a massive test particle when its  orbital
angular momentum vanishes. These type of circular orbits are called as static spheres, where one may realise the construction of a static Dyson-like sphere~\cite{Wei:2023bgp}.
We found that  a pair of stable and unstable static spheres exists in both neutral and charged black hole backgrounds.
Making use of a topological argument for static spheres,   we showed that, in a general static and spherically symmetric black hole (charged or neutral) background in massive gravity,  the stable and unstable static spheres always
come in pairs, if they exist. The stable and unstable static sphere possess the topological charge $+1$ and $-1$, respectively. These topological classifications are found to be in agreement with the analysis done for a dyonic black hole in quasitopological electromagnetism~\cite{Wei:2023bgp}. 
\vskip 0.2cm 
\noindent
Next, we searched for the existence of stable photon spheres. We observed that, for a given angular momentum $L$, there
exists a pair of stable and unstable photon spheres in both neutral and charged black hole backgrounds. In~\cite{Wei:2023fkn}, the presence of static spheres and stable photon spheres around a static and spherically symmetric
black hole, signalled  the occurrence of Aschenbach effect. Motivated by this, we focused on the properties of time-like circular orbits (TCOs) in order to see such effect in the massive gravity theory, with results distinct from Einstein gravity.
\vskip 0.2cm 
\noindent
In the case of dyonic black hole in quasi-topological electromagnetism, there exist two pairs of TCOs  for a given
angular momentum~\cite{Wei:2023fkn}. In contrast, for the neutral and charged black hole backgrounds in massive gravity considered here, we found only one pair of stable and unstable TCOs, such that the stable one
is the outer TCO, and the unstable one is the inner TCO. Moreover, it is observed that certain TCOs have imaginary angular momentum, due to which  the stable and unstable TCO regions are disconnected.
\vskip 0.2cm 
\noindent
We then proceeded by computing the angular velocity $\Omega_{\text{\tiny CO}}$ of the TCOs, measured by a static observer at rest. We found the monotonically increasing behaviour of $\Omega_{\text{\tiny CO}}$ with radius of the TCO, in the stable region of TCOs region, near the stable photon sphere. This then confirms the existence of Aschenbach effect for time-like particle motion around both neutral and charged black hole backgrounds in massive gravity. 
\vskip 0.2cm 
\noindent
Together with the dyonic black hole in quasi-topological electromagnetism~\cite{Wei:2023fkn,Wei:2023bgp},  the neutral and charged black hole backgrounds in massive gravity studied in this work are the other examples for existence of  static spheres and  the Aschenbach effect as well.
It would be interesting to find other systems, that show these properties.  Finding a general condition, possibly using the topological arguments, which could allow prediction of static spheres and Aschenbach effect, would be nice to explore. 
Checking the occurrence of Aschenbach effect with a charged test particle for our system and also for dyonic black hole in quasi-topological electromagnetism~\cite{Wei:2023fkn}, in addition to studying the effect of charge of the black hole, are possible directions for future work. \\

\vskip 0.2cm
\noindent
There are other interesting avenues for future work. From a theoretical standpoint, photon spheres play an important role in imaging Einstein rings of black holes in AdS~\cite{Hashimoto:2018okj,Hashimoto:2019jmw,Liu:2022cev}, in bringing out the conformal symmetries \cite{Perlick:2004tq,Grenzebach:2014fha,Grenzebach:2015oea} in asymptotically flat spacetime backgrounds\cite{Johnson:2019ljv,Gralla:2019drh,Himwich:2020msm,Hadar:2022xag}, as well as leading to interesting implications for warped AdS$_3$ black holes \cite{Kapec:2022dvc}, which can be explored in the massive gravity context. In particular, for the investigations of AdS/CMT (condensed matter theory) correspondence \cite{hartnoll2018holographic}, the techniques of optical imaging ~\cite{Hashimoto:2018okj,Hashimoto:2019jmw,Kaku:2021xqp,Liu:2022cev,Hashimoto:2022aso,Caron-Huot:2022lff,Zeng:2023zlf} can be used to find holographic gravitational duals to real materials~\cite{Kinoshita:2023hgc}, when there are impurities or defects in the lattice structure, using holographic massive gravity models~\cite{Vegh,Davison:2013jba,Blake:2013bqa,Cao:2015cza,Baggioli:2014roa}. 
There are further intriguing connections between photon spheres and quasinormal modes in the holographic set up~\cite{Bardeen:1972fi,Ching:1994bd,Ching:1995tj,Horowitz:1999jd,Cardoso:2008bp,Berti:2009kk,Daghigh:2022uws,PERLICK20221,Bardeen:2018omt,Qiao:2022hfv,Hashimoto:2018okj,Hashimoto:2019jmw}, as well as, ISCO's, anamalous dimensions of certain double twist operators in CFT, and the weak gravity conjecture~\cite{Arkani-Hamed:2006emk,Harlow:2022ich,Festuccia:2008zx,Riojas:2023pew,Berenstein:2020vlp,Dodelson:2023nnr,Moitra:2023yyc,Paul:2024khi}, which should also be explored for the class of massive gravity theories considered in this work. We leave these issues for feature work.\\

\section*{Acknowledgements}
We have benefitted from discussions with Chethan Gowdigere and Arindam Lala.  CB is grateful to CERN, Geneva for financial support and hospitality through Strings 2024 fellowship. CB also thanks Iosif Bena and Nick Warner for warm hospitality at IPhT, Saclay, where some part of this work was done, and Soumangsu Chakraborty for helpful discussions. 

\bibliographystyle{apsrev4-1}
\bibliography{da_massive}

\begin{thebibliography}{169}%
\makeatletter
\providecommand \@ifxundefined [1]{%
 \@ifx{#1\undefined}
}%
\providecommand \@ifnum [1]{%
 \ifnum #1\expandafter \@firstoftwo
 \else \expandafter \@secondoftwo
 \fi
}%
\providecommand \@ifx [1]{%
 \ifx #1\expandafter \@firstoftwo
 \else \expandafter \@secondoftwo
 \fi
}%
\providecommand \natexlab [1]{#1}%
\providecommand \enquote  [1]{``#1''}%
\providecommand \bibnamefont  [1]{#1}%
\providecommand \bibfnamefont [1]{#1}%
\providecommand \citenamefont [1]{#1}%
\providecommand \href@noop [0]{\@secondoftwo}%
\providecommand \href [0]{\begingroup \@sanitize@url \@href}%
\providecommand \@href[1]{\@@startlink{#1}\@@href}%
\providecommand \@@href[1]{\endgroup#1\@@endlink}%
\providecommand \@sanitize@url [0]{\catcode `\\12\catcode `\$12\catcode
  `\&12\catcode `\#12\catcode `\^12\catcode `\_12\catcode `\%12\relax}%
\providecommand \@@startlink[1]{}%
\providecommand \@@endlink[0]{}%
\providecommand \url  [0]{\begingroup\@sanitize@url \@url }%
\providecommand \@url [1]{\endgroup\@href {#1}{\urlprefix }}%
\providecommand \urlprefix  [0]{URL }%
\providecommand \Eprint [0]{\href }%
\providecommand \doibase [0]{http://dx.doi.org/}%
\providecommand \selectlanguage [0]{\@gobble}%
\providecommand \bibinfo  [0]{\@secondoftwo}%
\providecommand \bibfield  [0]{\@secondoftwo}%
\providecommand \translation [1]{[#1]}%
\providecommand \BibitemOpen [0]{}%
\providecommand \bibitemStop [0]{}%
\providecommand \bibitemNoStop [0]{.\EOS\space}%
\providecommand \EOS [0]{\spacefactor3000\relax}%
\providecommand \BibitemShut  [1]{\csname bibitem#1\endcsname}%
\let\auto@bib@innerbib\@empty
\bibitem [{\citenamefont {Abbott}\ \emph {et~al.}(2016)\citenamefont {Abbott}
  \emph {et~al.}}]{LIGOScientific:2016aoc}%
  \BibitemOpen
  \bibfield  {author} {\bibinfo {author} {\bibfnamefont {B.~P.}\ \bibnamefont
  {Abbott}} \emph {et~al.} (\bibinfo {collaboration} {LIGO Scientific,
  Virgo}),\ }\href {\doibase 10.1103/PhysRevLett.116.061102} {\bibfield
  {journal} {\bibinfo  {journal} {Phys. Rev. Lett.}\ }\textbf {\bibinfo
  {volume} {116}},\ \bibinfo {pages} {061102} (\bibinfo {year} {2016})},\
  \Eprint {http://arxiv.org/abs/1602.03837} {arXiv:1602.03837 [gr-qc]}
  \BibitemShut {NoStop}%
\bibitem [{\citenamefont {Akiyama}\ \emph {et~al.}(2022)\citenamefont {Akiyama}
  \emph {et~al.}}]{EventHorizonTelescope:2022wkp}%
  \BibitemOpen
  \bibfield  {author} {\bibinfo {author} {\bibfnamefont {K.}~\bibnamefont
  {Akiyama}} \emph {et~al.} (\bibinfo {collaboration} {Event Horizon
  Telescope}),\ }\href {\doibase 10.3847/2041-8213/ac6674} {\bibfield
  {journal} {\bibinfo  {journal} {Astrophys. J. Lett.}\ }\textbf {\bibinfo
  {volume} {930}},\ \bibinfo {pages} {L12} (\bibinfo {year} {2022})},\ \Eprint
  {http://arxiv.org/abs/2311.08680} {arXiv:2311.08680 [astro-ph.HE]}
  \BibitemShut {NoStop}%
\bibitem [{\citenamefont {Akiyama}\ \emph {et~al.}(2019)\citenamefont {Akiyama}
  \emph {et~al.}}]{EventHorizonTelescope:2019dse}%
  \BibitemOpen
  \bibfield  {author} {\bibinfo {author} {\bibfnamefont {K.}~\bibnamefont
  {Akiyama}} \emph {et~al.} (\bibinfo {collaboration} {Event Horizon
  Telescope}),\ }\href {\doibase 10.3847/2041-8213/ab0ec7} {\bibfield
  {journal} {\bibinfo  {journal} {Astrophys. J. Lett.}\ }\textbf {\bibinfo
  {volume} {875}},\ \bibinfo {pages} {L1} (\bibinfo {year} {2019})},\ \Eprint
  {http://arxiv.org/abs/1906.11238} {arXiv:1906.11238 [astro-ph.GA]}
  \BibitemShut {NoStop}%
\bibitem [{\citenamefont {Grandclement}\ \emph {et~al.}(2014)\citenamefont
  {Grandclement}, \citenamefont {Som\'e},\ and\ \citenamefont
  {Gourgoulhon}}]{Grandclement:2014msa}%
  \BibitemOpen
  \bibfield  {author} {\bibinfo {author} {\bibfnamefont {P.}~\bibnamefont
  {Grandclement}}, \bibinfo {author} {\bibfnamefont {C.}~\bibnamefont
  {Som\'e}}, \ and\ \bibinfo {author} {\bibfnamefont {E.}~\bibnamefont
  {Gourgoulhon}},\ }\href {\doibase 10.1103/PhysRevD.90.024068} {\bibfield
  {journal} {\bibinfo  {journal} {Phys. Rev. D}\ }\textbf {\bibinfo {volume}
  {90}},\ \bibinfo {pages} {024068} (\bibinfo {year} {2014})},\ \Eprint
  {http://arxiv.org/abs/1405.4837} {arXiv:1405.4837 [gr-qc]} \BibitemShut
  {NoStop}%
\bibitem [{\citenamefont {Grould}\ \emph {et~al.}(2017)\citenamefont {Grould},
  \citenamefont {Meliani}, \citenamefont {Vincent}, \citenamefont
  {Grandcl\'ement},\ and\ \citenamefont {Gourgoulhon}}]{Grould:2017rzz}%
  \BibitemOpen
  \bibfield  {author} {\bibinfo {author} {\bibfnamefont {M.}~\bibnamefont
  {Grould}}, \bibinfo {author} {\bibfnamefont {Z.}~\bibnamefont {Meliani}},
  \bibinfo {author} {\bibfnamefont {F.~H.}\ \bibnamefont {Vincent}}, \bibinfo
  {author} {\bibfnamefont {P.}~\bibnamefont {Grandcl\'ement}}, \ and\ \bibinfo
  {author} {\bibfnamefont {E.}~\bibnamefont {Gourgoulhon}},\ }\href {\doibase
  10.1088/1361-6382/aa8d39} {\bibfield  {journal} {\bibinfo  {journal} {Class.
  Quant. Grav.}\ }\textbf {\bibinfo {volume} {34}},\ \bibinfo {pages} {215007}
  (\bibinfo {year} {2017})},\ \Eprint {http://arxiv.org/abs/1709.05938}
  {arXiv:1709.05938 [astro-ph.HE]} \BibitemShut {NoStop}%
\bibitem [{\citenamefont {Teodoro}\ \emph
  {et~al.}(2021{\natexlab{a}})\citenamefont {Teodoro}, \citenamefont
  {Collodel},\ and\ \citenamefont {Kunz}}]{Teodoro:2020kok}%
  \BibitemOpen
  \bibfield  {author} {\bibinfo {author} {\bibfnamefont {M.~C.}\ \bibnamefont
  {Teodoro}}, \bibinfo {author} {\bibfnamefont {L.~G.}\ \bibnamefont
  {Collodel}}, \ and\ \bibinfo {author} {\bibfnamefont {J.}~\bibnamefont
  {Kunz}},\ }\href {\doibase 10.1088/1475-7516/2021/03/063} {\bibfield
  {journal} {\bibinfo  {journal} {JCAP}\ }\textbf {\bibinfo {volume} {03}},\
  \bibinfo {pages} {063} (\bibinfo {year} {2021}{\natexlab{a}})},\ \Eprint
  {http://arxiv.org/abs/2011.10288} {arXiv:2011.10288 [gr-qc]} \BibitemShut
  {NoStop}%
\bibitem [{\citenamefont {Teodoro}\ \emph
  {et~al.}(2021{\natexlab{b}})\citenamefont {Teodoro}, \citenamefont
  {Collodel}, \citenamefont {Doneva}, \citenamefont {Kunz}, \citenamefont
  {Nedkova},\ and\ \citenamefont {Yazadjiev}}]{Teodoro:2021ezj}%
  \BibitemOpen
  \bibfield  {author} {\bibinfo {author} {\bibfnamefont {M.~C.}\ \bibnamefont
  {Teodoro}}, \bibinfo {author} {\bibfnamefont {L.~G.}\ \bibnamefont
  {Collodel}}, \bibinfo {author} {\bibfnamefont {D.}~\bibnamefont {Doneva}},
  \bibinfo {author} {\bibfnamefont {J.}~\bibnamefont {Kunz}}, \bibinfo {author}
  {\bibfnamefont {P.}~\bibnamefont {Nedkova}}, \ and\ \bibinfo {author}
  {\bibfnamefont {S.}~\bibnamefont {Yazadjiev}},\ }\href {\doibase
  10.1103/PhysRevD.104.124047} {\bibfield  {journal} {\bibinfo  {journal}
  {Phys. Rev. D}\ }\textbf {\bibinfo {volume} {104}},\ \bibinfo {pages}
  {124047} (\bibinfo {year} {2021}{\natexlab{b}})},\ \Eprint
  {http://arxiv.org/abs/2108.08640} {arXiv:2108.08640 [gr-qc]} \BibitemShut
  {NoStop}%
\bibitem [{\citenamefont {Gibbons}\ and\ \citenamefont
  {Herdeiro}(1999)}]{Gibbons:1999uv}%
  \BibitemOpen
  \bibfield  {author} {\bibinfo {author} {\bibfnamefont {G.~W.}\ \bibnamefont
  {Gibbons}}\ and\ \bibinfo {author} {\bibfnamefont {C.~A.~R.}\ \bibnamefont
  {Herdeiro}},\ }\href {\doibase 10.1088/0264-9381/16/11/311} {\bibfield
  {journal} {\bibinfo  {journal} {Class. Quant. Grav.}\ }\textbf {\bibinfo
  {volume} {16}},\ \bibinfo {pages} {3619} (\bibinfo {year} {1999})},\ \Eprint
  {http://arxiv.org/abs/hep-th/9906098} {arXiv:hep-th/9906098} \BibitemShut
  {NoStop}%
\bibitem [{\citenamefont {Herdeiro}(2000)}]{Herdeiro:2000ap}%
  \BibitemOpen
  \bibfield  {author} {\bibinfo {author} {\bibfnamefont {C.~A.~R.}\
  \bibnamefont {Herdeiro}},\ }\href {\doibase 10.1016/S0550-3213(00)00335-7}
  {\bibfield  {journal} {\bibinfo  {journal} {Nucl. Phys. B}\ }\textbf
  {\bibinfo {volume} {582}},\ \bibinfo {pages} {363} (\bibinfo {year}
  {2000})},\ \Eprint {http://arxiv.org/abs/hep-th/0003063}
  {arXiv:hep-th/0003063} \BibitemShut {NoStop}%
\bibitem [{\citenamefont {Diemer}\ and\ \citenamefont
  {Kunz}(2014)}]{Diemer:2013fza}%
  \BibitemOpen
  \bibfield  {author} {\bibinfo {author} {\bibfnamefont {V.}~\bibnamefont
  {Diemer}}\ and\ \bibinfo {author} {\bibfnamefont {J.}~\bibnamefont {Kunz}},\
  }\href {\doibase 10.1103/PhysRevD.89.084001} {\bibfield  {journal} {\bibinfo
  {journal} {Phys. Rev. D}\ }\textbf {\bibinfo {volume} {89}},\ \bibinfo
  {pages} {084001} (\bibinfo {year} {2014})},\ \Eprint
  {http://arxiv.org/abs/1312.6540} {arXiv:1312.6540 [gr-qc]} \BibitemShut
  {NoStop}%
\bibitem [{\citenamefont {Delgado}\ \emph {et~al.}(2022)\citenamefont
  {Delgado}, \citenamefont {Herdeiro},\ and\ \citenamefont
  {Radu}}]{Delgado:2021jxd}%
  \BibitemOpen
  \bibfield  {author} {\bibinfo {author} {\bibfnamefont {J.~F.~M.}\
  \bibnamefont {Delgado}}, \bibinfo {author} {\bibfnamefont {C.~A.~R.}\
  \bibnamefont {Herdeiro}}, \ and\ \bibinfo {author} {\bibfnamefont
  {E.}~\bibnamefont {Radu}},\ }\href {\doibase 10.1103/PhysRevD.105.064026}
  {\bibfield  {journal} {\bibinfo  {journal} {Phys. Rev. D}\ }\textbf {\bibinfo
  {volume} {105}},\ \bibinfo {pages} {064026} (\bibinfo {year} {2022})},\
  \Eprint {http://arxiv.org/abs/2107.03404} {arXiv:2107.03404 [gr-qc]}
  \BibitemShut {NoStop}%
\bibitem [{\citenamefont {Stuchlik}\ and\ \citenamefont
  {Hledik}(1999)}]{Stuchlik:1999qk}%
  \BibitemOpen
  \bibfield  {author} {\bibinfo {author} {\bibfnamefont {Z.}~\bibnamefont
  {Stuchlik}}\ and\ \bibinfo {author} {\bibfnamefont {S.}~\bibnamefont
  {Hledik}},\ }\href {\doibase 10.1103/PhysRevD.60.044006} {\bibfield
  {journal} {\bibinfo  {journal} {Phys. Rev. D}\ }\textbf {\bibinfo {volume}
  {60}},\ \bibinfo {pages} {044006} (\bibinfo {year} {1999})}\BibitemShut
  {NoStop}%
\bibitem [{\citenamefont {Claudel}\ \emph {et~al.}(2001)\citenamefont
  {Claudel}, \citenamefont {Virbhadra},\ and\ \citenamefont
  {Ellis}}]{Claudel:2000yi}%
  \BibitemOpen
  \bibfield  {author} {\bibinfo {author} {\bibfnamefont {C.-M.}\ \bibnamefont
  {Claudel}}, \bibinfo {author} {\bibfnamefont {K.~S.}\ \bibnamefont
  {Virbhadra}}, \ and\ \bibinfo {author} {\bibfnamefont {G.~F.~R.}\
  \bibnamefont {Ellis}},\ }\href {\doibase 10.1063/1.1308507} {\bibfield
  {journal} {\bibinfo  {journal} {J. Math. Phys.}\ }\textbf {\bibinfo {volume}
  {42}},\ \bibinfo {pages} {818} (\bibinfo {year} {2001})},\ \Eprint
  {http://arxiv.org/abs/gr-qc/0005050} {arXiv:gr-qc/0005050} \BibitemShut
  {NoStop}%
\bibitem [{\citenamefont {Virbhadra}\ and\ \citenamefont
  {Ellis}(2000)}]{Virbhadra:1999nm}%
  \BibitemOpen
  \bibfield  {author} {\bibinfo {author} {\bibfnamefont {K.~S.}\ \bibnamefont
  {Virbhadra}}\ and\ \bibinfo {author} {\bibfnamefont {G.~F.~R.}\ \bibnamefont
  {Ellis}},\ }\href {\doibase 10.1103/PhysRevD.62.084003} {\bibfield  {journal}
  {\bibinfo  {journal} {Phys. Rev. D}\ }\textbf {\bibinfo {volume} {62}},\
  \bibinfo {pages} {084003} (\bibinfo {year} {2000})},\ \Eprint
  {http://arxiv.org/abs/astro-ph/9904193} {arXiv:astro-ph/9904193} \BibitemShut
  {NoStop}%
\bibitem [{\citenamefont {Virbhadra}\ and\ \citenamefont
  {Ellis}(2002)}]{Virbhadra:2002ju}%
  \BibitemOpen
  \bibfield  {author} {\bibinfo {author} {\bibfnamefont {K.~S.}\ \bibnamefont
  {Virbhadra}}\ and\ \bibinfo {author} {\bibfnamefont {G.~F.~R.}\ \bibnamefont
  {Ellis}},\ }\href {\doibase 10.1103/PhysRevD.65.103004} {\bibfield  {journal}
  {\bibinfo  {journal} {Phys. Rev. D}\ }\textbf {\bibinfo {volume} {65}},\
  \bibinfo {pages} {103004} (\bibinfo {year} {2002})}\BibitemShut {NoStop}%
\bibitem [{\citenamefont {Levin}\ and\ \citenamefont
  {Perez-Giz}(2008)}]{Levin:2008mq}%
  \BibitemOpen
  \bibfield  {author} {\bibinfo {author} {\bibfnamefont {J.}~\bibnamefont
  {Levin}}\ and\ \bibinfo {author} {\bibfnamefont {G.}~\bibnamefont
  {Perez-Giz}},\ }\href {\doibase 10.1103/PhysRevD.77.103005} {\bibfield
  {journal} {\bibinfo  {journal} {Phys. Rev. D}\ }\textbf {\bibinfo {volume}
  {77}},\ \bibinfo {pages} {103005} (\bibinfo {year} {2008})},\ \Eprint
  {http://arxiv.org/abs/0802.0459} {arXiv:0802.0459 [gr-qc]} \BibitemShut
  {NoStop}%
\bibitem [{\citenamefont {Pugliese}\ \emph {et~al.}(2011)\citenamefont
  {Pugliese}, \citenamefont {Quevedo},\ and\ \citenamefont
  {Ruffini}}]{Pugliese:2010ps}%
  \BibitemOpen
  \bibfield  {author} {\bibinfo {author} {\bibfnamefont {D.}~\bibnamefont
  {Pugliese}}, \bibinfo {author} {\bibfnamefont {H.}~\bibnamefont {Quevedo}}, \
  and\ \bibinfo {author} {\bibfnamefont {R.}~\bibnamefont {Ruffini}},\ }\href
  {\doibase 10.1103/PhysRevD.83.024021} {\bibfield  {journal} {\bibinfo
  {journal} {Phys. Rev. D}\ }\textbf {\bibinfo {volume} {83}},\ \bibinfo
  {pages} {024021} (\bibinfo {year} {2011})},\ \Eprint
  {http://arxiv.org/abs/1012.5411} {arXiv:1012.5411 [astro-ph.HE]} \BibitemShut
  {NoStop}%
\bibitem [{\citenamefont {Hackmann}\ \emph {et~al.}(2010)\citenamefont
  {Hackmann}, \citenamefont {Lammerzahl}, \citenamefont {Kagramanova},\ and\
  \citenamefont {Kunz}}]{Hackmann:2010zz}%
  \BibitemOpen
  \bibfield  {author} {\bibinfo {author} {\bibfnamefont {E.}~\bibnamefont
  {Hackmann}}, \bibinfo {author} {\bibfnamefont {C.}~\bibnamefont
  {Lammerzahl}}, \bibinfo {author} {\bibfnamefont {V.}~\bibnamefont
  {Kagramanova}}, \ and\ \bibinfo {author} {\bibfnamefont {J.}~\bibnamefont
  {Kunz}},\ }\href {\doibase 10.1103/PhysRevD.81.044020} {\bibfield  {journal}
  {\bibinfo  {journal} {Phys. Rev. D}\ }\textbf {\bibinfo {volume} {81}},\
  \bibinfo {pages} {044020} (\bibinfo {year} {2010})},\ \Eprint
  {http://arxiv.org/abs/1009.6117} {arXiv:1009.6117 [gr-qc]} \BibitemShut
  {NoStop}%
\bibitem [{\citenamefont {Villanueva}\ \emph {et~al.}(2013)\citenamefont
  {Villanueva}, \citenamefont {Saavedra}, \citenamefont {Olivares},\ and\
  \citenamefont {Cruz}}]{Villanueva:2013zta}%
  \BibitemOpen
  \bibfield  {author} {\bibinfo {author} {\bibfnamefont {J.~R.}\ \bibnamefont
  {Villanueva}}, \bibinfo {author} {\bibfnamefont {J.}~\bibnamefont
  {Saavedra}}, \bibinfo {author} {\bibfnamefont {M.}~\bibnamefont {Olivares}},
  \ and\ \bibinfo {author} {\bibfnamefont {N.}~\bibnamefont {Cruz}},\ }\href
  {\doibase 10.1007/s10509-012-1333-x} {\bibfield  {journal} {\bibinfo
  {journal} {Astrophys. Space Sci.}\ }\textbf {\bibinfo {volume} {344}},\
  \bibinfo {pages} {437} (\bibinfo {year} {2013})}\BibitemShut {NoStop}%
\bibitem [{\citenamefont {Wei}\ and\ \citenamefont {Liu}(2018)}]{Wei:2017mwc}%
  \BibitemOpen
  \bibfield  {author} {\bibinfo {author} {\bibfnamefont {S.-W.}\ \bibnamefont
  {Wei}}\ and\ \bibinfo {author} {\bibfnamefont {Y.-X.}\ \bibnamefont {Liu}},\
  }\href {\doibase 10.1103/PhysRevD.97.104027} {\bibfield  {journal} {\bibinfo
  {journal} {Phys. Rev. D}\ }\textbf {\bibinfo {volume} {97}},\ \bibinfo
  {pages} {104027} (\bibinfo {year} {2018})},\ \Eprint
  {http://arxiv.org/abs/1711.01522} {arXiv:1711.01522 [gr-qc]} \BibitemShut
  {NoStop}%
\bibitem [{\citenamefont {Chandrasekhar}\ and\ \citenamefont
  {Mohapatra}(2019)}]{Chandrasekhar:2018sjg}%
  \BibitemOpen
  \bibfield  {author} {\bibinfo {author} {\bibfnamefont {B.}~\bibnamefont
  {Chandrasekhar}}\ and\ \bibinfo {author} {\bibfnamefont {S.}~\bibnamefont
  {Mohapatra}},\ }\href {\doibase 10.1016/j.physletb.2019.02.042} {\bibfield
  {journal} {\bibinfo  {journal} {Phys. Lett. B}\ }\textbf {\bibinfo {volume}
  {791}},\ \bibinfo {pages} {367} (\bibinfo {year} {2019})},\ \Eprint
  {http://arxiv.org/abs/1805.05088} {arXiv:1805.05088 [hep-th]} \BibitemShut
  {NoStop}%
\bibitem [{\citenamefont {Leh\'ebel}\ and\ \citenamefont
  {Cardoso}(2022)}]{Lehebel:2022yyz}%
  \BibitemOpen
  \bibfield  {author} {\bibinfo {author} {\bibfnamefont {A.}~\bibnamefont
  {Leh\'ebel}}\ and\ \bibinfo {author} {\bibfnamefont {V.}~\bibnamefont
  {Cardoso}},\ }\href {\doibase 10.1103/PhysRevD.105.064014} {\bibfield
  {journal} {\bibinfo  {journal} {Phys. Rev. D}\ }\textbf {\bibinfo {volume}
  {105}},\ \bibinfo {pages} {064014} (\bibinfo {year} {2022})},\ \Eprint
  {http://arxiv.org/abs/2202.08850} {arXiv:2202.08850 [gr-qc]} \BibitemShut
  {NoStop}%
\bibitem [{\citenamefont {Bozza}(2002)}]{Bozza:2002zj}%
  \BibitemOpen
  \bibfield  {author} {\bibinfo {author} {\bibfnamefont {V.}~\bibnamefont
  {Bozza}},\ }\href {\doibase 10.1103/PhysRevD.66.103001} {\bibfield  {journal}
  {\bibinfo  {journal} {Phys. Rev. D}\ }\textbf {\bibinfo {volume} {66}},\
  \bibinfo {pages} {103001} (\bibinfo {year} {2002})},\ \Eprint
  {http://arxiv.org/abs/gr-qc/0208075} {arXiv:gr-qc/0208075} \BibitemShut
  {NoStop}%
\bibitem [{\citenamefont {Cardoso}\ \emph {et~al.}(2009)\citenamefont
  {Cardoso}, \citenamefont {Miranda}, \citenamefont {Berti}, \citenamefont
  {Witek},\ and\ \citenamefont {Zanchin}}]{Cardoso:2008bp}%
  \BibitemOpen
  \bibfield  {author} {\bibinfo {author} {\bibfnamefont {V.}~\bibnamefont
  {Cardoso}}, \bibinfo {author} {\bibfnamefont {A.~S.}\ \bibnamefont
  {Miranda}}, \bibinfo {author} {\bibfnamefont {E.}~\bibnamefont {Berti}},
  \bibinfo {author} {\bibfnamefont {H.}~\bibnamefont {Witek}}, \ and\ \bibinfo
  {author} {\bibfnamefont {V.~T.}\ \bibnamefont {Zanchin}},\ }\href {\doibase
  10.1103/PhysRevD.79.064016} {\bibfield  {journal} {\bibinfo  {journal} {Phys.
  Rev. D}\ }\textbf {\bibinfo {volume} {79}},\ \bibinfo {pages} {064016}
  (\bibinfo {year} {2009})},\ \Eprint {http://arxiv.org/abs/0812.1806}
  {arXiv:0812.1806 [hep-th]} \BibitemShut {NoStop}%
\bibitem [{\citenamefont {Hioki}\ and\ \citenamefont
  {Maeda}(2009)}]{Hioki:2009na}%
  \BibitemOpen
  \bibfield  {author} {\bibinfo {author} {\bibfnamefont {K.}~\bibnamefont
  {Hioki}}\ and\ \bibinfo {author} {\bibfnamefont {K.-i.}\ \bibnamefont
  {Maeda}},\ }\href {\doibase 10.1103/PhysRevD.80.024042} {\bibfield  {journal}
  {\bibinfo  {journal} {Phys. Rev. D}\ }\textbf {\bibinfo {volume} {80}},\
  \bibinfo {pages} {024042} (\bibinfo {year} {2009})},\ \Eprint
  {http://arxiv.org/abs/0904.3575} {arXiv:0904.3575 [astro-ph.HE]} \BibitemShut
  {NoStop}%
\bibitem [{\citenamefont {Collodel}\ \emph {et~al.}(2018)\citenamefont
  {Collodel}, \citenamefont {Kleihaus},\ and\ \citenamefont
  {Kunz}}]{Collodel:2017end}%
  \BibitemOpen
  \bibfield  {author} {\bibinfo {author} {\bibfnamefont {L.~G.}\ \bibnamefont
  {Collodel}}, \bibinfo {author} {\bibfnamefont {B.}~\bibnamefont {Kleihaus}},
  \ and\ \bibinfo {author} {\bibfnamefont {J.}~\bibnamefont {Kunz}},\ }\href
  {\doibase 10.1103/PhysRevLett.120.201103} {\bibfield  {journal} {\bibinfo
  {journal} {Phys. Rev. Lett.}\ }\textbf {\bibinfo {volume} {120}},\ \bibinfo
  {pages} {201103} (\bibinfo {year} {2018})},\ \Eprint
  {http://arxiv.org/abs/1711.05191} {arXiv:1711.05191 [gr-qc]} \BibitemShut
  {NoStop}%
\bibitem [{\citenamefont {Ye}\ and\ \citenamefont {Wei}(2023)}]{Ye:2023gmk}%
  \BibitemOpen
  \bibfield  {author} {\bibinfo {author} {\bibfnamefont {X.}~\bibnamefont
  {Ye}}\ and\ \bibinfo {author} {\bibfnamefont {S.-W.}\ \bibnamefont {Wei}},\
  }\href {\doibase 10.1088/1475-7516/2023/07/049} {\bibfield  {journal}
  {\bibinfo  {journal} {JCAP}\ }\textbf {\bibinfo {volume} {07}},\ \bibinfo
  {pages} {049} (\bibinfo {year} {2023})},\ \Eprint
  {http://arxiv.org/abs/2301.04786} {arXiv:2301.04786 [gr-qc]} \BibitemShut
  {NoStop}%
\bibitem [{\citenamefont {Cunha}\ and\ \citenamefont
  {Herdeiro}(2020)}]{Cunha:2020azh}%
  \BibitemOpen
  \bibfield  {author} {\bibinfo {author} {\bibfnamefont {P.~V.~P.}\
  \bibnamefont {Cunha}}\ and\ \bibinfo {author} {\bibfnamefont {C.~A.~R.}\
  \bibnamefont {Herdeiro}},\ }\href {\doibase 10.1103/PhysRevLett.124.181101}
  {\bibfield  {journal} {\bibinfo  {journal} {Phys. Rev. Lett.}\ }\textbf
  {\bibinfo {volume} {124}},\ \bibinfo {pages} {181101} (\bibinfo {year}
  {2020})},\ \Eprint {http://arxiv.org/abs/2003.06445} {arXiv:2003.06445
  [gr-qc]} \BibitemShut {NoStop}%
\bibitem [{\citenamefont {Cunha}\ \emph {et~al.}(2023)\citenamefont {Cunha},
  \citenamefont {Herdeiro}, \citenamefont {Radu},\ and\ \citenamefont
  {Sanchis-Gual}}]{Cunha:2022gde}%
  \BibitemOpen
  \bibfield  {author} {\bibinfo {author} {\bibfnamefont {P.~V.~P.}\
  \bibnamefont {Cunha}}, \bibinfo {author} {\bibfnamefont {C.}~\bibnamefont
  {Herdeiro}}, \bibinfo {author} {\bibfnamefont {E.}~\bibnamefont {Radu}}, \
  and\ \bibinfo {author} {\bibfnamefont {N.}~\bibnamefont {Sanchis-Gual}},\
  }\href {\doibase 10.1103/PhysRevLett.130.061401} {\bibfield  {journal}
  {\bibinfo  {journal} {Phys. Rev. Lett.}\ }\textbf {\bibinfo {volume} {130}},\
  \bibinfo {pages} {061401} (\bibinfo {year} {2023})},\ \Eprint
  {http://arxiv.org/abs/2207.13713} {arXiv:2207.13713 [gr-qc]} \BibitemShut
  {NoStop}%
\bibitem [{\citenamefont {Wei}\ and\ \citenamefont {Liu}(2023)}]{Wei:2022mzv}%
  \BibitemOpen
  \bibfield  {author} {\bibinfo {author} {\bibfnamefont {S.-W.}\ \bibnamefont
  {Wei}}\ and\ \bibinfo {author} {\bibfnamefont {Y.-X.}\ \bibnamefont {Liu}},\
  }\href {\doibase 10.1103/PhysRevD.107.064006} {\bibfield  {journal} {\bibinfo
   {journal} {Phys. Rev. D}\ }\textbf {\bibinfo {volume} {107}},\ \bibinfo
  {pages} {064006} (\bibinfo {year} {2023})},\ \Eprint
  {http://arxiv.org/abs/2207.08397} {arXiv:2207.08397 [gr-qc]} \BibitemShut
  {NoStop}%
\bibitem [{\citenamefont {Liu}\ \emph {et~al.}(2020)\citenamefont {Liu},
  \citenamefont {Mai}, \citenamefont {Li},\ and\ \citenamefont
  {L\"u}}]{Liu:2019rib}%
  \BibitemOpen
  \bibfield  {author} {\bibinfo {author} {\bibfnamefont {H.-S.}\ \bibnamefont
  {Liu}}, \bibinfo {author} {\bibfnamefont {Z.-F.}\ \bibnamefont {Mai}},
  \bibinfo {author} {\bibfnamefont {Y.-Z.}\ \bibnamefont {Li}}, \ and\ \bibinfo
  {author} {\bibfnamefont {H.}~\bibnamefont {L\"u}},\ }\href {\doibase
  10.1007/s11433-019-1446-1} {\bibfield  {journal} {\bibinfo  {journal} {Sci.
  China Phys. Mech. Astron.}\ }\textbf {\bibinfo {volume} {63}},\ \bibinfo
  {pages} {240411} (\bibinfo {year} {2020})},\ \Eprint
  {http://arxiv.org/abs/1907.10876} {arXiv:1907.10876 [hep-th]} \BibitemShut
  {NoStop}%
\bibitem [{\citenamefont {Wei}\ \emph {et~al.}(2023)\citenamefont {Wei},
  \citenamefont {Zhang}, \citenamefont {Liu},\ and\ \citenamefont
  {Mann}}]{Wei:2023bgp}%
  \BibitemOpen
  \bibfield  {author} {\bibinfo {author} {\bibfnamefont {S.-W.}\ \bibnamefont
  {Wei}}, \bibinfo {author} {\bibfnamefont {Y.-P.}\ \bibnamefont {Zhang}},
  \bibinfo {author} {\bibfnamefont {Y.-X.}\ \bibnamefont {Liu}}, \ and\
  \bibinfo {author} {\bibfnamefont {R.~B.}\ \bibnamefont {Mann}},\ }\href
  {\doibase 10.1103/PhysRevResearch.5.043050} {\bibfield  {journal} {\bibinfo
  {journal} {Phys. Rev. Res.}\ }\textbf {\bibinfo {volume} {5}},\ \bibinfo
  {pages} {043050} (\bibinfo {year} {2023})},\ \Eprint
  {http://arxiv.org/abs/2303.06814} {arXiv:2303.06814 [gr-qc]} \BibitemShut
  {NoStop}%
\bibitem [{\citenamefont {Dyson}(1960)}]{Dyson:1960xib}%
  \BibitemOpen
  \bibfield  {author} {\bibinfo {author} {\bibfnamefont {F.~J.}\ \bibnamefont
  {Dyson}},\ }\href {\doibase 10.1126/science.131.3414.1667} {\bibfield
  {journal} {\bibinfo  {journal} {Science}\ }\textbf {\bibinfo {volume}
  {131}},\ \bibinfo {pages} {1667} (\bibinfo {year} {1960})}\BibitemShut
  {NoStop}%
\bibitem [{\citenamefont {Papagiannis}(1985)}]{Papagiannis}%
  \BibitemOpen
  \bibfield  {author} {\bibinfo {author} {\bibfnamefont {M.~D.}\ \bibnamefont
  {Papagiannis}},\ }\href {\doibase 10.1007/978-94-009-5462-5_73} {\bibfield
  {journal} {\bibinfo  {journal} {IAU Symp.}\ }\textbf {\bibinfo {volume}
  {112}},\ \bibinfo {pages} {543} (\bibinfo {year} {1985})}\BibitemShut
  {NoStop}%
\bibitem [{\citenamefont {Wright}(2020)}]{Wright}%
  \BibitemOpen
  \bibfield  {author} {\bibinfo {author} {\bibfnamefont {J.}~\bibnamefont
  {Wright}},\ }\href {\doibase 10.2298/saj2000001w} {\bibfield  {journal}
  {\bibinfo  {journal} {Serb.Astron.J.}\ }\textbf {\bibinfo {volume} {200}},\
  \bibinfo {pages} {1} (\bibinfo {year} {2020})}\BibitemShut {NoStop}%
\bibitem [{\citenamefont {Aschenbach}(2004)}]{Aschenbach:2004kj}%
  \BibitemOpen
  \bibfield  {author} {\bibinfo {author} {\bibfnamefont {B.}~\bibnamefont
  {Aschenbach}},\ }\href {\doibase 10.1051/0004-6361:20041412} {\bibfield
  {journal} {\bibinfo  {journal} {Astron. Astrophys.}\ }\textbf {\bibinfo
  {volume} {425}},\ \bibinfo {pages} {1075} (\bibinfo {year} {2004})},\ \Eprint
  {http://arxiv.org/abs/astro-ph/0406545} {arXiv:astro-ph/0406545} \BibitemShut
  {NoStop}%
\bibitem [{\citenamefont {Thorne}(1974)}]{Thorne:1974ve}%
  \BibitemOpen
  \bibfield  {author} {\bibinfo {author} {\bibfnamefont {K.~S.}\ \bibnamefont
  {Thorne}},\ }\href {\doibase 10.1086/152991} {\bibfield  {journal} {\bibinfo
  {journal} {Astrophys. J.}\ }\textbf {\bibinfo {volume} {191}},\ \bibinfo
  {pages} {507} (\bibinfo {year} {1974})}\BibitemShut {NoStop}%
\bibitem [{\citenamefont {Stuchlik}\ \emph {et~al.}(2005)\citenamefont
  {Stuchlik}, \citenamefont {Slany}, \citenamefont {Torok},\ and\ \citenamefont
  {Abramowicz}}]{Stuchlik:2004wk}%
  \BibitemOpen
  \bibfield  {author} {\bibinfo {author} {\bibfnamefont {Z.}~\bibnamefont
  {Stuchlik}}, \bibinfo {author} {\bibfnamefont {P.}~\bibnamefont {Slany}},
  \bibinfo {author} {\bibfnamefont {G.}~\bibnamefont {Torok}}, \ and\ \bibinfo
  {author} {\bibfnamefont {M.~A.}\ \bibnamefont {Abramowicz}},\ }\href
  {\doibase 10.1103/PhysRevD.71.024037} {\bibfield  {journal} {\bibinfo
  {journal} {Phys. Rev. D}\ }\textbf {\bibinfo {volume} {71}},\ \bibinfo
  {pages} {024037} (\bibinfo {year} {2005})},\ \Eprint
  {http://arxiv.org/abs/gr-qc/0411091} {arXiv:gr-qc/0411091} \BibitemShut
  {NoStop}%
\bibitem [{\citenamefont {Mueller}\ and\ \citenamefont
  {Aschenbach}(2007)}]{Mueller:2007tf}%
  \BibitemOpen
  \bibfield  {author} {\bibinfo {author} {\bibfnamefont {A.}~\bibnamefont
  {Mueller}}\ and\ \bibinfo {author} {\bibfnamefont {B.}~\bibnamefont
  {Aschenbach}},\ }\href {\doibase 10.1088/0264-9381/24/10/009} {\bibfield
  {journal} {\bibinfo  {journal} {Class. Quant. Grav.}\ }\textbf {\bibinfo
  {volume} {24}},\ \bibinfo {pages} {2637} (\bibinfo {year} {2007})},\ \Eprint
  {http://arxiv.org/abs/0704.3963} {arXiv:0704.3963 [gr-qc]} \BibitemShut
  {NoStop}%
\bibitem [{\citenamefont {Stuchlik}\ \emph {et~al.}(2011)\citenamefont
  {Stuchlik}, \citenamefont {Blaschke},\ and\ \citenamefont
  {Slany}}]{Stuchlik:2011sw}%
  \BibitemOpen
  \bibfield  {author} {\bibinfo {author} {\bibfnamefont {Z.}~\bibnamefont
  {Stuchlik}}, \bibinfo {author} {\bibfnamefont {M.}~\bibnamefont {Blaschke}},
  \ and\ \bibinfo {author} {\bibfnamefont {P.}~\bibnamefont {Slany}},\ }\href
  {\doibase 10.1088/0264-9381/28/17/175002} {\bibfield  {journal} {\bibinfo
  {journal} {Class. Quant. Grav.}\ }\textbf {\bibinfo {volume} {28}},\ \bibinfo
  {pages} {175002} (\bibinfo {year} {2011})},\ \Eprint
  {http://arxiv.org/abs/1108.0191} {arXiv:1108.0191 [gr-qc]} \BibitemShut
  {NoStop}%
\bibitem [{\citenamefont {Tursunov}\ \emph {et~al.}(2016)\citenamefont
  {Tursunov}, \citenamefont {Stuchl\'\i{}k},\ and\ \citenamefont
  {Kolo\v{s}}}]{Tursunov:2016dss}%
  \BibitemOpen
  \bibfield  {author} {\bibinfo {author} {\bibfnamefont {A.}~\bibnamefont
  {Tursunov}}, \bibinfo {author} {\bibfnamefont {Z.}~\bibnamefont
  {Stuchl\'\i{}k}}, \ and\ \bibinfo {author} {\bibfnamefont {M.}~\bibnamefont
  {Kolo\v{s}}},\ }\href {\doibase 10.1103/PhysRevD.93.084012} {\bibfield
  {journal} {\bibinfo  {journal} {Phys. Rev. D}\ }\textbf {\bibinfo {volume}
  {93}},\ \bibinfo {pages} {084012} (\bibinfo {year} {2016})},\ \Eprint
  {http://arxiv.org/abs/1603.07264} {arXiv:1603.07264 [gr-qc]} \BibitemShut
  {NoStop}%
\bibitem [{\citenamefont {Khodagholizadeh}\ \emph {et~al.}(2020)\citenamefont
  {Khodagholizadeh}, \citenamefont {Perlick},\ and\ \citenamefont
  {Vahedi}}]{Khodagholizadeh:2020sex}%
  \BibitemOpen
  \bibfield  {author} {\bibinfo {author} {\bibfnamefont {J.}~\bibnamefont
  {Khodagholizadeh}}, \bibinfo {author} {\bibfnamefont {V.}~\bibnamefont
  {Perlick}}, \ and\ \bibinfo {author} {\bibfnamefont {A.}~\bibnamefont
  {Vahedi}},\ }\href {\doibase 10.1103/PhysRevD.102.024021} {\bibfield
  {journal} {\bibinfo  {journal} {Phys. Rev. D}\ }\textbf {\bibinfo {volume}
  {102}},\ \bibinfo {pages} {024021} (\bibinfo {year} {2020})},\ \Eprint
  {http://arxiv.org/abs/2002.04701} {arXiv:2002.04701 [gr-qc]} \BibitemShut
  {NoStop}%
\bibitem [{\citenamefont {Vahedi}\ \emph {et~al.}(2021)\citenamefont {Vahedi},
  \citenamefont {Khodagholizadeh},\ and\ \citenamefont
  {Tursunov}}]{Vahedi:2021ssf}%
  \BibitemOpen
  \bibfield  {author} {\bibinfo {author} {\bibfnamefont {A.}~\bibnamefont
  {Vahedi}}, \bibinfo {author} {\bibfnamefont {J.}~\bibnamefont
  {Khodagholizadeh}}, \ and\ \bibinfo {author} {\bibfnamefont {A.}~\bibnamefont
  {Tursunov}},\ }\href {\doibase 10.1140/epjc/s10052-021-09081-0} {\bibfield
  {journal} {\bibinfo  {journal} {Eur. Phys. J. C}\ }\textbf {\bibinfo {volume}
  {81}},\ \bibinfo {pages} {280} (\bibinfo {year} {2021})},\ \Eprint
  {http://arxiv.org/abs/2103.14912} {arXiv:2103.14912 [gr-qc]} \BibitemShut
  {NoStop}%
\bibitem [{\citenamefont {Hod}(2018)}]{Hod:2018kql}%
  \BibitemOpen
  \bibfield  {author} {\bibinfo {author} {\bibfnamefont {S.}~\bibnamefont
  {Hod}},\ }\href {\doibase 10.1140/epjc/s10052-018-5905-y} {\bibfield
  {journal} {\bibinfo  {journal} {Eur. Phys. J. C}\ }\textbf {\bibinfo {volume}
  {78}},\ \bibinfo {pages} {417} (\bibinfo {year} {2018})},\ \Eprint
  {http://arxiv.org/abs/1811.04948} {arXiv:1811.04948 [gr-qc]} \BibitemShut
  {NoStop}%
\bibitem [{\citenamefont {Abbott}\ and\ \citenamefont {et.
  al.}(2017)}]{LIGO2017}%
  \BibitemOpen
  \bibfield  {author} {\bibinfo {author} {\bibfnamefont {B.~P.}\ \bibnamefont
  {Abbott}}\ and\ \bibinfo {author} {\bibnamefont {et. al.}} (\bibinfo
  {collaboration} {LIGO Scientific and Virgo Collaboration}),\ }\href {\doibase
  10.1103/PhysRevLett.118.221101} {\bibfield  {journal} {\bibinfo  {journal}
  {Phys. Rev. Lett.}\ }\textbf {\bibinfo {volume} {118}},\ \bibinfo {pages}
  {221101} (\bibinfo {year} {2017})}\BibitemShut {NoStop}%
\bibitem [{\citenamefont {de~Rham}(2014{\natexlab{a}})}]{deRham2014Review}%
  \BibitemOpen
  \bibfield  {author} {\bibinfo {author} {\bibfnamefont {C.}~\bibnamefont
  {de~Rham}},\ }\href {\doibase 10.12942/lrr-2014-7} {\bibfield  {journal}
  {\bibinfo  {journal} {Living Rev. Rel.}\ }\textbf {\bibinfo {volume} {17}},\
  \bibinfo {pages} {7} (\bibinfo {year} {2014}{\natexlab{a}})},\ \Eprint
  {http://arxiv.org/abs/1401.4173} {arXiv:1401.4173 [hep-th]} \BibitemShut
  {NoStop}%
\bibitem [{\citenamefont {Dvali}\ \emph
  {et~al.}(2000{\natexlab{a}})\citenamefont {Dvali}, \citenamefont
  {Gabadadze},\ and\ \citenamefont {Porrati}}]{MassiveIb}%
  \BibitemOpen
  \bibfield  {author} {\bibinfo {author} {\bibfnamefont {G.}~\bibnamefont
  {Dvali}}, \bibinfo {author} {\bibfnamefont {G.}~\bibnamefont {Gabadadze}}, \
  and\ \bibinfo {author} {\bibfnamefont {M.}~\bibnamefont {Porrati}},\
  }\href@noop {} {\bibfield  {journal} {\bibinfo  {journal} {Physics Letters
  B}\ }\textbf {\bibinfo {volume} {484}},\ \bibinfo {pages} {112} (\bibinfo
  {year} {2000}{\natexlab{a}})}\BibitemShut {NoStop}%
\bibitem [{\citenamefont {Dvali}\ \emph
  {et~al.}(2000{\natexlab{b}})\citenamefont {Dvali}, \citenamefont
  {Gabadadze},\ and\ \citenamefont {Porrati}}]{MassiveIc}%
  \BibitemOpen
  \bibfield  {author} {\bibinfo {author} {\bibfnamefont {G.}~\bibnamefont
  {Dvali}}, \bibinfo {author} {\bibfnamefont {G.}~\bibnamefont {Gabadadze}}, \
  and\ \bibinfo {author} {\bibfnamefont {M.}~\bibnamefont {Porrati}},\
  }\href@noop {} {\bibfield  {journal} {\bibinfo  {journal} {Physics Letters
  B}\ }\textbf {\bibinfo {volume} {485}},\ \bibinfo {pages} {208} (\bibinfo
  {year} {2000}{\natexlab{b}})}\BibitemShut {NoStop}%
\bibitem [{\citenamefont {Abbott}\ and\ \citenamefont {et.
  al.}(2016)}]{Abbott}%
  \BibitemOpen
  \bibfield  {author} {\bibinfo {author} {\bibfnamefont {B.~P.}\ \bibnamefont
  {Abbott}}\ and\ \bibinfo {author} {\bibnamefont {et. al.}} (\bibinfo
  {collaboration} {LIGO Scientific Collaboration and Virgo Collaboration}),\
  }\href {\doibase 10.1103/PhysRevLett.116.061102} {\bibfield  {journal}
  {\bibinfo  {journal} {Phys. Rev. Lett.}\ }\textbf {\bibinfo {volume} {116}},\
  \bibinfo {pages} {061102} (\bibinfo {year} {2016})}\BibitemShut {NoStop}%
\bibitem [{\citenamefont {Fierz}\ and\ \citenamefont
  {Pauli}(1939)}]{Fierz1939}%
  \BibitemOpen
  \bibfield  {author} {\bibinfo {author} {\bibfnamefont {M.}~\bibnamefont
  {Fierz}}\ and\ \bibinfo {author} {\bibfnamefont {W.~E.}\ \bibnamefont
  {Pauli}},\ }\href@noop {} {\bibfield  {journal} {\bibinfo  {journal}
  {Proceedings of the Royal Society of London. Series A. Mathematical and
  Physical Sciences}\ }\textbf {\bibinfo {volume} {173}},\ \bibinfo {pages}
  {211} (\bibinfo {year} {1939})}\BibitemShut {NoStop}%
\bibitem [{\citenamefont {Boulware}\ and\ \citenamefont
  {Deser}(1972)}]{BDghost}%
  \BibitemOpen
  \bibfield  {author} {\bibinfo {author} {\bibfnamefont {D.~G.}\ \bibnamefont
  {Boulware}}\ and\ \bibinfo {author} {\bibfnamefont {S.}~\bibnamefont
  {Deser}},\ }\href {\doibase 10.1103/PhysRevD.6.3368} {\bibfield  {journal}
  {\bibinfo  {journal} {Phys. Rev. D}\ }\textbf {\bibinfo {volume} {6}},\
  \bibinfo {pages} {3368} (\bibinfo {year} {1972})}\BibitemShut {NoStop}%
\bibitem [{\citenamefont {Bergshoeff}\ \emph {et~al.}(2009)\citenamefont
  {Bergshoeff}, \citenamefont {Hohm},\ and\ \citenamefont
  {Townsend}}]{Newmasssive}%
  \BibitemOpen
  \bibfield  {author} {\bibinfo {author} {\bibfnamefont {E.~A.}\ \bibnamefont
  {Bergshoeff}}, \bibinfo {author} {\bibfnamefont {O.}~\bibnamefont {Hohm}}, \
  and\ \bibinfo {author} {\bibfnamefont {P.~K.}\ \bibnamefont {Townsend}},\
  }\href@noop {} {\bibfield  {journal} {\bibinfo  {journal} {Physical Review
  Letters}\ }\textbf {\bibinfo {volume} {102}},\ \bibinfo {pages} {201301}
  (\bibinfo {year} {2009})}\BibitemShut {NoStop}%
\bibitem [{\citenamefont {de~Rham}\ \emph
  {et~al.}(2011{\natexlab{a}})\citenamefont {de~Rham}, \citenamefont
  {Gabadadze},\ and\ \citenamefont {Tolley}}]{dRGTI}%
  \BibitemOpen
  \bibfield  {author} {\bibinfo {author} {\bibfnamefont {C.}~\bibnamefont
  {de~Rham}}, \bibinfo {author} {\bibfnamefont {G.}~\bibnamefont {Gabadadze}},
  \ and\ \bibinfo {author} {\bibfnamefont {A.~J.}\ \bibnamefont {Tolley}},\
  }\href {\doibase 10.1103/PhysRevLett.106.231101} {\bibfield  {journal}
  {\bibinfo  {journal} {Phys. Rev. Lett.}\ }\textbf {\bibinfo {volume} {106}},\
  \bibinfo {pages} {231101} (\bibinfo {year} {2011}{\natexlab{a}})}\BibitemShut
  {NoStop}%
\bibitem [{\citenamefont {de~Rham}\ \emph {et~al.}(2012)\citenamefont
  {de~Rham}, \citenamefont {Gabadadze},\ and\ \citenamefont {Tolley}}]{dRGTII}%
  \BibitemOpen
  \bibfield  {author} {\bibinfo {author} {\bibfnamefont {C.}~\bibnamefont
  {de~Rham}}, \bibinfo {author} {\bibfnamefont {G.}~\bibnamefont {Gabadadze}},
  \ and\ \bibinfo {author} {\bibfnamefont {A.~J.}\ \bibnamefont {Tolley}},\
  }\href@noop {} {\bibfield  {journal} {\bibinfo  {journal} {Physics Letters
  B}\ }\textbf {\bibinfo {volume} {711}},\ \bibinfo {pages} {190} (\bibinfo
  {year} {2012})}\BibitemShut {NoStop}%
\bibitem [{\citenamefont {Myung}\ \emph {et~al.}(2011)\citenamefont {Myung},
  \citenamefont {Kim}, \citenamefont {Moon},\ and\ \citenamefont
  {Park}}]{NewM1}%
  \BibitemOpen
  \bibfield  {author} {\bibinfo {author} {\bibfnamefont {Y.~S.}\ \bibnamefont
  {Myung}}, \bibinfo {author} {\bibfnamefont {Y.-W.}\ \bibnamefont {Kim}},
  \bibinfo {author} {\bibfnamefont {T.}~\bibnamefont {Moon}}, \ and\ \bibinfo
  {author} {\bibfnamefont {Y.-J.}\ \bibnamefont {Park}},\ }\href {\doibase
  10.1103/PhysRevD.84.024044} {\bibfield  {journal} {\bibinfo  {journal} {Phys.
  Rev. D}\ }\textbf {\bibinfo {volume} {84}},\ \bibinfo {pages} {024044}
  (\bibinfo {year} {2011})}\BibitemShut {NoStop}%
\bibitem [{\citenamefont {Bergshoeff}\ \emph {et~al.}(2011)\citenamefont
  {Bergshoeff}, \citenamefont {Hohm}, \citenamefont {Rosseel},\ and\
  \citenamefont {Townsend}}]{NewM2}%
  \BibitemOpen
  \bibfield  {author} {\bibinfo {author} {\bibfnamefont {E.~A.}\ \bibnamefont
  {Bergshoeff}}, \bibinfo {author} {\bibfnamefont {O.}~\bibnamefont {Hohm}},
  \bibinfo {author} {\bibfnamefont {J.}~\bibnamefont {Rosseel}}, \ and\
  \bibinfo {author} {\bibfnamefont {P.~K.}\ \bibnamefont {Townsend}},\ }\href
  {\doibase 10.1103/PhysRevD.83.104038} {\bibfield  {journal} {\bibinfo
  {journal} {Phys. Rev. D}\ }\textbf {\bibinfo {volume} {83}},\ \bibinfo
  {pages} {104038} (\bibinfo {year} {2011})}\BibitemShut {NoStop}%
\bibitem [{\citenamefont {Kim}\ \emph {et~al.}(2013)\citenamefont {Kim},
  \citenamefont {Kulkarni},\ and\ \citenamefont {Yi}}]{NewM3}%
  \BibitemOpen
  \bibfield  {author} {\bibinfo {author} {\bibfnamefont {W.}~\bibnamefont
  {Kim}}, \bibinfo {author} {\bibfnamefont {S.}~\bibnamefont {Kulkarni}}, \
  and\ \bibinfo {author} {\bibfnamefont {S.-H.}\ \bibnamefont {Yi}},\
  }\href@noop {} {\bibfield  {journal} {\bibinfo  {journal} {Journal of High
  Energy Physics}\ }\textbf {\bibinfo {volume} {2013}},\ \bibinfo {pages} {41}
  (\bibinfo {year} {2013})}\BibitemShut {NoStop}%
\bibitem [{\citenamefont {Ay\'on-Beato}\ \emph {et~al.}(2014)\citenamefont
  {Ay\'on-Beato}, \citenamefont {Hassa\"{\i}ne},\ and\ \citenamefont
  {Ju\'arez-Aubry}}]{NewM4}%
  \BibitemOpen
  \bibfield  {author} {\bibinfo {author} {\bibfnamefont {E.}~\bibnamefont
  {Ay\'on-Beato}}, \bibinfo {author} {\bibfnamefont {M.}~\bibnamefont
  {Hassa\"{\i}ne}}, \ and\ \bibinfo {author} {\bibfnamefont {M.~M.}\
  \bibnamefont {Ju\'arez-Aubry}},\ }\href {\doibase 10.1103/PhysRevD.90.044026}
  {\bibfield  {journal} {\bibinfo  {journal} {Phys. Rev. D}\ }\textbf {\bibinfo
  {volume} {90}},\ \bibinfo {pages} {044026} (\bibinfo {year}
  {2014})}\BibitemShut {NoStop}%
\bibitem [{\citenamefont {Myung}(2015)}]{NewM5}%
  \BibitemOpen
  \bibfield  {author} {\bibinfo {author} {\bibfnamefont {Y.~S.}\ \bibnamefont
  {Myung}},\ }\href@noop {} {\bibfield  {journal} {\bibinfo  {journal}
  {Advances in High Energy Physics}\ }\textbf {\bibinfo {volume} {2015}}
  (\bibinfo {year} {2015})}\BibitemShut {NoStop}%
\bibitem [{\citenamefont {Hassan}\ and\ \citenamefont {Rosen}(2012)}]{HassanI}%
  \BibitemOpen
  \bibfield  {author} {\bibinfo {author} {\bibfnamefont {S.~F.}\ \bibnamefont
  {Hassan}}\ and\ \bibinfo {author} {\bibfnamefont {R.~A.}\ \bibnamefont
  {Rosen}},\ }\href {\doibase 10.1103/PhysRevLett.108.041101} {\bibfield
  {journal} {\bibinfo  {journal} {Phys. Rev. Lett.}\ }\textbf {\bibinfo
  {volume} {108}},\ \bibinfo {pages} {041101} (\bibinfo {year}
  {2012})}\BibitemShut {NoStop}%
\bibitem [{\citenamefont {Hassan}\ \emph {et~al.}(2012)\citenamefont {Hassan},
  \citenamefont {Rosen},\ and\ \citenamefont {Schmidt-May}}]{HassanII}%
  \BibitemOpen
  \bibfield  {author} {\bibinfo {author} {\bibfnamefont {S.~F.}\ \bibnamefont
  {Hassan}}, \bibinfo {author} {\bibfnamefont {R.~A.}\ \bibnamefont {Rosen}}, \
  and\ \bibinfo {author} {\bibfnamefont {A.}~\bibnamefont {Schmidt-May}},\
  }\href@noop {} {\bibfield  {journal} {\bibinfo  {journal} {Journal of High
  Energy Physics}\ }\textbf {\bibinfo {volume} {2012}},\ \bibinfo {pages} {26}
  (\bibinfo {year} {2012})}\BibitemShut {NoStop}%
\bibitem [{\citenamefont {Cai}\ \emph {et~al.}(2013)\citenamefont {Cai},
  \citenamefont {Easson}, \citenamefont {Gao},\ and\ \citenamefont
  {Saridakis}}]{BHMassiveI}%
  \BibitemOpen
  \bibfield  {author} {\bibinfo {author} {\bibfnamefont {Y.-F.}\ \bibnamefont
  {Cai}}, \bibinfo {author} {\bibfnamefont {D.~A.}\ \bibnamefont {Easson}},
  \bibinfo {author} {\bibfnamefont {C.}~\bibnamefont {Gao}}, \ and\ \bibinfo
  {author} {\bibfnamefont {E.~N.}\ \bibnamefont {Saridakis}},\ }\href {\doibase
  10.1103/PhysRevD.87.064001} {\bibfield  {journal} {\bibinfo  {journal} {Phys.
  Rev. D}\ }\textbf {\bibinfo {volume} {87}},\ \bibinfo {pages} {064001}
  (\bibinfo {year} {2013})}\BibitemShut {NoStop}%
\bibitem [{\citenamefont {Kodama}\ and\ \citenamefont
  {Arraut}(2014)}]{BHMassiveII}%
  \BibitemOpen
  \bibfield  {author} {\bibinfo {author} {\bibfnamefont {H.}~\bibnamefont
  {Kodama}}\ and\ \bibinfo {author} {\bibfnamefont {I.}~\bibnamefont
  {Arraut}},\ }\href@noop {} {\bibfield  {journal} {\bibinfo  {journal}
  {Progress of Theoretical and Experimental Physics}\ }\textbf {\bibinfo
  {volume} {2014}} (\bibinfo {year} {2014})}\BibitemShut {NoStop}%
\bibitem [{\citenamefont {Zou}\ \emph {et~al.}(2017)\citenamefont {Zou},
  \citenamefont {Yue},\ and\ \citenamefont {Zhang}}]{BHMassiveIII}%
  \BibitemOpen
  \bibfield  {author} {\bibinfo {author} {\bibfnamefont {D.-C.}\ \bibnamefont
  {Zou}}, \bibinfo {author} {\bibfnamefont {R.}~\bibnamefont {Yue}}, \ and\
  \bibinfo {author} {\bibfnamefont {M.}~\bibnamefont {Zhang}},\ }\href@noop {}
  {\bibfield  {journal} {\bibinfo  {journal} {The European Physical Journal C}\
  }\textbf {\bibinfo {volume} {77}},\ \bibinfo {pages} {256} (\bibinfo {year}
  {2017})}\BibitemShut {NoStop}%
\bibitem [{\citenamefont {Tannukij}\ \emph {et~al.}(2017)\citenamefont
  {Tannukij}, \citenamefont {Wongjun},\ and\ \citenamefont
  {Ghosh}}]{BHMassiveIV}%
  \BibitemOpen
  \bibfield  {author} {\bibinfo {author} {\bibfnamefont {L.}~\bibnamefont
  {Tannukij}}, \bibinfo {author} {\bibfnamefont {P.}~\bibnamefont {Wongjun}}, \
  and\ \bibinfo {author} {\bibfnamefont {S.~G.}\ \bibnamefont {Ghosh}},\ }\href
  {\doibase 10.1140/epjc/s10052-017-5426-0} {\bibfield  {journal} {\bibinfo
  {journal} {Eur. Phys. J. C}\ }\textbf {\bibinfo {volume} {77}},\ \bibinfo
  {pages} {846} (\bibinfo {year} {2017})},\ \Eprint
  {http://arxiv.org/abs/1701.05332} {arXiv:1701.05332 [gr-qc]} \BibitemShut
  {NoStop}%
\bibitem [{\citenamefont {Katsuragawa}\ \emph {et~al.}(2016)\citenamefont
  {Katsuragawa}, \citenamefont {Nojiri}, \citenamefont {Odintsov},\ and\
  \citenamefont {Yamazaki}}]{Katsuragawa}%
  \BibitemOpen
  \bibfield  {author} {\bibinfo {author} {\bibfnamefont {T.}~\bibnamefont
  {Katsuragawa}}, \bibinfo {author} {\bibfnamefont {S.}~\bibnamefont {Nojiri}},
  \bibinfo {author} {\bibfnamefont {S.~D.}\ \bibnamefont {Odintsov}}, \ and\
  \bibinfo {author} {\bibfnamefont {M.}~\bibnamefont {Yamazaki}},\ }\href
  {\doibase 10.1103/PhysRevD.93.124013} {\bibfield  {journal} {\bibinfo
  {journal} {Phys. Rev. D}\ }\textbf {\bibinfo {volume} {93}},\ \bibinfo
  {pages} {124013} (\bibinfo {year} {2016})}\BibitemShut {NoStop}%
\bibitem [{\citenamefont {Saridakis}(2013)}]{Saridakis}%
  \BibitemOpen
  \bibfield  {author} {\bibinfo {author} {\bibfnamefont {E.~N.}\ \bibnamefont
  {Saridakis}},\ }\href@noop {} {\bibfield  {journal} {\bibinfo  {journal}
  {Classical and Quantum Gravity}\ }\textbf {\bibinfo {volume} {30}},\ \bibinfo
  {pages} {075003} (\bibinfo {year} {2013})}\BibitemShut {NoStop}%
\bibitem [{\citenamefont {Cai}\ \emph {et~al.}(2012)\citenamefont {Cai},
  \citenamefont {Gao},\ and\ \citenamefont {Saridakis}}]{YFCai}%
  \BibitemOpen
  \bibfield  {author} {\bibinfo {author} {\bibfnamefont {Y.-F.}\ \bibnamefont
  {Cai}}, \bibinfo {author} {\bibfnamefont {C.}~\bibnamefont {Gao}}, \ and\
  \bibinfo {author} {\bibfnamefont {E.~N.}\ \bibnamefont {Saridakis}},\ }\href
  {\doibase 10.1088/1475-7516/2012/10/048} {\bibfield  {journal} {\bibinfo
  {journal} {JCAP}\ }\textbf {\bibinfo {volume} {10}},\ \bibinfo {pages} {048}
  (\bibinfo {year} {2012})},\ \Eprint {http://arxiv.org/abs/1207.3786}
  {arXiv:1207.3786 [astro-ph.CO]} \BibitemShut {NoStop}%
\bibitem [{\citenamefont {Leon}\ \emph {et~al.}(2013)\citenamefont {Leon},
  \citenamefont {Saavedra},\ and\ \citenamefont {Saridakis}}]{Leon}%
  \BibitemOpen
  \bibfield  {author} {\bibinfo {author} {\bibfnamefont {G.}~\bibnamefont
  {Leon}}, \bibinfo {author} {\bibfnamefont {J.}~\bibnamefont {Saavedra}}, \
  and\ \bibinfo {author} {\bibfnamefont {E.~N.}\ \bibnamefont {Saridakis}},\
  }\href@noop {} {\bibfield  {journal} {\bibinfo  {journal} {Classical and
  Quantum Gravity}\ }\textbf {\bibinfo {volume} {30}},\ \bibinfo {pages}
  {135001} (\bibinfo {year} {2013})}\BibitemShut {NoStop}%
\bibitem [{\citenamefont {Hinterbichler}\ \emph {et~al.}(2013)\citenamefont
  {Hinterbichler}, \citenamefont {Stokes},\ and\ \citenamefont
  {Trodden}}]{Hinterbichler}%
  \BibitemOpen
  \bibfield  {author} {\bibinfo {author} {\bibfnamefont {K.}~\bibnamefont
  {Hinterbichler}}, \bibinfo {author} {\bibfnamefont {J.}~\bibnamefont
  {Stokes}}, \ and\ \bibinfo {author} {\bibfnamefont {M.}~\bibnamefont
  {Trodden}},\ }\href {\doibase https://doi.org/10.1016/j.physletb.2013.07.009}
  {\bibfield  {journal} {\bibinfo  {journal} {Physics Letters B}\ }\textbf
  {\bibinfo {volume} {725}},\ \bibinfo {pages} {1 } (\bibinfo {year}
  {2013})}\BibitemShut {NoStop}%
\bibitem [{\citenamefont {Fasiello}\ and\ \citenamefont
  {Tolley}(2013)}]{Fasiello}%
  \BibitemOpen
  \bibfield  {author} {\bibinfo {author} {\bibfnamefont {M.}~\bibnamefont
  {Fasiello}}\ and\ \bibinfo {author} {\bibfnamefont {A.~J.}\ \bibnamefont
  {Tolley}},\ }\href@noop {} {\bibfield  {journal} {\bibinfo  {journal}
  {Journal of Cosmology and Astroparticle Physics}\ }\textbf {\bibinfo {volume}
  {2013}},\ \bibinfo {pages} {002} (\bibinfo {year} {2013})}\BibitemShut
  {NoStop}%
\bibitem [{\citenamefont {Bamba}\ \emph {et~al.}(2014)\citenamefont {Bamba},
  \citenamefont {Hossain}, \citenamefont {Myrzakulov}, \citenamefont {Nojiri},\
  and\ \citenamefont {Sami}}]{Bamba}%
  \BibitemOpen
  \bibfield  {author} {\bibinfo {author} {\bibfnamefont {K.}~\bibnamefont
  {Bamba}}, \bibinfo {author} {\bibfnamefont {M.~W.}\ \bibnamefont {Hossain}},
  \bibinfo {author} {\bibfnamefont {R.}~\bibnamefont {Myrzakulov}}, \bibinfo
  {author} {\bibfnamefont {S.}~\bibnamefont {Nojiri}}, \ and\ \bibinfo {author}
  {\bibfnamefont {M.}~\bibnamefont {Sami}},\ }\href {\doibase
  10.1103/PhysRevD.89.083518} {\bibfield  {journal} {\bibinfo  {journal} {Phys.
  Rev. D}\ }\textbf {\bibinfo {volume} {89}},\ \bibinfo {pages} {083518}
  (\bibinfo {year} {2014})}\BibitemShut {NoStop}%
\bibitem [{\citenamefont {Vegh}(2013)}]{Vegh}%
  \BibitemOpen
  \bibfield  {author} {\bibinfo {author} {\bibfnamefont {D.}~\bibnamefont
  {Vegh}},\ }\href@noop {} {\  (\bibinfo {year} {2013})},\ \Eprint
  {http://arxiv.org/abs/1301.0537} {arXiv:1301.0537 [hep-th]} \BibitemShut
  {NoStop}%
\bibitem [{\citenamefont {Hinterbichler}(2012)}]{Hinterbichler:2011tt}%
  \BibitemOpen
  \bibfield  {author} {\bibinfo {author} {\bibfnamefont {K.}~\bibnamefont
  {Hinterbichler}},\ }\href {\doibase 10.1103/RevModPhys.84.671} {\bibfield
  {journal} {\bibinfo  {journal} {Rev. Mod. Phys.}\ }\textbf {\bibinfo {volume}
  {84}},\ \bibinfo {pages} {671} (\bibinfo {year} {2012})},\ \Eprint
  {http://arxiv.org/abs/1105.3735} {arXiv:1105.3735 [hep-th]} \BibitemShut
  {NoStop}%
\bibitem [{\citenamefont {de~Rham}(2014{\natexlab{b}})}]{deRham:2014zqa}%
  \BibitemOpen
  \bibfield  {author} {\bibinfo {author} {\bibfnamefont {C.}~\bibnamefont
  {de~Rham}},\ }\href {\doibase 10.12942/lrr-2014-7} {\bibfield  {journal}
  {\bibinfo  {journal} {Living Rev. Rel.}\ }\textbf {\bibinfo {volume} {17}},\
  \bibinfo {pages} {7} (\bibinfo {year} {2014}{\natexlab{b}})},\ \Eprint
  {http://arxiv.org/abs/1401.4173} {arXiv:1401.4173 [hep-th]} \BibitemShut
  {NoStop}%
\bibitem [{\citenamefont {G{\"u}mr{\"u}k{\c{c}}{\"u}o{\u{g}}lu}\ \emph
  {et~al.}(2011)\citenamefont {G{\"u}mr{\"u}k{\c{c}}{\"u}o{\u{g}}lu},
  \citenamefont {Lin},\ and\ \citenamefont {Mukohyama}}]{Gumrukcuoglu}%
  \BibitemOpen
  \bibfield  {author} {\bibinfo {author} {\bibfnamefont {A.~E.}\ \bibnamefont
  {G{\"u}mr{\"u}k{\c{c}}{\"u}o{\u{g}}lu}}, \bibinfo {author} {\bibfnamefont
  {C.}~\bibnamefont {Lin}}, \ and\ \bibinfo {author} {\bibfnamefont
  {S.}~\bibnamefont {Mukohyama}},\ }\href@noop {} {\bibfield  {journal}
  {\bibinfo  {journal} {Journal of Cosmology and Astroparticle Physics}\
  }\textbf {\bibinfo {volume} {2011}},\ \bibinfo {pages} {030} (\bibinfo {year}
  {2011})}\BibitemShut {NoStop}%
\bibitem [{\citenamefont {Gratia}\ \emph {et~al.}(2012)\citenamefont {Gratia},
  \citenamefont {Hu},\ and\ \citenamefont {Wyman}}]{Gratia}%
  \BibitemOpen
  \bibfield  {author} {\bibinfo {author} {\bibfnamefont {P.}~\bibnamefont
  {Gratia}}, \bibinfo {author} {\bibfnamefont {W.}~\bibnamefont {Hu}}, \ and\
  \bibinfo {author} {\bibfnamefont {M.}~\bibnamefont {Wyman}},\ }\href
  {\doibase 10.1103/PhysRevD.86.061504} {\bibfield  {journal} {\bibinfo
  {journal} {Phys. Rev. D}\ }\textbf {\bibinfo {volume} {86}},\ \bibinfo
  {pages} {061504} (\bibinfo {year} {2012})}\BibitemShut {NoStop}%
\bibitem [{\citenamefont {Kobayashi}\ \emph {et~al.}(2012)\citenamefont
  {Kobayashi}, \citenamefont {Siino}, \citenamefont {Yamaguchi},\ and\
  \citenamefont {Yoshida}}]{Kobayash}%
  \BibitemOpen
  \bibfield  {author} {\bibinfo {author} {\bibfnamefont {T.}~\bibnamefont
  {Kobayashi}}, \bibinfo {author} {\bibfnamefont {M.}~\bibnamefont {Siino}},
  \bibinfo {author} {\bibfnamefont {M.}~\bibnamefont {Yamaguchi}}, \ and\
  \bibinfo {author} {\bibfnamefont {D.}~\bibnamefont {Yoshida}},\ }\href
  {\doibase 10.1103/PhysRevD.86.061505} {\bibfield  {journal} {\bibinfo
  {journal} {Phys. Rev. D}\ }\textbf {\bibinfo {volume} {86}},\ \bibinfo
  {pages} {061505} (\bibinfo {year} {2012})}\BibitemShut {NoStop}%
\bibitem [{\citenamefont {Deffayet}(2001)}]{DeffayetI}%
  \BibitemOpen
  \bibfield  {author} {\bibinfo {author} {\bibfnamefont {C.}~\bibnamefont
  {Deffayet}},\ }\href {\doibase 10.1016/S0370-2693(01)00160-5} {\bibfield
  {journal} {\bibinfo  {journal} {Phys. Lett. B}\ }\textbf {\bibinfo {volume}
  {502}},\ \bibinfo {pages} {199} (\bibinfo {year} {2001})},\ \Eprint
  {http://arxiv.org/abs/hep-th/0010186} {arXiv:hep-th/0010186} \BibitemShut
  {NoStop}%
\bibitem [{\citenamefont {Deffayet}\ \emph {et~al.}(2002)\citenamefont
  {Deffayet}, \citenamefont {Dvali},\ and\ \citenamefont
  {Gabadadze}}]{DeffayetII}%
  \BibitemOpen
  \bibfield  {author} {\bibinfo {author} {\bibfnamefont {C.}~\bibnamefont
  {Deffayet}}, \bibinfo {author} {\bibfnamefont {G.}~\bibnamefont {Dvali}}, \
  and\ \bibinfo {author} {\bibfnamefont {G.}~\bibnamefont {Gabadadze}},\ }\href
  {\doibase 10.1103/PhysRevD.65.044023} {\bibfield  {journal} {\bibinfo
  {journal} {Phys. Rev. D}\ }\textbf {\bibinfo {volume} {65}},\ \bibinfo
  {pages} {044023} (\bibinfo {year} {2002})}\BibitemShut {NoStop}%
\bibitem [{\citenamefont {Dvali}\ \emph {et~al.}(2003)\citenamefont {Dvali},
  \citenamefont {Gabadadze},\ and\ \citenamefont {Shifman}}]{DvaliI}%
  \BibitemOpen
  \bibfield  {author} {\bibinfo {author} {\bibfnamefont {G.}~\bibnamefont
  {Dvali}}, \bibinfo {author} {\bibfnamefont {G.}~\bibnamefont {Gabadadze}}, \
  and\ \bibinfo {author} {\bibfnamefont {M.}~\bibnamefont {Shifman}},\ }\href
  {\doibase 10.1103/PhysRevD.67.044020} {\bibfield  {journal} {\bibinfo
  {journal} {Phys. Rev. D}\ }\textbf {\bibinfo {volume} {67}},\ \bibinfo
  {pages} {044020} (\bibinfo {year} {2003})}\BibitemShut {NoStop}%
\bibitem [{\citenamefont {Dvali}\ \emph {et~al.}(2007)\citenamefont {Dvali},
  \citenamefont {Hofmann},\ and\ \citenamefont {Khoury}}]{DvaliII}%
  \BibitemOpen
  \bibfield  {author} {\bibinfo {author} {\bibfnamefont {G.}~\bibnamefont
  {Dvali}}, \bibinfo {author} {\bibfnamefont {S.}~\bibnamefont {Hofmann}}, \
  and\ \bibinfo {author} {\bibfnamefont {J.}~\bibnamefont {Khoury}},\ }\href
  {\doibase 10.1103/PhysRevD.76.084006} {\bibfield  {journal} {\bibinfo
  {journal} {Phys. Rev. D}\ }\textbf {\bibinfo {volume} {76}},\ \bibinfo
  {pages} {084006} (\bibinfo {year} {2007})}\BibitemShut {NoStop}%
\bibitem [{\citenamefont {Will}(2014)}]{Will}%
  \BibitemOpen
  \bibfield  {author} {\bibinfo {author} {\bibfnamefont {C.~M.}\ \bibnamefont
  {Will}},\ }\href@noop {} {\bibfield  {journal} {\bibinfo  {journal} {Living
  reviews in relativity}\ }\textbf {\bibinfo {volume} {17}},\ \bibinfo {pages}
  {4} (\bibinfo {year} {2014})}\BibitemShut {NoStop}%
\bibitem [{\citenamefont {Mohseni}(2011)}]{Mohseni}%
  \BibitemOpen
  \bibfield  {author} {\bibinfo {author} {\bibfnamefont {M.}~\bibnamefont
  {Mohseni}},\ }\href {\doibase 10.1103/PhysRevD.84.064026} {\bibfield
  {journal} {\bibinfo  {journal} {Phys. Rev. D}\ }\textbf {\bibinfo {volume}
  {84}},\ \bibinfo {pages} {064026} (\bibinfo {year} {2011})}\BibitemShut
  {NoStop}%
\bibitem [{\citenamefont {Gumrukcuoglu}\ \emph {et~al.}(2012)\citenamefont
  {Gumrukcuoglu}, \citenamefont {Kuroyanagi}, \citenamefont {Lin},
  \citenamefont {Mukohyama},\ and\ \citenamefont {Tanahashi}}]{GumrukcuogluII}%
  \BibitemOpen
  \bibfield  {author} {\bibinfo {author} {\bibfnamefont {A.}~\bibnamefont
  {Gumrukcuoglu}}, \bibinfo {author} {\bibfnamefont {S.}~\bibnamefont
  {Kuroyanagi}}, \bibinfo {author} {\bibfnamefont {C.}~\bibnamefont {Lin}},
  \bibinfo {author} {\bibfnamefont {S.}~\bibnamefont {Mukohyama}}, \ and\
  \bibinfo {author} {\bibfnamefont {N.}~\bibnamefont {Tanahashi}},\ }\href
  {\doibase 10.1088/0264-9381/29/23/235026} {\bibfield  {journal} {\bibinfo
  {journal} {Class. Quant. Grav.}\ }\textbf {\bibinfo {volume} {29}},\ \bibinfo
  {pages} {235026} (\bibinfo {year} {2012})},\ \Eprint
  {http://arxiv.org/abs/1208.5975} {arXiv:1208.5975 [hep-th]} \BibitemShut
  {NoStop}%
\bibitem [{\citenamefont {Hendi}\ \emph
  {et~al.}(2017{\natexlab{a}})\citenamefont {Hendi}, \citenamefont {Bordbar},
  \citenamefont {Eslam~Panah},\ and\ \citenamefont {Panahiyan}}]{NeutronMass}%
  \BibitemOpen
  \bibfield  {author} {\bibinfo {author} {\bibfnamefont {S.}~\bibnamefont
  {Hendi}}, \bibinfo {author} {\bibfnamefont {G.}~\bibnamefont {Bordbar}},
  \bibinfo {author} {\bibfnamefont {B.}~\bibnamefont {Eslam~Panah}}, \ and\
  \bibinfo {author} {\bibfnamefont {S.}~\bibnamefont {Panahiyan}},\ }\href
  {\doibase 10.1088/1475-7516/2017/07/004} {\bibfield  {journal} {\bibinfo
  {journal} {JCAP}\ }\textbf {\bibinfo {volume} {07}},\ \bibinfo {pages} {004}
  (\bibinfo {year} {2017}{\natexlab{a}})},\ \Eprint
  {http://arxiv.org/abs/1701.01039} {arXiv:1701.01039 [gr-qc]} \BibitemShut
  {NoStop}%
\bibitem [{\citenamefont {Rhoades}\ and\ \citenamefont
  {Ruffini}(1974)}]{Ruffini}%
  \BibitemOpen
  \bibfield  {author} {\bibinfo {author} {\bibfnamefont {C.~E.}\ \bibnamefont
  {Rhoades}}\ and\ \bibinfo {author} {\bibfnamefont {R.}~\bibnamefont
  {Ruffini}},\ }\href {\doibase 10.1103/PhysRevLett.32.324} {\bibfield
  {journal} {\bibinfo  {journal} {Phys. Rev. Lett.}\ }\textbf {\bibinfo
  {volume} {32}},\ \bibinfo {pages} {324} (\bibinfo {year} {1974})}\BibitemShut
  {NoStop}%
\bibitem [{\citenamefont {Eslam~Panah}\ and\ \citenamefont
  {Liu}(2019)}]{EslamPanah:2018evk}%
  \BibitemOpen
  \bibfield  {author} {\bibinfo {author} {\bibfnamefont {B.}~\bibnamefont
  {Eslam~Panah}}\ and\ \bibinfo {author} {\bibfnamefont {H.~L.}\ \bibnamefont
  {Liu}},\ }\href {\doibase 10.1103/PhysRevD.99.104074} {\bibfield  {journal}
  {\bibinfo  {journal} {Phys. Rev. D}\ }\textbf {\bibinfo {volume} {99}},\
  \bibinfo {pages} {104074} (\bibinfo {year} {2019})},\ \Eprint
  {http://arxiv.org/abs/1805.10650} {arXiv:1805.10650 [gr-qc]} \BibitemShut
  {NoStop}%
\bibitem [{\citenamefont {Babichev}\ \emph {et~al.}(2016)\citenamefont
  {Babichev}, \citenamefont {Marzola}, \citenamefont {Raidal}, \citenamefont
  {Schmidt-May}, \citenamefont {Urban}, \citenamefont {Veermäe},\ and\
  \citenamefont {von Strauss}}]{Schmidt-May2016DarkMatter}%
  \BibitemOpen
  \bibfield  {author} {\bibinfo {author} {\bibfnamefont {E.}~\bibnamefont
  {Babichev}}, \bibinfo {author} {\bibfnamefont {L.}~\bibnamefont {Marzola}},
  \bibinfo {author} {\bibfnamefont {M.}~\bibnamefont {Raidal}}, \bibinfo
  {author} {\bibfnamefont {A.}~\bibnamefont {Schmidt-May}}, \bibinfo {author}
  {\bibfnamefont {F.}~\bibnamefont {Urban}}, \bibinfo {author} {\bibfnamefont
  {H.}~\bibnamefont {Veermäe}}, \ and\ \bibinfo {author} {\bibfnamefont
  {M.}~\bibnamefont {von Strauss}},\ }\href {\doibase
  10.1088/1475-7516/2016/09/016} {\bibfield  {journal} {\bibinfo  {journal}
  {JCAP}\ }\textbf {\bibinfo {volume} {09}},\ \bibinfo {pages} {016} (\bibinfo
  {year} {2016})},\ \Eprint {http://arxiv.org/abs/1607.03497} {arXiv:1607.03497
  [hep-th]} \BibitemShut {NoStop}%
\bibitem [{\citenamefont {Akrami}\ \emph {et~al.}(2013)\citenamefont {Akrami},
  \citenamefont {Koivisto},\ and\ \citenamefont
  {Sandstad}}]{MassiveCosmology2013}%
  \BibitemOpen
  \bibfield  {author} {\bibinfo {author} {\bibfnamefont {Y.}~\bibnamefont
  {Akrami}}, \bibinfo {author} {\bibfnamefont {T.~S.}\ \bibnamefont
  {Koivisto}}, \ and\ \bibinfo {author} {\bibfnamefont {M.}~\bibnamefont
  {Sandstad}},\ }\href@noop {} {\bibfield  {journal} {\bibinfo  {journal}
  {Journal of High Energy Physics}\ }\textbf {\bibinfo {volume} {2013}},\
  \bibinfo {pages} {99} (\bibinfo {year} {2013})}\BibitemShut {NoStop}%
\bibitem [{\citenamefont {Akrami}\ \emph {et~al.}(2015)\citenamefont {Akrami},
  \citenamefont {Hassan}, \citenamefont {K{\"o}nnig}, \citenamefont
  {Schmidt-May},\ and\ \citenamefont {Solomon}}]{MassiveCosmology2015}%
  \BibitemOpen
  \bibfield  {author} {\bibinfo {author} {\bibfnamefont {Y.}~\bibnamefont
  {Akrami}}, \bibinfo {author} {\bibfnamefont {S.~F.}\ \bibnamefont {Hassan}},
  \bibinfo {author} {\bibfnamefont {F.}~\bibnamefont {K{\"o}nnig}}, \bibinfo
  {author} {\bibfnamefont {A.}~\bibnamefont {Schmidt-May}}, \ and\ \bibinfo
  {author} {\bibfnamefont {A.~R.}\ \bibnamefont {Solomon}},\ }\href@noop {}
  {\bibfield  {journal} {\bibinfo  {journal} {Physics Letters B}\ }\textbf
  {\bibinfo {volume} {748}},\ \bibinfo {pages} {37} (\bibinfo {year}
  {2015})}\BibitemShut {NoStop}%
\bibitem [{\citenamefont {Bachas}\ and\ \citenamefont
  {Lavdas}(2018)}]{MGinString2018}%
  \BibitemOpen
  \bibfield  {author} {\bibinfo {author} {\bibfnamefont {C.}~\bibnamefont
  {Bachas}}\ and\ \bibinfo {author} {\bibfnamefont {I.}~\bibnamefont
  {Lavdas}},\ }\href@noop {} {\bibfield  {journal} {\bibinfo  {journal}
  {Journal of High Energy Physics}\ }\textbf {\bibinfo {volume} {2018}},\
  \bibinfo {pages} {3} (\bibinfo {year} {2018})}\BibitemShut {NoStop}%
\bibitem [{\citenamefont {Geng}\ and\ \citenamefont
  {Karch}(2020)}]{Geng:2020qvw}%
  \BibitemOpen
  \bibfield  {author} {\bibinfo {author} {\bibfnamefont {H.}~\bibnamefont
  {Geng}}\ and\ \bibinfo {author} {\bibfnamefont {A.}~\bibnamefont {Karch}},\
  }\href {\doibase 10.1007/JHEP09(2020)121} {\bibfield  {journal} {\bibinfo
  {journal} {JHEP}\ }\textbf {\bibinfo {volume} {09}},\ \bibinfo {pages} {121}
  (\bibinfo {year} {2020})},\ \Eprint {http://arxiv.org/abs/2006.02438}
  {arXiv:2006.02438 [hep-th]} \BibitemShut {NoStop}%
\bibitem [{\citenamefont {Geng}\ \emph {et~al.}(2021)\citenamefont {Geng},
  \citenamefont {Karch}, \citenamefont {Perez-Pardavila}, \citenamefont {Raju},
  \citenamefont {Randall}, \citenamefont {Riojas},\ and\ \citenamefont
  {Shashi}}]{Geng:2020fxl}%
  \BibitemOpen
  \bibfield  {author} {\bibinfo {author} {\bibfnamefont {H.}~\bibnamefont
  {Geng}}, \bibinfo {author} {\bibfnamefont {A.}~\bibnamefont {Karch}},
  \bibinfo {author} {\bibfnamefont {C.}~\bibnamefont {Perez-Pardavila}},
  \bibinfo {author} {\bibfnamefont {S.}~\bibnamefont {Raju}}, \bibinfo {author}
  {\bibfnamefont {L.}~\bibnamefont {Randall}}, \bibinfo {author} {\bibfnamefont
  {M.}~\bibnamefont {Riojas}}, \ and\ \bibinfo {author} {\bibfnamefont
  {S.}~\bibnamefont {Shashi}},\ }\href {\doibase 10.21468/SciPostPhys.10.5.103}
  {\bibfield  {journal} {\bibinfo  {journal} {SciPost Phys.}\ }\textbf
  {\bibinfo {volume} {10}},\ \bibinfo {pages} {103} (\bibinfo {year} {2021})},\
  \Eprint {http://arxiv.org/abs/2012.04671} {arXiv:2012.04671 [hep-th]}
  \BibitemShut {NoStop}%
\bibitem [{\citenamefont {Ahmed}\ \emph {et~al.}(2023)\citenamefont {Ahmed},
  \citenamefont {Cong}, \citenamefont {Kubiz\v{n}\'ak}, \citenamefont {Mann},\
  and\ \citenamefont {Visser}}]{Ahmed:2023snm}%
  \BibitemOpen
  \bibfield  {author} {\bibinfo {author} {\bibfnamefont {M.~B.}\ \bibnamefont
  {Ahmed}}, \bibinfo {author} {\bibfnamefont {W.}~\bibnamefont {Cong}},
  \bibinfo {author} {\bibfnamefont {D.}~\bibnamefont {Kubiz\v{n}\'ak}},
  \bibinfo {author} {\bibfnamefont {R.~B.}\ \bibnamefont {Mann}}, \ and\
  \bibinfo {author} {\bibfnamefont {M.~R.}\ \bibnamefont {Visser}},\ }\href
  {\doibase 10.1103/PhysRevLett.130.181401} {\bibfield  {journal} {\bibinfo
  {journal} {Phys. Rev. Lett.}\ }\textbf {\bibinfo {volume} {130}},\ \bibinfo
  {pages} {181401} (\bibinfo {year} {2023})},\ \Eprint
  {http://arxiv.org/abs/2302.08163} {arXiv:2302.08163 [hep-th]} \BibitemShut
  {NoStop}%
\bibitem [{\citenamefont {Cai}\ \emph {et~al.}(2015)\citenamefont {Cai},
  \citenamefont {Hu}, \citenamefont {Pan},\ and\ \citenamefont
  {Zhang}}]{Cai2015}%
  \BibitemOpen
  \bibfield  {author} {\bibinfo {author} {\bibfnamefont {R.-G.}\ \bibnamefont
  {Cai}}, \bibinfo {author} {\bibfnamefont {Y.-P.}\ \bibnamefont {Hu}},
  \bibinfo {author} {\bibfnamefont {Q.-Y.}\ \bibnamefont {Pan}}, \ and\
  \bibinfo {author} {\bibfnamefont {Y.-L.}\ \bibnamefont {Zhang}},\ }\href
  {\doibase 10.1103/PhysRevD.91.024032} {\bibfield  {journal} {\bibinfo
  {journal} {Phys. Rev. D}\ }\textbf {\bibinfo {volume} {91}},\ \bibinfo
  {pages} {024032} (\bibinfo {year} {2015})}\BibitemShut {NoStop}%
\bibitem [{\citenamefont {Hendi}\ \emph
  {et~al.}(2017{\natexlab{b}})\citenamefont {Hendi}, \citenamefont {Panah},\
  and\ \citenamefont {Panahiyan}}]{PVMassV}%
  \BibitemOpen
  \bibfield  {author} {\bibinfo {author} {\bibfnamefont {S.}~\bibnamefont
  {Hendi}}, \bibinfo {author} {\bibfnamefont {B.~E.}\ \bibnamefont {Panah}}, \
  and\ \bibinfo {author} {\bibfnamefont {S.}~\bibnamefont {Panahiyan}},\
  }\href@noop {} {\bibfield  {journal} {\bibinfo  {journal} {Physics Letters
  B}\ }\textbf {\bibinfo {volume} {769}},\ \bibinfo {pages} {191} (\bibinfo
  {year} {2017}{\natexlab{b}})}\BibitemShut {NoStop}%
\bibitem [{\citenamefont {Hendi}\ \emph
  {et~al.}(2017{\natexlab{c}})\citenamefont {Hendi}, \citenamefont {Mann},
  \citenamefont {Panahiyan},\ and\ \citenamefont {Eslam~Panah}}]{PVMassIV}%
  \BibitemOpen
  \bibfield  {author} {\bibinfo {author} {\bibfnamefont {S.~H.}\ \bibnamefont
  {Hendi}}, \bibinfo {author} {\bibfnamefont {R.~B.}\ \bibnamefont {Mann}},
  \bibinfo {author} {\bibfnamefont {S.}~\bibnamefont {Panahiyan}}, \ and\
  \bibinfo {author} {\bibfnamefont {B.}~\bibnamefont {Eslam~Panah}},\ }\href
  {\doibase 10.1103/PhysRevD.95.021501} {\bibfield  {journal} {\bibinfo
  {journal} {Phys. Rev. D}\ }\textbf {\bibinfo {volume} {95}},\ \bibinfo
  {pages} {021501} (\bibinfo {year} {2017}{\natexlab{c}})}\BibitemShut
  {NoStop}%
\bibitem [{\citenamefont {Alberte}\ and\ \citenamefont
  {Khmelnitsky}(2015)}]{Alberte}%
  \BibitemOpen
  \bibfield  {author} {\bibinfo {author} {\bibfnamefont {L.}~\bibnamefont
  {Alberte}}\ and\ \bibinfo {author} {\bibfnamefont {A.}~\bibnamefont
  {Khmelnitsky}},\ }\href {\doibase 10.1103/PhysRevD.91.046006} {\bibfield
  {journal} {\bibinfo  {journal} {Phys. Rev. D}\ }\textbf {\bibinfo {volume}
  {91}},\ \bibinfo {pages} {046006} (\bibinfo {year} {2015})}\BibitemShut
  {NoStop}%
\bibitem [{\citenamefont {Zhou}\ \emph {et~al.}(2015)\citenamefont {Zhou},
  \citenamefont {Wu},\ and\ \citenamefont {Ling}}]{Zhou}%
  \BibitemOpen
  \bibfield  {author} {\bibinfo {author} {\bibfnamefont {Z.}~\bibnamefont
  {Zhou}}, \bibinfo {author} {\bibfnamefont {J.-P.}\ \bibnamefont {Wu}}, \ and\
  \bibinfo {author} {\bibfnamefont {Y.}~\bibnamefont {Ling}},\ }\href@noop {}
  {\bibfield  {journal} {\bibinfo  {journal} {Journal of High Energy Physics}\
  }\textbf {\bibinfo {volume} {2015}},\ \bibinfo {pages} {67} (\bibinfo {year}
  {2015})}\BibitemShut {NoStop}%
\bibitem [{\citenamefont {Dehyadegari}\ \emph {et~al.}(2017)\citenamefont
  {Dehyadegari}, \citenamefont {Kord~Zangeneh},\ and\ \citenamefont
  {Sheykhi}}]{Dehyadegari}%
  \BibitemOpen
  \bibfield  {author} {\bibinfo {author} {\bibfnamefont {A.}~\bibnamefont
  {Dehyadegari}}, \bibinfo {author} {\bibfnamefont {M.}~\bibnamefont
  {Kord~Zangeneh}}, \ and\ \bibinfo {author} {\bibfnamefont {A.}~\bibnamefont
  {Sheykhi}},\ }\href {\doibase 10.1016/j.physletb.2017.08.029} {\bibfield
  {journal} {\bibinfo  {journal} {Phys. Lett. B}\ }\textbf {\bibinfo {volume}
  {773}},\ \bibinfo {pages} {344} (\bibinfo {year} {2017})},\ \Eprint
  {http://arxiv.org/abs/1703.00975} {arXiv:1703.00975 [hep-th]} \BibitemShut
  {NoStop}%
\bibitem [{\citenamefont {Hendi}\ \emph
  {et~al.}(2017{\natexlab{d}})\citenamefont {Hendi}, \citenamefont
  {Eslam~Panah}, \citenamefont {Panahiyan},\ and\ \citenamefont
  {Momennia}}]{Magmass}%
  \BibitemOpen
  \bibfield  {author} {\bibinfo {author} {\bibfnamefont {S.}~\bibnamefont
  {Hendi}}, \bibinfo {author} {\bibfnamefont {B.}~\bibnamefont {Eslam~Panah}},
  \bibinfo {author} {\bibfnamefont {S.}~\bibnamefont {Panahiyan}}, \ and\
  \bibinfo {author} {\bibfnamefont {M.}~\bibnamefont {Momennia}},\ }\href
  {\doibase 10.1016/j.physletb.2017.10.053} {\bibfield  {journal} {\bibinfo
  {journal} {Phys. Lett. B}\ }\textbf {\bibinfo {volume} {775}},\ \bibinfo
  {pages} {251} (\bibinfo {year} {2017}{\natexlab{d}})},\ \Eprint
  {http://arxiv.org/abs/1704.00996} {arXiv:1704.00996 [gr-qc]} \BibitemShut
  {NoStop}%
\bibitem [{\citenamefont {Dehghani}\ and\ \citenamefont
  {Hendi}(2020)}]{Dehghani:2019thq}%
  \BibitemOpen
  \bibfield  {author} {\bibinfo {author} {\bibfnamefont {A.}~\bibnamefont
  {Dehghani}}\ and\ \bibinfo {author} {\bibfnamefont {S.~H.}\ \bibnamefont
  {Hendi}},\ }\href {\doibase 10.1088/1361-6382/ab5eb4} {\bibfield  {journal}
  {\bibinfo  {journal} {Class. Quant. Grav.}\ }\textbf {\bibinfo {volume}
  {37}},\ \bibinfo {pages} {024001} (\bibinfo {year} {2020})},\ \Eprint
  {http://arxiv.org/abs/1909.00956} {arXiv:1909.00956 [hep-th]} \BibitemShut
  {NoStop}%
\bibitem [{\citenamefont {Hendi}\ \emph
  {et~al.}(2016{\natexlab{a}})\citenamefont {Hendi}, \citenamefont {Panahiyan},
  \citenamefont {Eslam~Panah},\ and\ \citenamefont {Momennia}}]{Hendi:2015eca}%
  \BibitemOpen
  \bibfield  {author} {\bibinfo {author} {\bibfnamefont {S.~H.}\ \bibnamefont
  {Hendi}}, \bibinfo {author} {\bibfnamefont {S.}~\bibnamefont {Panahiyan}},
  \bibinfo {author} {\bibfnamefont {B.}~\bibnamefont {Eslam~Panah}}, \ and\
  \bibinfo {author} {\bibfnamefont {M.}~\bibnamefont {Momennia}},\ }\href
  {\doibase 10.1002/andp.201600180} {\bibfield  {journal} {\bibinfo  {journal}
  {Annalen Phys.}\ }\textbf {\bibinfo {volume} {528}},\ \bibinfo {pages} {819}
  (\bibinfo {year} {2016}{\natexlab{a}})},\ \Eprint
  {http://arxiv.org/abs/1506.07262} {arXiv:1506.07262 [hep-th]} \BibitemShut
  {NoStop}%
\bibitem [{\citenamefont {Akbarieh}\ \emph {et~al.}(2021)\citenamefont
  {Akbarieh}, \citenamefont {Kazempour},\ and\ \citenamefont
  {Shao}}]{Akbarieh:2021vhv}%
  \BibitemOpen
  \bibfield  {author} {\bibinfo {author} {\bibfnamefont {A.~R.}\ \bibnamefont
  {Akbarieh}}, \bibinfo {author} {\bibfnamefont {S.}~\bibnamefont {Kazempour}},
  \ and\ \bibinfo {author} {\bibfnamefont {L.}~\bibnamefont {Shao}},\ }\href
  {\doibase 10.1103/PhysRevD.103.123518} {\bibfield  {journal} {\bibinfo
  {journal} {Phys. Rev. D}\ }\textbf {\bibinfo {volume} {103}},\ \bibinfo
  {pages} {123518} (\bibinfo {year} {2021})},\ \Eprint
  {http://arxiv.org/abs/2105.03744} {arXiv:2105.03744 [gr-qc]} \BibitemShut
  {NoStop}%
\bibitem [{\citenamefont {Yerra}\ and\ \citenamefont
  {Bhamidipati}(2021{\natexlab{a}})}]{Yerra:2020tzg}%
  \BibitemOpen
  \bibfield  {author} {\bibinfo {author} {\bibfnamefont {P.~K.}\ \bibnamefont
  {Yerra}}\ and\ \bibinfo {author} {\bibfnamefont {C.}~\bibnamefont
  {Bhamidipati}},\ }\href {\doibase 10.1016/j.physletb.2021.136450} {\bibfield
  {journal} {\bibinfo  {journal} {Phys. Lett. B}\ }\textbf {\bibinfo {volume}
  {819}},\ \bibinfo {pages} {136450} (\bibinfo {year} {2021}{\natexlab{a}})},\
  \Eprint {http://arxiv.org/abs/2007.11515} {arXiv:2007.11515 [hep-th]}
  \BibitemShut {NoStop}%
\bibitem [{\citenamefont {Yerra}\ and\ \citenamefont
  {Bhamidipati}(2020)}]{Yerra:2020oph}%
  \BibitemOpen
  \bibfield  {author} {\bibinfo {author} {\bibfnamefont {P.~K.}\ \bibnamefont
  {Yerra}}\ and\ \bibinfo {author} {\bibfnamefont {C.}~\bibnamefont
  {Bhamidipati}},\ }\href {\doibase 10.1142/S0217751X20501201} {\bibfield
  {journal} {\bibinfo  {journal} {Int. J. Mod. Phys. A}\ }\textbf {\bibinfo
  {volume} {35}},\ \bibinfo {pages} {2050120} (\bibinfo {year} {2020})},\
  \Eprint {http://arxiv.org/abs/2006.07775} {arXiv:2006.07775 [hep-th]}
  \BibitemShut {NoStop}%
\bibitem [{\citenamefont {Yerra}\ and\ \citenamefont
  {Bhamidipati}(2021{\natexlab{b}})}]{Yerra:2021hnh}%
  \BibitemOpen
  \bibfield  {author} {\bibinfo {author} {\bibfnamefont {P.~K.}\ \bibnamefont
  {Yerra}}\ and\ \bibinfo {author} {\bibfnamefont {C.}~\bibnamefont
  {Bhamidipati}},\ }\href {\doibase 10.1103/PhysRevD.104.104049} {\bibfield
  {journal} {\bibinfo  {journal} {Phys. Rev. D}\ }\textbf {\bibinfo {volume}
  {104}},\ \bibinfo {pages} {104049} (\bibinfo {year} {2021}{\natexlab{b}})},\
  \Eprint {http://arxiv.org/abs/2107.04504} {arXiv:2107.04504 [gr-qc]}
  \BibitemShut {NoStop}%
\bibitem [{\citenamefont {H\"og\r{a}s}\ and\ \citenamefont
  {M\"ortsell}(2021)}]{Hogas:2021saw}%
  \BibitemOpen
  \bibfield  {author} {\bibinfo {author} {\bibfnamefont {M.}~\bibnamefont
  {H\"og\r{a}s}}\ and\ \bibinfo {author} {\bibfnamefont {E.}~\bibnamefont
  {M\"ortsell}},\ }\href@noop {} {\  (\bibinfo {year} {2021})},\ \Eprint
  {http://arxiv.org/abs/2106.09030} {arXiv:2106.09030 [astro-ph.CO]}
  \BibitemShut {NoStop}%
\bibitem [{\citenamefont {Caravano}\ \emph {et~al.}(2021)\citenamefont
  {Caravano}, \citenamefont {L\"uben},\ and\ \citenamefont
  {Weller}}]{Caravano:2021aum}%
  \BibitemOpen
  \bibfield  {author} {\bibinfo {author} {\bibfnamefont {A.}~\bibnamefont
  {Caravano}}, \bibinfo {author} {\bibfnamefont {M.}~\bibnamefont {L\"uben}}, \
  and\ \bibinfo {author} {\bibfnamefont {J.}~\bibnamefont {Weller}},\
  }\href@noop {} {\  (\bibinfo {year} {2021})},\ \Eprint
  {http://arxiv.org/abs/2101.08791} {arXiv:2101.08791 [gr-qc]} \BibitemShut
  {NoStop}%
\bibitem [{\citenamefont {Chabab}\ \emph {et~al.}(2019)\citenamefont {Chabab},
  \citenamefont {El~Moumni}, \citenamefont {Iraoui},\ and\ \citenamefont
  {Masmar}}]{Chabab:2019mlu}%
  \BibitemOpen
  \bibfield  {author} {\bibinfo {author} {\bibfnamefont {M.}~\bibnamefont
  {Chabab}}, \bibinfo {author} {\bibfnamefont {H.}~\bibnamefont {El~Moumni}},
  \bibinfo {author} {\bibfnamefont {S.}~\bibnamefont {Iraoui}}, \ and\ \bibinfo
  {author} {\bibfnamefont {K.}~\bibnamefont {Masmar}},\ }\href {\doibase
  10.1140/epjc/s10052-019-6850-0} {\bibfield  {journal} {\bibinfo  {journal}
  {Eur. Phys. J. C}\ }\textbf {\bibinfo {volume} {79}},\ \bibinfo {pages} {342}
  (\bibinfo {year} {2019})},\ \Eprint {http://arxiv.org/abs/1904.03532}
  {arXiv:1904.03532 [hep-th]} \BibitemShut {NoStop}%
\bibitem [{\citenamefont {Wu}\ \emph {et~al.}(2020)\citenamefont {Wu},
  \citenamefont {Wang}, \citenamefont {Xu},\ and\ \citenamefont
  {Yang}}]{Wu:2020fij}%
  \BibitemOpen
  \bibfield  {author} {\bibinfo {author} {\bibfnamefont {B.}~\bibnamefont
  {Wu}}, \bibinfo {author} {\bibfnamefont {C.}~\bibnamefont {Wang}}, \bibinfo
  {author} {\bibfnamefont {Z.-M.}\ \bibnamefont {Xu}}, \ and\ \bibinfo {author}
  {\bibfnamefont {W.-L.}\ \bibnamefont {Yang}},\ }\href@noop {} {\  (\bibinfo
  {year} {2020})},\ \Eprint {http://arxiv.org/abs/2006.09021} {arXiv:2006.09021
  [gr-qc]} \BibitemShut {NoStop}%
\bibitem [{\citenamefont {de~Rham}\ \emph
  {et~al.}(2011{\natexlab{b}})\citenamefont {de~Rham}, \citenamefont
  {Gabadadze},\ and\ \citenamefont {Tolley}}]{deRham:2010kj}%
  \BibitemOpen
  \bibfield  {author} {\bibinfo {author} {\bibfnamefont {C.}~\bibnamefont
  {de~Rham}}, \bibinfo {author} {\bibfnamefont {G.}~\bibnamefont {Gabadadze}},
  \ and\ \bibinfo {author} {\bibfnamefont {A.~J.}\ \bibnamefont {Tolley}},\
  }\href {\doibase 10.1103/PhysRevLett.106.231101} {\bibfield  {journal}
  {\bibinfo  {journal} {Phys. Rev. Lett.}\ }\textbf {\bibinfo {volume} {106}},\
  \bibinfo {pages} {231101} (\bibinfo {year} {2011}{\natexlab{b}})},\ \Eprint
  {http://arxiv.org/abs/1011.1232} {arXiv:1011.1232 [hep-th]} \BibitemShut
  {NoStop}%
\bibitem [{\citenamefont {Ghosh}\ \emph {et~al.}(2016)\citenamefont {Ghosh},
  \citenamefont {Tannukij},\ and\ \citenamefont {Wongjun}}]{Ghosh:2015cva}%
  \BibitemOpen
  \bibfield  {author} {\bibinfo {author} {\bibfnamefont {S.~G.}\ \bibnamefont
  {Ghosh}}, \bibinfo {author} {\bibfnamefont {L.}~\bibnamefont {Tannukij}}, \
  and\ \bibinfo {author} {\bibfnamefont {P.}~\bibnamefont {Wongjun}},\ }\href
  {\doibase 10.1140/epjc/s10052-016-3943-x} {\bibfield  {journal} {\bibinfo
  {journal} {Eur. Phys. J. C}\ }\textbf {\bibinfo {volume} {76}},\ \bibinfo
  {pages} {119} (\bibinfo {year} {2016})},\ \Eprint
  {http://arxiv.org/abs/1506.07119} {arXiv:1506.07119 [gr-qc]} \BibitemShut
  {NoStop}%
\bibitem [{\citenamefont {Hendi}\ \emph {et~al.}(2023)\citenamefont {Hendi},
  \citenamefont {Jafarzade},\ and\ \citenamefont
  {Eslam~Panah}}]{Hendi:2022qgi}%
  \BibitemOpen
  \bibfield  {author} {\bibinfo {author} {\bibfnamefont {S.~H.}\ \bibnamefont
  {Hendi}}, \bibinfo {author} {\bibfnamefont {K.}~\bibnamefont {Jafarzade}}, \
  and\ \bibinfo {author} {\bibfnamefont {B.}~\bibnamefont {Eslam~Panah}},\
  }\href {\doibase 10.1088/1475-7516/2023/02/022} {\bibfield  {journal}
  {\bibinfo  {journal} {JCAP}\ }\textbf {\bibinfo {volume} {02}},\ \bibinfo
  {pages} {022} (\bibinfo {year} {2023})},\ \Eprint
  {http://arxiv.org/abs/2206.05132} {arXiv:2206.05132 [gr-qc]} \BibitemShut
  {NoStop}%
\bibitem [{\citenamefont {Zhang}\ and\ \citenamefont {Li}(2016)}]{HZhang}%
  \BibitemOpen
  \bibfield  {author} {\bibinfo {author} {\bibfnamefont {H.}~\bibnamefont
  {Zhang}}\ and\ \bibinfo {author} {\bibfnamefont {X.-Z.}\ \bibnamefont {Li}},\
  }\href {\doibase 10.1103/PhysRevD.93.124039} {\bibfield  {journal} {\bibinfo
  {journal} {Phys. Rev. D}\ }\textbf {\bibinfo {volume} {93}},\ \bibinfo
  {pages} {124039} (\bibinfo {year} {2016})}\BibitemShut {NoStop}%
\bibitem [{\citenamefont {Xu}\ \emph {et~al.}(2015)\citenamefont {Xu},
  \citenamefont {Cao},\ and\ \citenamefont {Hu}}]{PVMassI}%
  \BibitemOpen
  \bibfield  {author} {\bibinfo {author} {\bibfnamefont {J.}~\bibnamefont
  {Xu}}, \bibinfo {author} {\bibfnamefont {L.-M.}\ \bibnamefont {Cao}}, \ and\
  \bibinfo {author} {\bibfnamefont {Y.-P.}\ \bibnamefont {Hu}},\ }\href
  {\doibase 10.1103/PhysRevD.91.124033} {\bibfield  {journal} {\bibinfo
  {journal} {Phys. Rev. D}\ }\textbf {\bibinfo {volume} {91}},\ \bibinfo
  {pages} {124033} (\bibinfo {year} {2015})}\BibitemShut {NoStop}%
\bibitem [{\citenamefont {Hendi}\ \emph {et~al.}(2015)\citenamefont {Hendi},
  \citenamefont {Panah},\ and\ \citenamefont {Panahiyan}}]{PVMassII}%
  \BibitemOpen
  \bibfield  {author} {\bibinfo {author} {\bibfnamefont {S.~H.}\ \bibnamefont
  {Hendi}}, \bibinfo {author} {\bibfnamefont {B.~E.}\ \bibnamefont {Panah}}, \
  and\ \bibinfo {author} {\bibfnamefont {S.}~\bibnamefont {Panahiyan}},\
  }\href@noop {} {\bibfield  {journal} {\bibinfo  {journal} {Journal of High
  Energy Physics}\ }\textbf {\bibinfo {volume} {2015}},\ \bibinfo {pages} {157}
  (\bibinfo {year} {2015})}\BibitemShut {NoStop}%
\bibitem [{\citenamefont {Hendi}\ \emph
  {et~al.}(2016{\natexlab{b}})\citenamefont {Hendi}, \citenamefont {Panah},\
  and\ \citenamefont {Panahiyan}}]{PVMassIII}%
  \BibitemOpen
  \bibfield  {author} {\bibinfo {author} {\bibfnamefont {S.}~\bibnamefont
  {Hendi}}, \bibinfo {author} {\bibfnamefont {B.~E.}\ \bibnamefont {Panah}}, \
  and\ \bibinfo {author} {\bibfnamefont {S.}~\bibnamefont {Panahiyan}},\
  }\href@noop {} {\bibfield  {journal} {\bibinfo  {journal} {Classical and
  Quantum Gravity}\ }\textbf {\bibinfo {volume} {33}},\ \bibinfo {pages}
  {235007} (\bibinfo {year} {2016}{\natexlab{b}})}\BibitemShut {NoStop}%
\bibitem [{\citenamefont {Davison}(2013)}]{Davison:2013jba}%
  \BibitemOpen
  \bibfield  {author} {\bibinfo {author} {\bibfnamefont {R.~A.}\ \bibnamefont
  {Davison}},\ }\href {\doibase 10.1103/PhysRevD.88.086003} {\bibfield
  {journal} {\bibinfo  {journal} {Phys. Rev. D}\ }\textbf {\bibinfo {volume}
  {88}},\ \bibinfo {pages} {086003} (\bibinfo {year} {2013})},\ \Eprint
  {http://arxiv.org/abs/1306.5792} {arXiv:1306.5792 [hep-th]} \BibitemShut
  {NoStop}%
\bibitem [{\citenamefont {Hartnoll}\ \emph
  {et~al.}(2008{\natexlab{a}})\citenamefont {Hartnoll}, \citenamefont
  {Herzog},\ and\ \citenamefont {Horowitz}}]{Hartnoll:2008vx}%
  \BibitemOpen
  \bibfield  {author} {\bibinfo {author} {\bibfnamefont {S.~A.}\ \bibnamefont
  {Hartnoll}}, \bibinfo {author} {\bibfnamefont {C.~P.}\ \bibnamefont
  {Herzog}}, \ and\ \bibinfo {author} {\bibfnamefont {G.~T.}\ \bibnamefont
  {Horowitz}},\ }\href {\doibase 10.1103/PhysRevLett.101.031601} {\bibfield
  {journal} {\bibinfo  {journal} {Phys. Rev. Lett.}\ }\textbf {\bibinfo
  {volume} {101}},\ \bibinfo {pages} {031601} (\bibinfo {year}
  {2008}{\natexlab{a}})},\ \Eprint {http://arxiv.org/abs/0803.3295}
  {arXiv:0803.3295 [hep-th]} \BibitemShut {NoStop}%
\bibitem [{\citenamefont {Hartnoll}\ \emph
  {et~al.}(2008{\natexlab{b}})\citenamefont {Hartnoll}, \citenamefont
  {Herzog},\ and\ \citenamefont {Horowitz}}]{Hartnoll:2008kx}%
  \BibitemOpen
  \bibfield  {author} {\bibinfo {author} {\bibfnamefont {S.~A.}\ \bibnamefont
  {Hartnoll}}, \bibinfo {author} {\bibfnamefont {C.~P.}\ \bibnamefont
  {Herzog}}, \ and\ \bibinfo {author} {\bibfnamefont {G.~T.}\ \bibnamefont
  {Horowitz}},\ }\href {\doibase 10.1088/1126-6708/2008/12/015} {\bibfield
  {journal} {\bibinfo  {journal} {JHEP}\ }\textbf {\bibinfo {volume} {12}},\
  \bibinfo {pages} {015} (\bibinfo {year} {2008}{\natexlab{b}})},\ \Eprint
  {http://arxiv.org/abs/0810.1563} {arXiv:0810.1563 [hep-th]} \BibitemShut
  {NoStop}%
\bibitem [{\citenamefont {Gregory}\ \emph {et~al.}(2009)\citenamefont
  {Gregory}, \citenamefont {Kanno},\ and\ \citenamefont
  {Soda}}]{Gregory:2009fj}%
  \BibitemOpen
  \bibfield  {author} {\bibinfo {author} {\bibfnamefont {R.}~\bibnamefont
  {Gregory}}, \bibinfo {author} {\bibfnamefont {S.}~\bibnamefont {Kanno}}, \
  and\ \bibinfo {author} {\bibfnamefont {J.}~\bibnamefont {Soda}},\ }\href
  {\doibase 10.1088/1126-6708/2009/10/010} {\bibfield  {journal} {\bibinfo
  {journal} {JHEP}\ }\textbf {\bibinfo {volume} {10}},\ \bibinfo {pages} {010}
  (\bibinfo {year} {2009})},\ \Eprint {http://arxiv.org/abs/0907.3203}
  {arXiv:0907.3203 [hep-th]} \BibitemShut {NoStop}%
\bibitem [{\citenamefont {Barclay}\ \emph {et~al.}(2010)\citenamefont
  {Barclay}, \citenamefont {Gregory}, \citenamefont {Kanno},\ and\
  \citenamefont {Sutcliffe}}]{Barclay:2010up}%
  \BibitemOpen
  \bibfield  {author} {\bibinfo {author} {\bibfnamefont {L.}~\bibnamefont
  {Barclay}}, \bibinfo {author} {\bibfnamefont {R.}~\bibnamefont {Gregory}},
  \bibinfo {author} {\bibfnamefont {S.}~\bibnamefont {Kanno}}, \ and\ \bibinfo
  {author} {\bibfnamefont {P.}~\bibnamefont {Sutcliffe}},\ }\href {\doibase
  10.1007/JHEP12(2010)029} {\bibfield  {journal} {\bibinfo  {journal} {JHEP}\
  }\textbf {\bibinfo {volume} {12}},\ \bibinfo {pages} {029} (\bibinfo {year}
  {2010})},\ \Eprint {http://arxiv.org/abs/1009.1991} {arXiv:1009.1991
  [hep-th]} \BibitemShut {NoStop}%
\bibitem [{\citenamefont {Alberte}\ \emph {et~al.}(2016)\citenamefont
  {Alberte}, \citenamefont {Baggioli}, \citenamefont {Khmelnitsky},\ and\
  \citenamefont {Pujolas}}]{Alberte:2015isw}%
  \BibitemOpen
  \bibfield  {author} {\bibinfo {author} {\bibfnamefont {L.}~\bibnamefont
  {Alberte}}, \bibinfo {author} {\bibfnamefont {M.}~\bibnamefont {Baggioli}},
  \bibinfo {author} {\bibfnamefont {A.}~\bibnamefont {Khmelnitsky}}, \ and\
  \bibinfo {author} {\bibfnamefont {O.}~\bibnamefont {Pujolas}},\ }\href
  {\doibase 10.1007/JHEP02(2016)114} {\bibfield  {journal} {\bibinfo  {journal}
  {JHEP}\ }\textbf {\bibinfo {volume} {02}},\ \bibinfo {pages} {114} (\bibinfo
  {year} {2016})},\ \Eprint {http://arxiv.org/abs/1510.09089} {arXiv:1510.09089
  [hep-th]} \BibitemShut {NoStop}%
\bibitem [{\citenamefont {Alberte}\ \emph {et~al.}(2018)\citenamefont
  {Alberte}, \citenamefont {Ammon}, \citenamefont {Jim\'enez-Alba},
  \citenamefont {Baggioli},\ and\ \citenamefont {Pujol\`as}}]{Alberte:2017oqx}%
  \BibitemOpen
  \bibfield  {author} {\bibinfo {author} {\bibfnamefont {L.}~\bibnamefont
  {Alberte}}, \bibinfo {author} {\bibfnamefont {M.}~\bibnamefont {Ammon}},
  \bibinfo {author} {\bibfnamefont {A.}~\bibnamefont {Jim\'enez-Alba}},
  \bibinfo {author} {\bibfnamefont {M.}~\bibnamefont {Baggioli}}, \ and\
  \bibinfo {author} {\bibfnamefont {O.}~\bibnamefont {Pujol\`as}},\ }\href
  {\doibase 10.1103/PhysRevLett.120.171602} {\bibfield  {journal} {\bibinfo
  {journal} {Phys. Rev. Lett.}\ }\textbf {\bibinfo {volume} {120}},\ \bibinfo
  {pages} {171602} (\bibinfo {year} {2018})},\ \Eprint
  {http://arxiv.org/abs/1711.03100} {arXiv:1711.03100 [hep-th]} \BibitemShut
  {NoStop}%
\bibitem [{\citenamefont {Zhang}\ \emph {et~al.}(2019)\citenamefont {Zhang},
  \citenamefont {Hu},\ and\ \citenamefont {Zhang}}]{Zhang:2019oes}%
  \BibitemOpen
  \bibfield  {author} {\bibinfo {author} {\bibfnamefont {H.}~\bibnamefont
  {Zhang}}, \bibinfo {author} {\bibfnamefont {Y.-p.}\ \bibnamefont {Hu}}, \
  and\ \bibinfo {author} {\bibfnamefont {Y.}~\bibnamefont {Zhang}},\ }\href
  {\doibase 10.1016/j.dark.2018.100257} {\bibfield  {journal} {\bibinfo
  {journal} {Phys. Dark Univ.}\ }\textbf {\bibinfo {volume} {23}},\ \bibinfo
  {pages} {100257} (\bibinfo {year} {2019})},\ \Eprint
  {http://arxiv.org/abs/1901.09331} {arXiv:1901.09331 [gr-qc]} \BibitemShut
  {NoStop}%
\bibitem [{\citenamefont {Blake}\ and\ \citenamefont
  {Tong}(2013)}]{Blake:2013bqa}%
  \BibitemOpen
  \bibfield  {author} {\bibinfo {author} {\bibfnamefont {M.}~\bibnamefont
  {Blake}}\ and\ \bibinfo {author} {\bibfnamefont {D.}~\bibnamefont {Tong}},\
  }\href {\doibase 10.1103/PhysRevD.88.106004} {\bibfield  {journal} {\bibinfo
  {journal} {Phys. Rev. D}\ }\textbf {\bibinfo {volume} {88}},\ \bibinfo
  {pages} {106004} (\bibinfo {year} {2013})},\ \Eprint
  {http://arxiv.org/abs/1308.4970} {arXiv:1308.4970 [hep-th]} \BibitemShut
  {NoStop}%
\bibitem [{\citenamefont {Cao}\ and\ \citenamefont {Peng}(2015)}]{Cao:2015cza}%
  \BibitemOpen
  \bibfield  {author} {\bibinfo {author} {\bibfnamefont {L.-M.}\ \bibnamefont
  {Cao}}\ and\ \bibinfo {author} {\bibfnamefont {Y.}~\bibnamefont {Peng}},\
  }\href {\doibase 10.1103/PhysRevD.92.124052} {\bibfield  {journal} {\bibinfo
  {journal} {Phys. Rev. D}\ }\textbf {\bibinfo {volume} {92}},\ \bibinfo
  {pages} {124052} (\bibinfo {year} {2015})},\ \Eprint
  {http://arxiv.org/abs/1509.08738} {arXiv:1509.08738 [hep-th]} \BibitemShut
  {NoStop}%
\bibitem [{\citenamefont {Baggioli}\ and\ \citenamefont
  {Pujolas}(2015)}]{Baggioli:2014roa}%
  \BibitemOpen
  \bibfield  {author} {\bibinfo {author} {\bibfnamefont {M.}~\bibnamefont
  {Baggioli}}\ and\ \bibinfo {author} {\bibfnamefont {O.}~\bibnamefont
  {Pujolas}},\ }\href {\doibase 10.1103/PhysRevLett.114.251602} {\bibfield
  {journal} {\bibinfo  {journal} {Phys. Rev. Lett.}\ }\textbf {\bibinfo
  {volume} {114}},\ \bibinfo {pages} {251602} (\bibinfo {year} {2015})},\
  \Eprint {http://arxiv.org/abs/1411.1003} {arXiv:1411.1003 [hep-th]}
  \BibitemShut {NoStop}%
\bibitem [{\citenamefont {Gialamas}\ and\ \citenamefont
  {Tamvakis}(2023{\natexlab{a}})}]{Gialamas:2023aim}%
  \BibitemOpen
  \bibfield  {author} {\bibinfo {author} {\bibfnamefont {I.~D.}\ \bibnamefont
  {Gialamas}}\ and\ \bibinfo {author} {\bibfnamefont {K.}~\bibnamefont
  {Tamvakis}},\ }\href {\doibase 10.1103/PhysRevD.107.104012} {\bibfield
  {journal} {\bibinfo  {journal} {Phys. Rev. D}\ }\textbf {\bibinfo {volume}
  {107}},\ \bibinfo {pages} {104012} (\bibinfo {year} {2023}{\natexlab{a}})},\
  \Eprint {http://arxiv.org/abs/2303.11353} {arXiv:2303.11353 [gr-qc]}
  \BibitemShut {NoStop}%
\bibitem [{\citenamefont {Gialamas}\ and\ \citenamefont
  {Tamvakis}(2023{\natexlab{b}})}]{Gialamas:2023lxj}%
  \BibitemOpen
  \bibfield  {author} {\bibinfo {author} {\bibfnamefont {I.~D.}\ \bibnamefont
  {Gialamas}}\ and\ \bibinfo {author} {\bibfnamefont {K.}~\bibnamefont
  {Tamvakis}},\ }\href {\doibase 10.1103/PhysRevD.108.104023} {\bibfield
  {journal} {\bibinfo  {journal} {Phys. Rev. D}\ }\textbf {\bibinfo {volume}
  {108}},\ \bibinfo {pages} {104023} (\bibinfo {year} {2023}{\natexlab{b}})},\
  \Eprint {http://arxiv.org/abs/2307.05673} {arXiv:2307.05673 [gr-qc]}
  \BibitemShut {NoStop}%
\bibitem [{\citenamefont {Gialamas}\ and\ \citenamefont
  {Tamvakis}(2024)}]{Gialamas:2023fly}%
  \BibitemOpen
  \bibfield  {author} {\bibinfo {author} {\bibfnamefont {I.~D.}\ \bibnamefont
  {Gialamas}}\ and\ \bibinfo {author} {\bibfnamefont {K.}~\bibnamefont
  {Tamvakis}},\ }\href {\doibase 10.1088/1475-7516/2024/03/016} {\bibfield
  {journal} {\bibinfo  {journal} {JCAP}\ }\textbf {\bibinfo {volume} {03}},\
  \bibinfo {pages} {016} (\bibinfo {year} {2024})},\ \Eprint
  {http://arxiv.org/abs/2311.14799} {arXiv:2311.14799 [gr-qc]} \BibitemShut
  {NoStop}%
\bibitem [{\citenamefont {Wei}\ and\ \citenamefont {Liu}(2024)}]{Wei:2023fkn}%
  \BibitemOpen
  \bibfield  {author} {\bibinfo {author} {\bibfnamefont {S.-W.}\ \bibnamefont
  {Wei}}\ and\ \bibinfo {author} {\bibfnamefont {Y.-X.}\ \bibnamefont {Liu}},\
  }\href {\doibase 10.1016/j.dark.2023.101409} {\bibfield  {journal} {\bibinfo
  {journal} {Phys. Dark Univ.}\ }\textbf {\bibinfo {volume} {43}},\ \bibinfo
  {pages} {101409} (\bibinfo {year} {2024})},\ \Eprint
  {http://arxiv.org/abs/2308.11883} {arXiv:2308.11883 [gr-qc]} \BibitemShut
  {NoStop}%
\bibitem [{\citenamefont {Panpanich}\ \emph {et~al.}(2019)\citenamefont
  {Panpanich}, \citenamefont {Ponglertsakul},\ and\ \citenamefont
  {Tannukij}}]{Panpanich:2019mll}%
  \BibitemOpen
  \bibfield  {author} {\bibinfo {author} {\bibfnamefont {S.}~\bibnamefont
  {Panpanich}}, \bibinfo {author} {\bibfnamefont {S.}~\bibnamefont
  {Ponglertsakul}}, \ and\ \bibinfo {author} {\bibfnamefont {L.}~\bibnamefont
  {Tannukij}},\ }\href {\doibase 10.1103/PhysRevD.100.044031} {\bibfield
  {journal} {\bibinfo  {journal} {Phys. Rev. D}\ }\textbf {\bibinfo {volume}
  {100}},\ \bibinfo {pages} {044031} (\bibinfo {year} {2019})},\ \Eprint
  {http://arxiv.org/abs/1904.02915} {arXiv:1904.02915 [gr-qc]} \BibitemShut
  {NoStop}%
\bibitem [{\citenamefont {Hashimoto}\ \emph {et~al.}(2020)\citenamefont
  {Hashimoto}, \citenamefont {Kinoshita},\ and\ \citenamefont
  {Murata}}]{Hashimoto:2018okj}%
  \BibitemOpen
  \bibfield  {author} {\bibinfo {author} {\bibfnamefont {K.}~\bibnamefont
  {Hashimoto}}, \bibinfo {author} {\bibfnamefont {S.}~\bibnamefont
  {Kinoshita}}, \ and\ \bibinfo {author} {\bibfnamefont {K.}~\bibnamefont
  {Murata}},\ }\href {\doibase 10.1103/PhysRevD.101.066018} {\bibfield
  {journal} {\bibinfo  {journal} {Phys. Rev. D}\ }\textbf {\bibinfo {volume}
  {101}},\ \bibinfo {pages} {066018} (\bibinfo {year} {2020})},\ \Eprint
  {http://arxiv.org/abs/1811.12617} {arXiv:1811.12617 [hep-th]} \BibitemShut
  {NoStop}%
\bibitem [{\citenamefont {Hashimoto}\ \emph {et~al.}(2019)\citenamefont
  {Hashimoto}, \citenamefont {Kinoshita},\ and\ \citenamefont
  {Murata}}]{Hashimoto:2019jmw}%
  \BibitemOpen
  \bibfield  {author} {\bibinfo {author} {\bibfnamefont {K.}~\bibnamefont
  {Hashimoto}}, \bibinfo {author} {\bibfnamefont {S.}~\bibnamefont
  {Kinoshita}}, \ and\ \bibinfo {author} {\bibfnamefont {K.}~\bibnamefont
  {Murata}},\ }\href {\doibase 10.1103/PhysRevLett.123.031602} {\bibfield
  {journal} {\bibinfo  {journal} {Phys. Rev. Lett.}\ }\textbf {\bibinfo
  {volume} {123}},\ \bibinfo {pages} {031602} (\bibinfo {year} {2019})},\
  \Eprint {http://arxiv.org/abs/1906.09113} {arXiv:1906.09113 [hep-th]}
  \BibitemShut {NoStop}%
\bibitem [{\citenamefont {Liu}\ \emph {et~al.}(2022)\citenamefont {Liu},
  \citenamefont {Chen}, \citenamefont {Zeng}, \citenamefont {Zhang},
  \citenamefont {Zhang},\ and\ \citenamefont {Zhang}}]{Liu:2022cev}%
  \BibitemOpen
  \bibfield  {author} {\bibinfo {author} {\bibfnamefont {Y.}~\bibnamefont
  {Liu}}, \bibinfo {author} {\bibfnamefont {Q.}~\bibnamefont {Chen}}, \bibinfo
  {author} {\bibfnamefont {X.-X.}\ \bibnamefont {Zeng}}, \bibinfo {author}
  {\bibfnamefont {H.}~\bibnamefont {Zhang}}, \bibinfo {author} {\bibfnamefont
  {W.-L.}\ \bibnamefont {Zhang}}, \ and\ \bibinfo {author} {\bibfnamefont
  {W.}~\bibnamefont {Zhang}},\ }\href {\doibase 10.1007/JHEP10(2022)189}
  {\bibfield  {journal} {\bibinfo  {journal} {JHEP}\ }\textbf {\bibinfo
  {volume} {10}},\ \bibinfo {pages} {189} (\bibinfo {year} {2022})},\ \Eprint
  {http://arxiv.org/abs/2201.03161} {arXiv:2201.03161 [hep-th]} \BibitemShut
  {NoStop}%
\bibitem [{\citenamefont {Perlick}(2004)}]{Perlick:2004tq}%
  \BibitemOpen
  \bibfield  {author} {\bibinfo {author} {\bibfnamefont {V.}~\bibnamefont
  {Perlick}},\ }\href@noop {} {\bibfield  {journal} {\bibinfo  {journal}
  {Living Rev. Rel.}\ }\textbf {\bibinfo {volume} {7}},\ \bibinfo {pages} {9}
  (\bibinfo {year} {2004})}\BibitemShut {NoStop}%
\bibitem [{\citenamefont {Grenzebach}\ \emph {et~al.}(2014)\citenamefont
  {Grenzebach}, \citenamefont {Perlick},\ and\ \citenamefont
  {L\"ammerzahl}}]{Grenzebach:2014fha}%
  \BibitemOpen
  \bibfield  {author} {\bibinfo {author} {\bibfnamefont {A.}~\bibnamefont
  {Grenzebach}}, \bibinfo {author} {\bibfnamefont {V.}~\bibnamefont {Perlick}},
  \ and\ \bibinfo {author} {\bibfnamefont {C.}~\bibnamefont {L\"ammerzahl}},\
  }\href {\doibase 10.1103/PhysRevD.89.124004} {\bibfield  {journal} {\bibinfo
  {journal} {Phys. Rev. D}\ }\textbf {\bibinfo {volume} {89}},\ \bibinfo
  {pages} {124004} (\bibinfo {year} {2014})},\ \Eprint
  {http://arxiv.org/abs/1403.5234} {arXiv:1403.5234 [gr-qc]} \BibitemShut
  {NoStop}%
\bibitem [{\citenamefont {Grenzebach}\ \emph {et~al.}(2015)\citenamefont
  {Grenzebach}, \citenamefont {Perlick},\ and\ \citenamefont
  {L\"ammerzahl}}]{Grenzebach:2015oea}%
  \BibitemOpen
  \bibfield  {author} {\bibinfo {author} {\bibfnamefont {A.}~\bibnamefont
  {Grenzebach}}, \bibinfo {author} {\bibfnamefont {V.}~\bibnamefont {Perlick}},
  \ and\ \bibinfo {author} {\bibfnamefont {C.}~\bibnamefont {L\"ammerzahl}},\
  }\href {\doibase 10.1142/S0218271815420249} {\bibfield  {journal} {\bibinfo
  {journal} {Int. J. Mod. Phys. D}\ }\textbf {\bibinfo {volume} {24}},\
  \bibinfo {pages} {1542024} (\bibinfo {year} {2015})},\ \Eprint
  {http://arxiv.org/abs/1503.03036} {arXiv:1503.03036 [gr-qc]} \BibitemShut
  {NoStop}%
\bibitem [{\citenamefont {Johnson}\ \emph {et~al.}(2020)\citenamefont {Johnson}
  \emph {et~al.}}]{Johnson:2019ljv}%
  \BibitemOpen
  \bibfield  {author} {\bibinfo {author} {\bibfnamefont {M.~D.}\ \bibnamefont
  {Johnson}} \emph {et~al.},\ }\href {\doibase 10.1126/sciadv.aaz1310}
  {\bibfield  {journal} {\bibinfo  {journal} {Sci. Adv.}\ }\textbf {\bibinfo
  {volume} {6}},\ \bibinfo {pages} {eaaz1310} (\bibinfo {year} {2020})},\
  \Eprint {http://arxiv.org/abs/1907.04329} {arXiv:1907.04329 [astro-ph.IM]}
  \BibitemShut {NoStop}%
\bibitem [{\citenamefont {Gralla}\ and\ \citenamefont
  {Lupsasca}(2020)}]{Gralla:2019drh}%
  \BibitemOpen
  \bibfield  {author} {\bibinfo {author} {\bibfnamefont {S.~E.}\ \bibnamefont
  {Gralla}}\ and\ \bibinfo {author} {\bibfnamefont {A.}~\bibnamefont
  {Lupsasca}},\ }\href {\doibase 10.1103/PhysRevD.101.044031} {\bibfield
  {journal} {\bibinfo  {journal} {Phys. Rev. D}\ }\textbf {\bibinfo {volume}
  {101}},\ \bibinfo {pages} {044031} (\bibinfo {year} {2020})},\ \Eprint
  {http://arxiv.org/abs/1910.12873} {arXiv:1910.12873 [gr-qc]} \BibitemShut
  {NoStop}%
\bibitem [{\citenamefont {Himwich}\ \emph {et~al.}(2020)\citenamefont
  {Himwich}, \citenamefont {Johnson}, \citenamefont {Lupsasca},\ and\
  \citenamefont {Strominger}}]{Himwich:2020msm}%
  \BibitemOpen
  \bibfield  {author} {\bibinfo {author} {\bibfnamefont {E.}~\bibnamefont
  {Himwich}}, \bibinfo {author} {\bibfnamefont {M.~D.}\ \bibnamefont
  {Johnson}}, \bibinfo {author} {\bibfnamefont {A.}~\bibnamefont {Lupsasca}}, \
  and\ \bibinfo {author} {\bibfnamefont {A.}~\bibnamefont {Strominger}},\
  }\href {\doibase 10.1103/PhysRevD.101.084020} {\bibfield  {journal} {\bibinfo
   {journal} {Phys. Rev. D}\ }\textbf {\bibinfo {volume} {101}},\ \bibinfo
  {pages} {084020} (\bibinfo {year} {2020})},\ \Eprint
  {http://arxiv.org/abs/2001.08750} {arXiv:2001.08750 [gr-qc]} \BibitemShut
  {NoStop}%
\bibitem [{\citenamefont {Hadar}\ \emph {et~al.}(2022)\citenamefont {Hadar},
  \citenamefont {Kapec}, \citenamefont {Lupsasca},\ and\ \citenamefont
  {Strominger}}]{Hadar:2022xag}%
  \BibitemOpen
  \bibfield  {author} {\bibinfo {author} {\bibfnamefont {S.}~\bibnamefont
  {Hadar}}, \bibinfo {author} {\bibfnamefont {D.}~\bibnamefont {Kapec}},
  \bibinfo {author} {\bibfnamefont {A.}~\bibnamefont {Lupsasca}}, \ and\
  \bibinfo {author} {\bibfnamefont {A.}~\bibnamefont {Strominger}},\ }\href
  {\doibase 10.1088/1361-6382/ac8d43} {\bibfield  {journal} {\bibinfo
  {journal} {Class. Quant. Grav.}\ }\textbf {\bibinfo {volume} {39}},\ \bibinfo
  {pages} {215001} (\bibinfo {year} {2022})},\ \Eprint
  {http://arxiv.org/abs/2205.05064} {arXiv:2205.05064 [gr-qc]} \BibitemShut
  {NoStop}%
\bibitem [{\citenamefont {Kapec}\ \emph {et~al.}(2023)\citenamefont {Kapec},
  \citenamefont {Lupsasca},\ and\ \citenamefont {Strominger}}]{Kapec:2022dvc}%
  \BibitemOpen
  \bibfield  {author} {\bibinfo {author} {\bibfnamefont {D.}~\bibnamefont
  {Kapec}}, \bibinfo {author} {\bibfnamefont {A.}~\bibnamefont {Lupsasca}}, \
  and\ \bibinfo {author} {\bibfnamefont {A.}~\bibnamefont {Strominger}},\
  }\href {\doibase 10.1088/1361-6382/acc164} {\bibfield  {journal} {\bibinfo
  {journal} {Class. Quant. Grav.}\ }\textbf {\bibinfo {volume} {40}},\ \bibinfo
  {pages} {095006} (\bibinfo {year} {2023})},\ \Eprint
  {http://arxiv.org/abs/2211.01674} {arXiv:2211.01674 [gr-qc]} \BibitemShut
  {NoStop}%
\bibitem [{\citenamefont {Hartnoll}\ \emph {et~al.}(2018)\citenamefont
  {Hartnoll}, \citenamefont {Lucas},\ and\ \citenamefont
  {Sachdev}}]{hartnoll2018holographic}%
  \BibitemOpen
  \bibfield  {author} {\bibinfo {author} {\bibfnamefont {S.~A.}\ \bibnamefont
  {Hartnoll}}, \bibinfo {author} {\bibfnamefont {A.}~\bibnamefont {Lucas}}, \
  and\ \bibinfo {author} {\bibfnamefont {S.}~\bibnamefont {Sachdev}},\
  }\href@noop {} {\emph {\bibinfo {title} {Holographic quantum matter}}}\
  (\bibinfo  {publisher} {MIT press},\ \bibinfo {year} {2018})\BibitemShut
  {NoStop}%
\bibitem [{\citenamefont {Kaku}\ \emph {et~al.}(2021)\citenamefont {Kaku},
  \citenamefont {Murata},\ and\ \citenamefont {Tsujimura}}]{Kaku:2021xqp}%
  \BibitemOpen
  \bibfield  {author} {\bibinfo {author} {\bibfnamefont {Y.}~\bibnamefont
  {Kaku}}, \bibinfo {author} {\bibfnamefont {K.}~\bibnamefont {Murata}}, \ and\
  \bibinfo {author} {\bibfnamefont {J.}~\bibnamefont {Tsujimura}},\ }\href
  {\doibase 10.1007/JHEP09(2021)138} {\bibfield  {journal} {\bibinfo  {journal}
  {JHEP}\ }\textbf {\bibinfo {volume} {09}},\ \bibinfo {pages} {138} (\bibinfo
  {year} {2021})},\ \Eprint {http://arxiv.org/abs/2106.00304} {arXiv:2106.00304
  [hep-th]} \BibitemShut {NoStop}%
\bibitem [{\citenamefont {Hashimoto}\ \emph {et~al.}(2023)\citenamefont
  {Hashimoto}, \citenamefont {Takeda}, \citenamefont {Tanaka},\ and\
  \citenamefont {Yonezawa}}]{Hashimoto:2022aso}%
  \BibitemOpen
  \bibfield  {author} {\bibinfo {author} {\bibfnamefont {K.}~\bibnamefont
  {Hashimoto}}, \bibinfo {author} {\bibfnamefont {D.}~\bibnamefont {Takeda}},
  \bibinfo {author} {\bibfnamefont {K.}~\bibnamefont {Tanaka}}, \ and\ \bibinfo
  {author} {\bibfnamefont {S.}~\bibnamefont {Yonezawa}},\ }\href {\doibase
  10.1103/PhysRevResearch.5.023168} {\bibfield  {journal} {\bibinfo  {journal}
  {Phys. Rev. Res.}\ }\textbf {\bibinfo {volume} {5}},\ \bibinfo {pages}
  {023168} (\bibinfo {year} {2023})},\ \Eprint
  {http://arxiv.org/abs/2211.13863} {arXiv:2211.13863 [hep-th]} \BibitemShut
  {NoStop}%
\bibitem [{\citenamefont {Caron-Huot}(2023)}]{Caron-Huot:2022lff}%
  \BibitemOpen
  \bibfield  {author} {\bibinfo {author} {\bibfnamefont {S.}~\bibnamefont
  {Caron-Huot}},\ }\href {\doibase 10.1007/JHEP03(2023)047} {\bibfield
  {journal} {\bibinfo  {journal} {JHEP}\ }\textbf {\bibinfo {volume} {03}},\
  \bibinfo {pages} {047} (\bibinfo {year} {2023})},\ \Eprint
  {http://arxiv.org/abs/2211.11791} {arXiv:2211.11791 [hep-th]} \BibitemShut
  {NoStop}%
\bibitem [{\citenamefont {Zeng}\ \emph {et~al.}(2023)\citenamefont {Zeng},
  \citenamefont {He}, \citenamefont {Pu}, \citenamefont {Li},\ and\
  \citenamefont {Jiang}}]{Zeng:2023zlf}%
  \BibitemOpen
  \bibfield  {author} {\bibinfo {author} {\bibfnamefont {X.-X.}\ \bibnamefont
  {Zeng}}, \bibinfo {author} {\bibfnamefont {K.-J.}\ \bibnamefont {He}},
  \bibinfo {author} {\bibfnamefont {J.}~\bibnamefont {Pu}}, \bibinfo {author}
  {\bibfnamefont {G.-p.}\ \bibnamefont {Li}}, \ and\ \bibinfo {author}
  {\bibfnamefont {Q.-Q.}\ \bibnamefont {Jiang}},\ }\href {\doibase
  10.1140/epjc/s10052-023-12079-5} {\bibfield  {journal} {\bibinfo  {journal}
  {Eur. Phys. J. C}\ }\textbf {\bibinfo {volume} {83}},\ \bibinfo {pages} {897}
  (\bibinfo {year} {2023})},\ \Eprint {http://arxiv.org/abs/2302.03692}
  {arXiv:2302.03692 [gr-qc]} \BibitemShut {NoStop}%
\bibitem [{\citenamefont {Kinoshita}\ \emph {et~al.}(2023)\citenamefont
  {Kinoshita}, \citenamefont {Murata},\ and\ \citenamefont
  {Takeda}}]{Kinoshita:2023hgc}%
  \BibitemOpen
  \bibfield  {author} {\bibinfo {author} {\bibfnamefont {S.}~\bibnamefont
  {Kinoshita}}, \bibinfo {author} {\bibfnamefont {K.}~\bibnamefont {Murata}}, \
  and\ \bibinfo {author} {\bibfnamefont {D.}~\bibnamefont {Takeda}},\ }\href
  {\doibase 10.1007/JHEP10(2023)074} {\bibfield  {journal} {\bibinfo  {journal}
  {JHEP}\ }\textbf {\bibinfo {volume} {10}},\ \bibinfo {pages} {074} (\bibinfo
  {year} {2023})},\ \Eprint {http://arxiv.org/abs/2304.01936} {arXiv:2304.01936
  [hep-th]} \BibitemShut {NoStop}%
\bibitem [{\citenamefont {Bardeen}\ \emph {et~al.}(1972)\citenamefont
  {Bardeen}, \citenamefont {Press},\ and\ \citenamefont
  {Teukolsky}}]{Bardeen:1972fi}%
  \BibitemOpen
  \bibfield  {author} {\bibinfo {author} {\bibfnamefont {J.~M.}\ \bibnamefont
  {Bardeen}}, \bibinfo {author} {\bibfnamefont {W.~H.}\ \bibnamefont {Press}},
  \ and\ \bibinfo {author} {\bibfnamefont {S.~A.}\ \bibnamefont {Teukolsky}},\
  }\href {\doibase 10.1086/151796} {\bibfield  {journal} {\bibinfo  {journal}
  {Astrophys. J.}\ }\textbf {\bibinfo {volume} {178}},\ \bibinfo {pages} {347}
  (\bibinfo {year} {1972})}\BibitemShut {NoStop}%
\bibitem [{\citenamefont {Ching}\ \emph
  {et~al.}(1995{\natexlab{a}})\citenamefont {Ching}, \citenamefont {Leung},
  \citenamefont {Suen},\ and\ \citenamefont {Young}}]{Ching:1994bd}%
  \BibitemOpen
  \bibfield  {author} {\bibinfo {author} {\bibfnamefont {E.~S.~C.}\
  \bibnamefont {Ching}}, \bibinfo {author} {\bibfnamefont {P.~T.}\ \bibnamefont
  {Leung}}, \bibinfo {author} {\bibfnamefont {W.~M.}\ \bibnamefont {Suen}}, \
  and\ \bibinfo {author} {\bibfnamefont {K.}~\bibnamefont {Young}},\ }\href
  {\doibase 10.1103/PhysRevLett.74.2414} {\bibfield  {journal} {\bibinfo
  {journal} {Phys. Rev. Lett.}\ }\textbf {\bibinfo {volume} {74}},\ \bibinfo
  {pages} {2414} (\bibinfo {year} {1995}{\natexlab{a}})},\ \Eprint
  {http://arxiv.org/abs/gr-qc/9410044} {arXiv:gr-qc/9410044} \BibitemShut
  {NoStop}%
\bibitem [{\citenamefont {Ching}\ \emph
  {et~al.}(1995{\natexlab{b}})\citenamefont {Ching}, \citenamefont {Leung},
  \citenamefont {Suen},\ and\ \citenamefont {Young}}]{Ching:1995tj}%
  \BibitemOpen
  \bibfield  {author} {\bibinfo {author} {\bibfnamefont {E.~S.~C.}\
  \bibnamefont {Ching}}, \bibinfo {author} {\bibfnamefont {P.~T.}\ \bibnamefont
  {Leung}}, \bibinfo {author} {\bibfnamefont {W.~M.}\ \bibnamefont {Suen}}, \
  and\ \bibinfo {author} {\bibfnamefont {K.}~\bibnamefont {Young}},\ }\href
  {\doibase 10.1103/PhysRevD.52.2118} {\bibfield  {journal} {\bibinfo
  {journal} {Phys. Rev. D}\ }\textbf {\bibinfo {volume} {52}},\ \bibinfo
  {pages} {2118} (\bibinfo {year} {1995}{\natexlab{b}})},\ \Eprint
  {http://arxiv.org/abs/gr-qc/9507035} {arXiv:gr-qc/9507035} \BibitemShut
  {NoStop}%
\bibitem [{\citenamefont {Horowitz}\ and\ \citenamefont
  {Hubeny}(2000)}]{Horowitz:1999jd}%
  \BibitemOpen
  \bibfield  {author} {\bibinfo {author} {\bibfnamefont {G.~T.}\ \bibnamefont
  {Horowitz}}\ and\ \bibinfo {author} {\bibfnamefont {V.~E.}\ \bibnamefont
  {Hubeny}},\ }\href {\doibase 10.1103/PhysRevD.62.024027} {\bibfield
  {journal} {\bibinfo  {journal} {Phys. Rev. D}\ }\textbf {\bibinfo {volume}
  {62}},\ \bibinfo {pages} {024027} (\bibinfo {year} {2000})},\ \Eprint
  {http://arxiv.org/abs/hep-th/9909056} {arXiv:hep-th/9909056} \BibitemShut
  {NoStop}%
\bibitem [{\citenamefont {Berti}\ \emph {et~al.}(2009)\citenamefont {Berti},
  \citenamefont {Cardoso},\ and\ \citenamefont {Starinets}}]{Berti:2009kk}%
  \BibitemOpen
  \bibfield  {author} {\bibinfo {author} {\bibfnamefont {E.}~\bibnamefont
  {Berti}}, \bibinfo {author} {\bibfnamefont {V.}~\bibnamefont {Cardoso}}, \
  and\ \bibinfo {author} {\bibfnamefont {A.~O.}\ \bibnamefont {Starinets}},\
  }\href {\doibase 10.1088/0264-9381/26/16/163001} {\bibfield  {journal}
  {\bibinfo  {journal} {Class. Quant. Grav.}\ }\textbf {\bibinfo {volume}
  {26}},\ \bibinfo {pages} {163001} (\bibinfo {year} {2009})},\ \Eprint
  {http://arxiv.org/abs/0905.2975} {arXiv:0905.2975 [gr-qc]} \BibitemShut
  {NoStop}%
\bibitem [{\citenamefont {Daghigh}\ \emph {et~al.}(2023)\citenamefont
  {Daghigh}, \citenamefont {Green},\ and\ \citenamefont
  {Morey}}]{Daghigh:2022uws}%
  \BibitemOpen
  \bibfield  {author} {\bibinfo {author} {\bibfnamefont {R.~G.}\ \bibnamefont
  {Daghigh}}, \bibinfo {author} {\bibfnamefont {M.~D.}\ \bibnamefont {Green}},
  \ and\ \bibinfo {author} {\bibfnamefont {J.~C.}\ \bibnamefont {Morey}},\
  }\href {\doibase 10.1103/PhysRevD.107.024023} {\bibfield  {journal} {\bibinfo
   {journal} {Phys. Rev. D}\ }\textbf {\bibinfo {volume} {107}},\ \bibinfo
  {pages} {024023} (\bibinfo {year} {2023})},\ \Eprint
  {http://arxiv.org/abs/2209.09324} {arXiv:2209.09324 [gr-qc]} \BibitemShut
  {NoStop}%
\bibitem [{\citenamefont {Perlick}\ and\ \citenamefont
  {Tsupko}(2022)}]{PERLICK20221}%
  \BibitemOpen
  \bibfield  {author} {\bibinfo {author} {\bibfnamefont {V.}~\bibnamefont
  {Perlick}}\ and\ \bibinfo {author} {\bibfnamefont {O.~Y.}\ \bibnamefont
  {Tsupko}},\ }\href {\doibase https://doi.org/10.1016/j.physrep.2021.10.004}
  {\bibfield  {journal} {\bibinfo  {journal} {Physics Reports}\ }\textbf
  {\bibinfo {volume} {947}},\ \bibinfo {pages} {1} (\bibinfo {year} {2022})},\
  \bibinfo {note} {calculating black hole shadows: Review of analytical
  studies}\BibitemShut {NoStop}%
\bibitem [{\citenamefont {Bardeen}(2018)}]{Bardeen:2018omt}%
  \BibitemOpen
  \bibfield  {author} {\bibinfo {author} {\bibfnamefont {J.~M.}\ \bibnamefont
  {Bardeen}},\ }\href@noop {} {\  (\bibinfo {year} {2018})},\ \Eprint
  {http://arxiv.org/abs/1808.08638} {arXiv:1808.08638 [gr-qc]} \BibitemShut
  {NoStop}%
\bibitem [{\citenamefont {Qiao}(2022)}]{Qiao:2022hfv}%
  \BibitemOpen
  \bibfield  {author} {\bibinfo {author} {\bibfnamefont {C.-K.}\ \bibnamefont
  {Qiao}},\ }\href {\doibase 10.1103/PhysRevD.106.084060} {\bibfield  {journal}
  {\bibinfo  {journal} {Phys. Rev. D}\ }\textbf {\bibinfo {volume} {106}},\
  \bibinfo {pages} {084060} (\bibinfo {year} {2022})},\ \Eprint
  {http://arxiv.org/abs/2208.01771} {arXiv:2208.01771 [gr-qc]} \BibitemShut
  {NoStop}%
\bibitem [{\citenamefont {Arkani-Hamed}\ \emph {et~al.}(2007)\citenamefont
  {Arkani-Hamed}, \citenamefont {Motl}, \citenamefont {Nicolis},\ and\
  \citenamefont {Vafa}}]{Arkani-Hamed:2006emk}%
  \BibitemOpen
  \bibfield  {author} {\bibinfo {author} {\bibfnamefont {N.}~\bibnamefont
  {Arkani-Hamed}}, \bibinfo {author} {\bibfnamefont {L.}~\bibnamefont {Motl}},
  \bibinfo {author} {\bibfnamefont {A.}~\bibnamefont {Nicolis}}, \ and\
  \bibinfo {author} {\bibfnamefont {C.}~\bibnamefont {Vafa}},\ }\href {\doibase
  10.1088/1126-6708/2007/06/060} {\bibfield  {journal} {\bibinfo  {journal}
  {JHEP}\ }\textbf {\bibinfo {volume} {06}},\ \bibinfo {pages} {060} (\bibinfo
  {year} {2007})},\ \Eprint {http://arxiv.org/abs/hep-th/0601001}
  {arXiv:hep-th/0601001} \BibitemShut {NoStop}%
\bibitem [{\citenamefont {Harlow}\ \emph {et~al.}(2023)\citenamefont {Harlow},
  \citenamefont {Heidenreich}, \citenamefont {Reece},\ and\ \citenamefont
  {Rudelius}}]{Harlow:2022ich}%
  \BibitemOpen
  \bibfield  {author} {\bibinfo {author} {\bibfnamefont {D.}~\bibnamefont
  {Harlow}}, \bibinfo {author} {\bibfnamefont {B.}~\bibnamefont {Heidenreich}},
  \bibinfo {author} {\bibfnamefont {M.}~\bibnamefont {Reece}}, \ and\ \bibinfo
  {author} {\bibfnamefont {T.}~\bibnamefont {Rudelius}},\ }\href {\doibase
  10.1103/RevModPhys.95.035003} {\bibfield  {journal} {\bibinfo  {journal}
  {Rev. Mod. Phys.}\ }\textbf {\bibinfo {volume} {95}},\ \bibinfo {pages}
  {035003} (\bibinfo {year} {2023})},\ \Eprint
  {http://arxiv.org/abs/2201.08380} {arXiv:2201.08380 [hep-th]} \BibitemShut
  {NoStop}%
\bibitem [{\citenamefont {Festuccia}\ and\ \citenamefont
  {Liu}(2009)}]{Festuccia:2008zx}%
  \BibitemOpen
  \bibfield  {author} {\bibinfo {author} {\bibfnamefont {G.}~\bibnamefont
  {Festuccia}}\ and\ \bibinfo {author} {\bibfnamefont {H.}~\bibnamefont
  {Liu}},\ }\href {\doibase 10.1166/asl.2009.1029} {\bibfield  {journal}
  {\bibinfo  {journal} {Adv. Sci. Lett.}\ }\textbf {\bibinfo {volume} {2}},\
  \bibinfo {pages} {221} (\bibinfo {year} {2009})},\ \Eprint
  {http://arxiv.org/abs/0811.1033} {arXiv:0811.1033 [gr-qc]} \BibitemShut
  {NoStop}%
\bibitem [{\citenamefont {Riojas}\ and\ \citenamefont
  {Sun}(2023)}]{Riojas:2023pew}%
  \BibitemOpen
  \bibfield  {author} {\bibinfo {author} {\bibfnamefont {M.}~\bibnamefont
  {Riojas}}\ and\ \bibinfo {author} {\bibfnamefont {H.-Y.}\ \bibnamefont
  {Sun}},\ }\href@noop {} {\  (\bibinfo {year} {2023})},\ \Eprint
  {http://arxiv.org/abs/2307.06415} {arXiv:2307.06415 [hep-th]} \BibitemShut
  {NoStop}%
\bibitem [{\citenamefont {Berenstein}\ \emph {et~al.}(2021)\citenamefont
  {Berenstein}, \citenamefont {Li},\ and\ \citenamefont
  {Simon}}]{Berenstein:2020vlp}%
  \BibitemOpen
  \bibfield  {author} {\bibinfo {author} {\bibfnamefont {D.}~\bibnamefont
  {Berenstein}}, \bibinfo {author} {\bibfnamefont {Z.}~\bibnamefont {Li}}, \
  and\ \bibinfo {author} {\bibfnamefont {J.}~\bibnamefont {Simon}},\ }\href
  {\doibase 10.1088/1361-6382/abcaeb} {\bibfield  {journal} {\bibinfo
  {journal} {Class. Quant. Grav.}\ }\textbf {\bibinfo {volume} {38}},\ \bibinfo
  {pages} {045009} (\bibinfo {year} {2021})},\ \Eprint
  {http://arxiv.org/abs/2009.04500} {arXiv:2009.04500 [hep-th]} \BibitemShut
  {NoStop}%
\bibitem [{\citenamefont {Dodelson}\ \emph {et~al.}(2024)\citenamefont
  {Dodelson}, \citenamefont {Iossa}, \citenamefont {Karlsson}, \citenamefont
  {Lupsasca},\ and\ \citenamefont {Zhiboedov}}]{Dodelson:2023nnr}%
  \BibitemOpen
  \bibfield  {author} {\bibinfo {author} {\bibfnamefont {M.}~\bibnamefont
  {Dodelson}}, \bibinfo {author} {\bibfnamefont {C.}~\bibnamefont {Iossa}},
  \bibinfo {author} {\bibfnamefont {R.}~\bibnamefont {Karlsson}}, \bibinfo
  {author} {\bibfnamefont {A.}~\bibnamefont {Lupsasca}}, \ and\ \bibinfo
  {author} {\bibfnamefont {A.}~\bibnamefont {Zhiboedov}},\ }\href {\doibase
  10.1007/JHEP07(2024)046} {\bibfield  {journal} {\bibinfo  {journal} {JHEP}\
  }\textbf {\bibinfo {volume} {07}},\ \bibinfo {pages} {046} (\bibinfo {year}
  {2024})},\ \Eprint {http://arxiv.org/abs/2310.15236} {arXiv:2310.15236
  [hep-th]} \BibitemShut {NoStop}%
\bibitem [{\citenamefont {Moitra}(2024)}]{Moitra:2023yyc}%
  \BibitemOpen
  \bibfield  {author} {\bibinfo {author} {\bibfnamefont {U.}~\bibnamefont
  {Moitra}},\ }\href {\doibase 10.1103/PhysRevD.109.L041903} {\bibfield
  {journal} {\bibinfo  {journal} {Phys. Rev. D}\ }\textbf {\bibinfo {volume}
  {109}},\ \bibinfo {pages} {L041903} (\bibinfo {year} {2024})},\ \Eprint
  {http://arxiv.org/abs/2305.08907} {arXiv:2305.08907 [hep-th]} \BibitemShut
  {NoStop}%
\bibitem [{\citenamefont {Paul}\ and\ \citenamefont
  {Bhamidipati}(2024)}]{Paul:2024khi}%
  \BibitemOpen
  \bibfield  {author} {\bibinfo {author} {\bibfnamefont {A.}~\bibnamefont
  {Paul}}\ and\ \bibinfo {author} {\bibfnamefont {C.}~\bibnamefont
  {Bhamidipati}},\ }\href@noop {} {\  (\bibinfo {year} {2024})},\ \Eprint
  {http://arxiv.org/abs/2404.07980} {arXiv:2404.07980 [hep-th]} \BibitemShut
  {NoStop}%
\end{thebibliography}%
\end{document}